\documentclass[12pt]{article}
\usepackage{natbib}
\usepackage{amssymb}
\usepackage{epsfig}
\usepackage{amsthm}
\usepackage{amsmath}
\usepackage{amsfonts}
\usepackage{color}
\usepackage{multirow}
\usepackage{todonotes}

\newcommand{\field}[1]{\mathbb{#1}}
\newcommand{\R}{\field{R}}

\newcommand{\Z}{\field{Z}}

\theoremstyle{example}
\theoremstyle{remark}
\theoremstyle{lemma}
\theoremstyle{definition}
\theoremstyle{corol}
\theoremstyle{proposition}
\theoremstyle{condition}
\theoremstyle{assumption}

\newtheorem{theorem}{\n{Theorem}}[section]
\newtheorem{example}{\n{Example}}[section]
\newtheorem{remark}{\n{Remark}}[section]

\newtheorem{algorithm}{\n{Algorithm}}[section]
\def\cod{\stackrel{\cal D}{\longrightarrow}}
\def\cop{\stackrel{\cal P}{\longrightarrow}}

\def\lf{\lfloor}
\def\rf{\rfloor}

\font\n=cmcsc10
\def\cov{{\mbox{cov}}}

\begin{document}
\begin{center} {\bf\Large Bootstrap-Assisted Unit Root Testing With Piecewise Locally Stationary Errors}\end{center}
\centerline{\textsc{Yeonwoo Rho\footnote{Yeonwoo Rho is Assistant Professor in Statistics at the Department of Mathematical Sciences, Michigan Technological University, Houghton, MI 49931. (Email: yrho@mtu.edu)} and Xiaofeng Shao\footnote{
Xiaofeng Shao is  Professor at the Department of Statistics, University of  Illinois, at Urbana-Champaign,  Champaign, IL 61820 (Email: xshao@illinois.edu).
}}}
\centerline {\it  Michigan Technological University$^1$}
\centerline {\it  University of Illinois at Urbana-Champaign$^2$}
\bigskip
\centerline{\today}
\bigskip

\begin{abstract}
In unit root testing, a piecewise locally stationary process is adopted to accommodate nonstationary errors that can have both smooth and abrupt changes in second- or higher-order properties.
Under this framework, the limiting null distributions of the conventional unit root test statistics are derived and shown to contain a number of unknown parameters. 
To circumvent the difficulty of direct consistent estimation, we propose to use the dependent wild bootstrap to approximate the non-pivotal limiting null distributions and provide a rigorous theoretical justification for  bootstrap consistency. The proposed method is compared through finite sample simulations with the recolored wild bootstrap procedure, which was developed for errors that follow a heteroscedastic linear process. Further, a combination of autoregressive sieve recoloring with the dependent wild bootstrap is shown to perform well.
The validity of the dependent wild bootstrap in a nonstationary setting is demonstrated for the first time,
showing the possibility of extensions
to other inference problems associated with locally stationary processes.
\\[5mm] JEL classification: C12, C22
\\[5mm] \noindent{KEY WORDS:} Heteroscedasticity; Locally stationary; Unit root testing; Wild bootstrap.
\end{abstract}

\newpage

\section{Introduction}\label{Introduction}
Unit root testing has received a lot of attention in econometrics since the seminal work by \cite{Dickey:Fuller:1979, Dickey:Fuller:1981}.
In their papers,  unit root tests were developed under the assumption of independent and identically distributed (i.i.d.) Gaussian errors.
Many variants of the Dickey-Fuller test have been proposed, when the error processes are stationary, weakly dependent, and free of the Gaussian assumption.
Most of the variants rely on  two fundamental approaches to accommodate  weak dependence in the error.
One approach is the Phillips-Perron test \citep{Phillips:1987a, Phillips:Perron:1988},
where the longrun variance of the error process is consistently estimated in a nonparametric way, using
heteroscedasticity and autocorrelation consistent estimators
\citep{Newey:West:1987, Andrews:1991}.
The other approach is the augmented Dickey-Fuller test \citep{Said:Dickey:1984}, which
approximates dependence structure in error processes with an AR($p$) model, where $p$ can grow with respect to the sample size.
In addition to these two popular methods and their variants, bootstrap-based tests were also proposed by \cite{Paparoditis:Politis:2002, Paparoditis:Politis:2003}, \cite{Chang:Park:2003},
\cite{Parker:Paparoditis:Politis:2006}, and
\cite{Cavaliere:Taylor:2009a}, among others.
For reviews and comparisons of some of these bootstrap-based methods, we refer to \cite{Paparoditis:Politis:2005} and \cite{Palm:Smeekes:Urbain:2008}.

Recently, it has been argued that many macroeconomic and financial series exhibit nonstationary 
behavior in the error.
In particular, heteroscedastic behavior in the error is well known in the unit root testing literature.
For instance, the U.S. gross domestic product series was observed to have less variability since the 1980s;
see \cite{Kim:Nelson:1999},  \cite{McConnell:Perez-Quiros:2000}, \cite{Busetti:Taylor:2003}, and references therein.
Also the majority of macroeconomic data in \cite{Stock:Watson:1999} exhibit heteroscedasticity in unconditional variances, as pointed out by \cite{Sensier:vanDijk:2004}.
If there are breaks in the error structure, it is known that  traditional unit root tests
are biased towards rejecting the unit root assumption \citep{Busetti:Taylor:2003}.
For this reason, a number of unit root tests that are robust to heteroscedasticity have been developed in the literature,
such as in \cite{Busetti:Taylor:2003}, \cite{Cavaliere:Taylor:2007, Cavaliere:Taylor:2008a, Cavaliere:Taylor:2008b, Cavaliere:Taylor:2009a}, and \cite{Smeekes:Urbain:2014}.
Such tests allow for smooth and abrupt changes in the unconditional or conditional variance in error processes.
However, changes in underlying dynamics do not have to be limited to  heteroscedastic behavior.
For example, in financial econometrics, it has been argued that a slow decay in sample autocorrelation in squared or absolute stock returns
might be due to a smooth change in its dynamics rather than a long-memory behavior.
Accordingly, \cite{Starica:Granger:2005} and \cite{Fryzlewicz_etal:2008}, among others, proposed to model stock returns series with locally stationary models.
A unit root test that is robust to changes in general  error dynamics would be useful in identifying a long-memory behavior in such financial series.

In a series of papers by Taylor and coauthors, heteroscedasticity in the error is accommodated by assuming that the error is generated from
linear processes with heteroscedastic innovations, i.e.,
\begin{equation}\label{GLP}
u_t=\sum_{j=0}^\infty c_je_{t-j},~~~~~~e_{t}=\omega_t\varepsilon_t.
\end{equation}
Here, $\varepsilon_t$ is assumed to be i.i.d. or to be a martingale difference sequence,
and  $\omega_t$ is a sequence of deterministic numbers that account for heteroscedasticity.
The process (\ref{GLP}) can be considered as a generalized linear process.
Although this generalization allows for some departures from stationarity,
it is still restrictive in the following three aspects.
First, this kind of linear process cannot accommodate nonlinearity in the error process and does not include nonlinear models that are popular in time series analysis,
such as threshold, bilinear and nonlinear moving average models.
Second, this error structure is somewhat special in that  temporal dependence and heteroscedasticity can be separated,
and it seems that most  methods developed to account for heteroscedasticity in the error take advantage of this special error structure.
Third,
changes in second- or higher-order properties in the error process $u_t$ are not as flexibly accommodated as those in $e_t$.
Recently, \cite{Smeekes:Urbain:2014} proposed a unit root test for a piecewise modulated stationary process $u_t=\omega_t v_t$,
where $v_t$ is a weakly stationary process and $\omega_t$ is a sequence of deterministic numbers that accounts for heteroscedasticity.
While a modulated stationary processes is more flexible in handling heteroscedasticity in $u_t$, it is still restrictive  in the sense that time dependence and  heteroscedasticity can be separated, similarly to (\ref{GLP}).
For more discussion on  the separability of (\ref{GLP}) and modulated stationary processes in the context of linear regression models with fixed
regressors and nonstationary errors, see \cite{Rho:Shao:2015}.

In this paper, a general framework of nonstationarity is adapted to capture
both smooth and abrupt changes in  second- or higher-order properties of the error process.
Specifically, the error process is assumed to follow a piecewise locally stationary  (PLS) process,
which was recently proposed by \cite{Zhou:2013},
as a generalization of a locally stationary process.
Locally stationary processes have received a lot of attention since the seminal works of \cite{Priestley:1965} and \cite{Dahlhaus:1997}.
Local stationarity naturally expands the notion of stationarity by allowing a change of the second-order properties of a time series;
see \cite{Dahlhaus:1997}, \cite{Mallat:Papanicolaou:Zhang:1998}, \cite{Giurcanu:Spokoiny:2004}, and \cite{Zhou:Wu:2009}, among others, for more related work.
However, locally stationary processes exclude abrupt changes in  second- or higher-order properties, which is often observed in real data.
To accommodate abrupt changes, PLS processes were proposed to allow for a finite number of breaks in addition to smooth changes.
For example, \cite{Adak:1998} proposed a PLS model in frequency domain, generalizing the  local stationary model due to \cite{Dahlhaus:1997}.
\cite{Zhou:2013}  proposed another PLS model in  time domain as an extension
of the framework of \cite{Zhou:Wu:2009} and \cite{Draghicescu:Guillas:Wu:2009}.
This PLS process allows for both nonlinearity and piecewise local stationarity and covers a wide range of processes;
see \cite{Wu:2005}, \cite{Zhou:Wu:2009}, and \cite{Zhou:2013} for more details.

Under the general PLS framework for the error, the limiting null distributions of the conventional unit root test statistics are not pivotal;
they depend on the local longrun variance
of the PLS error and some other nuisance parameters.
A direct consistent estimation of the unknown parameters in the limiting null distributions is very involved,
unlike the case of stationary errors \citep{Phillips:1987a}.
To overcome this difficulty, we propose to apply the dependent wild bootstrap (DWB) proposed in \cite{Shao:2010b}  to approximate the limiting null distributions.
We also provide a rigorous theoretical justification by establishing the  functional central limit theorem for the standardized partial sum process of the bootstrapped residuals.
This seems to be the first time  DWB is justified for  PLS processes and in the unit root setting.
It suggests the ability of  DWB to accommodate both piecewise local stationarity and weak dependence, which can be
potentially used for other inference problems related to locally stationary processes.

The rest of the paper is organized as follows.
Section \ref{Method} presents the model, along with the test statistics and their limiting distributions under the null and local alternatives.
In Section \ref{sec:boot},  DWB is described and its consistency is justified. The power behavior under local alternatives is also presented.
Section \ref{Simulation} presents some simulation results.
Section \ref{Conclusion} summarizes the paper.
Technical details are relegated to the supplementary material.

We now set some standard notation.
Throughout the paper,  $\cod$ is used for convergence in distribution and $\Rightarrow$ signifies weak convergence in $D[0,1]$,
the space of functions on [0,1] which are right continuous and have left limits, endowed with the Skorohod metric \citep{Billingsley:1968}.
Let $a_n\asymp c_n$ indicate $a_n/c_n\to1$ as $n\to\infty$.
For $a\in\field{R}$, $\lf a\rf$ denotes the largest integer smaller than or equal to $a$.
$B(\cdot)$  denotes a standard Brownian motion, and
$N(\mu,\Sigma)$  the (multivariate) normal distribution with mean $\mu$ and covariance matrix $\Sigma$.
Set  $||X||_p=(E|X|^p)^{1/p}$.
Let ${\bf 1}(\mathcal{E})$ be the indicator function, being 1 if the event $\mathcal{E}$ occurs and 0 otherwise.

\section{Unit root testing under piecewise locally stationary errors}\label{Method}
Following the framework in \cite{Phillips:Xiao:1998}, consider data $\{y_{1,n},\ldots,y_{n,n}\}$ generated from
\begin{equation}\label{truemodel1}
 y_{t,n}=X_{t,n}+\beta'z_{t,n},~~~~~~~t=0,1,\ldots,n,
 \end{equation}
 and
 \begin{equation}\label{truemodel2}
X_{t,n}=\rho X_{t-1,n}+u_{t,n},~~~t=1,2,\ldots,n.
\end{equation}
Here, $\beta$ is a $p\times 1$ vector of coefficients and $z_{t,n}$ is a $p\times 1$ vector of deterministic
trend functions, which satisfies
the following conditions:
\begin{itemize}
\item[(Z1)] There exists a sequence of scaling matrices $D_n$ and a piecewise continuous function $Z(r)$ such that
$D_n^{-1}z_{\lf nr\rf,n}\Rightarrow Z(r)$ as $n\to\infty$.
\item[(Z2)] $\int_0^1Z(r)Z(r)'dr$ is positive definite.
\end{itemize}
These assumptions include some popular trend functions, such as $(p-1)$st order polynomial trends,
and are quite standard in the literature;
see Section 2.1 of \cite{Phillips:Xiao:1998} and Section 2 of \cite{Cavaliere:Taylor:2007}.
The initial condition, $X_{0,n}=0$ is assumed in order to simplify the argument.
This assumption can be relaxed to allow, e.g.,  $X_{0,n}$ to be bounded in probability, which does not alter our asymptotic results.

Following the framework introduced by \cite{Zhou:2013}, the error process $\{u_{t,n}\}_{t=1}^n$ is assumed to be mean-zero piecewise locally stationary (PLS) with a finite number of  break points.
Let the break points be denoted by $b_1,b_2,\ldots,b_\tau$, where $0=b_0<b_1<\ldots<b_\tau<b_{\tau+1}=1$.
The process $\{u_{t,n}\}_{t=1}^n$ is considered as a concatenation of $\tau+1$ measurable functions
$G_j(s,\mathcal{F}_t):[0,1]\times\R^\infty\to\R$, $j=0,1,\ldots,\tau$, where
$$u_{t,n}=G_j(s_t,\mathcal{F}_t)~~~~{\rm if}~b_j\leq s_t< b_{j+1},$$
$s_t=t/n$, $\mathcal{F}_t=(\ldots,\varepsilon_0,\ldots,\varepsilon_{t-1},\varepsilon_t)$, and  the $\varepsilon_t$ are i.i.d. random variables
with mean 0 and variance 1.
The following is  further assumed:
\begin{itemize}
\item[(A1)] For each $j=0,1,\ldots,\tau$, the function $G_j(s,\mathcal{F}_t)$ is stochastically Lipschitz continuous with respect to  $s$.
That is, there exists a finite constant $C$ such that, for $s,s'\in[b_j,b_{j+1}],~s\neq s'$,
$$\frac{||G_j(s,\mathcal{F}_0)-G_j(s',\mathcal{F}_0)||_2}{|s-s'|}\leq C.$$
\item[(A2)] $\displaystyle\max_{j\in\{0,1,\ldots,\tau\}}\sup_{s\in[b_j,b_{j+1}]}||G_j(s,\mathcal{F}_0)||_4<\infty$.
\item[(A3)] $\delta_4(k)=O(\chi^k)$ for some $\chi\in(0,1)$, where $\delta_p(k)$ is the physical dependence measure defined as
$$\delta_p(k)=\max_{j\in\{0,1,\ldots,\tau\}}\sup_{s\in[b_j,b_{j+1}]}||G_j(s,\mathcal{F}_k)-G_j\{s,(\mathcal{F}_{-1},\varepsilon_0',\varepsilon_1,\ldots,\varepsilon_k)\}||_p$$
if $k\geq 0$, and $\delta_p(k)=0$ if $k<0$.
Here, $\varepsilon_0'$ is an i.i.d. copy of $\varepsilon_0$.
\item[(A4)] $\inf_{s\in[0,1]}\sigma^2(s)>0$, where $\sigma^2(s)=\sum_{h=-\infty}^\infty c_j(s;h)$ is the longrun variance function,
$c_j(s;h)=\cov\{G_j(s,\mathcal{F}_0),G_j(s,\mathcal{F}_h)\}$ for  $s\in[b_j,b_{j+1})$,
and $\sigma^2(1)=\lim_{s\uparrow1}\sigma^2(s)$.
\end{itemize}
If there is no break point and the function $G$ does not depend on its first argument, then the PLS process reduces to
a nonlinear causal process $G(\mathcal{F}_t)$, which can accommodate a wide range of stationary processes.
A special example is when $G$ is a linear function, in which case $G(\mathcal{F}_t)=\sum_{j=0}^\infty c_j\varepsilon_{t-j}$ is the commonly used linear process.
For nonlinear time series models that fall into the above framework of a nonlinear causal process, see \cite{Shao:Wu:2007}.

By introducing the dependence on the relative location $t/n$, the PLS series naturally extends this stationary causal process $G(\mathcal{F}_t)$ to a locally stationary one; see \cite{Zhou:Wu:2009}. In particular, assumption (A1) states that if $t/n$ and $t'/n$ are close and there is no break point in between, then $u_{t,n}$ and $u_{t',n}$
are expected to be stochastically close.
In other words, the second- or higher-order property of  $u_{t,n}$ should be smoothly changing, except for a finite number of break points.
This ensures the local stationarity between break points. As stated in \cite{Zhou:2013}, the goal of extending from locally stationary processes to PLS processes is to
allow for abrupt changes in both second- and high-order properties, and to accommodate both nonlinearity and nonstationarity in a broad fashion.

The physical dependence measure in (A3) was introduced in \cite{Zhou:Wu:2009} as an extension of its stationary counterpart first introduced by \cite{Wu:2005}.
Assumption (A3) implies that $u_{t,n}$ is locally short-range dependent and that the dependence decays exponentially fast.
When the location $s\in[0,1]$ is fixed, the process $\{G_j(s,\mathcal{F}_t)\}_{t\in\Z}$ is stationary for each $j$, and
(A4) introduces the time-varying longrun variance parameter $\sigma^2(s)$.
Assumptions (A1)--(A4) are similar to those of \cite{Zhou:Wu:2009}, \cite{Wu:Zhou:2011}, and \cite{Zhou:2013}.
These assumptions are not the weakest possible for our theoretical results to hold but are satisfied by a wide class of time series models.
See the following examples:

\begin{example}\label{ex1}{\rm 
The PLS framework includes a time-varying linear process as a special case.
Suppose  $s_t=t/n$ lies in the $j$th segment of [0,1], i.e., $s_t\in[b_j,b_{j+1})$.
Let the measurable functions $G_j(s_t,\mathcal{F}_t)$ be linear functions with respect to $\mathcal{F}_t=(\ldots,\varepsilon_{t-1},\varepsilon_t)$.
Then we can write
\begin{equation}\label{PLSLP}
u_{t,n}=G_j(s_t,\mathcal{F}_t)=\sum_{i=0}^\infty\psi_{i,j}(s_t)\varepsilon_{t-i},~~~~~~s_t\in[b_j,b_{j+1}),
\end{equation}
where  the $\varepsilon_t$ are i.i.d. (0,1).
If the following assumptions (LP1)--(LP4) are satisfied, the process in (\ref{PLSLP}) is a PLS process.
\begin{itemize}
\item[(LP1)] There exists a finite constant $C$ such that 
$$\displaystyle\max_{j=0,\cdots,\tau}\sup_{s\not=s'\in [b_j,b_{j+1}]}|\psi_{i,j}(s)-\psi_{i,j}(s')|/|s-s'|\le C~~~~{\rm for}~i=0,1,\dots.$$
\item[(LP2)] $||\varepsilon_{t}||_4<\infty$.
\item[(LP3)] $\displaystyle\max_{j=0,\cdots,\tau}\sup_{s\not=s'\in [b_j,b_{j+1}]}|\psi_{k,j}(s)|=O(\chi^k)$
for some $\chi\in(0,1)$.
\item[(LP4)] $\displaystyle\inf_{s\in[0,1]}\sigma^2(s)=\inf_{j=0,\cdots,\tau}\inf_{s\in[b_j,b_{j+1}]}\sum_{h=-\infty}^\infty\sum_{i=0}^\infty\psi_{i,j}(s)\psi_{i+h,j}(s)>0$.
\end{itemize}
Under (\ref{PLSLP}), it is not difficult to show that (LP1) implies (A1); (LP2) and (LP3) imply (A2) and (A3); and (LP4) implies (A4).
It is worth noting that the heteroscedastic linear process  in (\ref{GLP}) can be  expressed as a special example of
 our framework in (\ref{PLSLP}) by letting
$$\psi_{i,j}(s_t)=c_i \omega(s_t),$$
where $\omega(s_t)$ is a non-stochastic and strictly positive function on [0,1],
representing the standard deviation of the error term of the linear process at the relative location $s_t=t/n$.
\cite{Cavaliere:Taylor:2008b} further assumed
\begin{itemize}
\item[(CT1)] $\sum_{i=0}^\infty i|c_i|<\infty$.
\item[(CT2)] $C(z)=\sum_{i=0}^\infty c_iz^i\neq 0$ for all $|z|\leq 1$.
\end{itemize}
As long as $\omega(s)$ is bounded for $s\in[0,1]$, (CT1) can be equivalently written as
\begin{equation*}\label{cmt:eq1}
\max_{0\leq j\leq \tau}\sup_{b_j\leq s\leq b_{j+1}} \sum_{i=0}^\infty i|\psi_{i,j}(s)|<\infty,
\end{equation*}
which is weaker than (LP3). Note that our sufficient assumptions (A1)--(A4) are not necessary and can be relaxed at the expense of
 lengthy technical details. The PLS framework allows for locally stationary nonlinear processes as detailed below, and our technical argument is considerably
different from that in \cite{Cavaliere:Taylor:2008b}.
The latter relies on  theoretical results in \cite{Chang:Park:2002} and the  Beveridge-Nelson Decomposition [\cite{Phillips:Solo:1992}], which are tailored to
 linear processes.
}
\end{example}

\begin{example}\label{ex2}{\rm
The PLS process accommodates time-varying nonlinear models.
For example, the autoregressive conditional heteroscedasticity (ARCH(1)) model with time-varying coefficients \citep{Dahlhaus:Rao:2006} can be represented in the PLS form.
For simplicity, assume there are no break points.
Define $G(s,\mathcal{F}_t)=H\{s,G(s,\mathcal{F}_{t-1}),\varepsilon_t\}$, where $H:[0,1]\times\R\times\R\to\R$ satisfies
$$H\{s,G(s,\mathcal{F}_{t-1}),\varepsilon_{t}\}=\{w(s)+\alpha(s)G(s,\mathcal{F}_{t-1})^2\}^{1/2}\varepsilon_{t}$$
with $\varepsilon_t\stackrel{i.i.d.}{\sim}(0,1)$. Under  standard assumptions on the smoothness and boundedness of $w(\cdot)$ and
 $\alpha(\cdot)$, and moment assumptions on $\varepsilon_t$, it can be shown that the time-varying ARCH(1) satisfies (A1)--(A4); see Proposition 5.1 in \cite{Shao:Wu:2007}
  and Assumption 2 in \cite{Dahlhaus:Rao:2006}. 
  Additionally, as mentioned in Zhou (2013), many stationary nonlinear time series models naturally fall into the framework of $G({\cal F}_t)$ and can be extended
  to piecewise stationary nonlinear models by introducing break points and allowing different nonlinear models within each segment. This flexibility of modeling
  complex dynamics of time series is automatically built into the PLS process.

}
\end{example}

\begin{remark}{\rm
The PLS framework is based on the physical dependence measure in (A3), whereas the mixing conditions can be understood as being based on a more abstract probabilistic dependence measure, and neither is inclusive of the other.
On the one hand, special cases of PLS processes can be shown to satisfy a mixing condition.
For example, the time-varying ARCH process introduced in Example 2.2 can be shown to
be $\alpha$-mixing  under some appropriate conditions on $\omega(s)$ and $\alpha(s)$ and the smoothness of the density of the $\varepsilon_t$, using Theorem 3.1 of \cite{Fryzlewicz:SubbaRao:2011}.
On the other hand, examples that are PLS but not strong-mixing can also be found.
For instance,  an autoregressive (AR) process of order 1 with  AR parameter 0.5 and Bernoulli innovations is known to violate the strong-mixing conditions    but
can easily fit into the physical dependence measure framework \citep{Andrews:1984,Wu:2005}.
}
\end{remark}

Given the observations $\{y_{t,n},z_{t,n}\}_{t=1}^n$, consider testing the unit root hypothesis
\begin{equation*}\label{hyp}
H_0:~\rho=1~~~~~{\rm vs}~~~~~H_1:~|\rho|< 1.\end{equation*}
The ordinary least squares (OLS) estimator $\widehat{\rho}_n=\left(\sum_{t=1}^n\widehat{X}_{t,n}\widehat{X}_{t-1,n}\right)/\left(\sum_{t=1}^n\widehat{X}_{t-1,n}^2\right)$ of $\rho$ is considered,
where $\widehat{X}_{t,n}=y_{t,n}-\widehat{\beta}_n'z_{t,n}$ are the OLS residuals of $y_{t,n}$ regressed on $z_{t,n}$.\footnote{It is worth mentioning
that the generalized least squares (GLS) detrending can be used instead of the OLS detrending to make the tests more powerful.
This  is quite straightforward and well established in the literature \citep{Elliott_etal:1996,Muller:Elliott:2003,Smeekes:2013}, and will not be pursued in this paper.}
We proceed to define  two test statistics  that are popular in the literature, namely
$$\mathbf{T}_n=n(\widehat{\rho}_n-1)~~~~{\rm and}~~~~
\mathbf{t}_n
=\frac{\left(\sum_{t=1}^n\widehat{X}_{t-1,n}^2\right)^{1/2}\left(\widehat{\rho}_n-1\right)}{(s_n^2)^{1/2}},
$$
where $s_n^2=(n-2)^{-1}\sum_{t=1}^n\left(\widehat{X}_{t,n}-\widehat{\rho}_n\widehat{X}_{t-1,n}\right)^2$.

In the unit root testing literature, local alternatives, i.e., $\rho_n=1+c/n$, $c<0$, are often considered
to examine the behavior of the test when the true $\rho$ is close to the unity.
The Ornstein-Uhlenbeck process, $J_c(r)=\int_0^re^{(r-s)c}dB(s)$, is usually involved in the limiting distributions of
$\mathbf{T}_n$ and $\mathbf{t}_n$ for this near-integrated case.
Under our error assumptions, define a similar process, $J_{c,\sigma}(r):=\int_0^re^{(r-s)c}\sigma(s)dB(s)$.
Theorem \ref{thm:NI} below states the limiting distributions of the two test statistics under the null hypothesis, $\rho=1$, and under  local alternatives, $\rho_n=1+c/n$, $c<0$.
\begin{theorem}\label{thm:NI}
Assume (A1)--(A4) and (Z1)--(Z2). When $\rho=\rho_n=1+c/n$, $c\leq0$,
\begin{equation}\label{thm:NI:ut}
n^{-1/2}\widehat{X}_{\lf nr\rf,n}\Rightarrow J_{c,\sigma|Z}(r),
\end{equation}
where
$J_{c,\sigma|Z}(r)=
J_{c,\sigma}(r)-\{\int_0^1J_{c,\sigma}(s) Z(s)'ds\}\{\int_0^1 Z(s)Z(s)'ds\}^{-1}Z(r)$
is the Hilbert projection of $J_{c,\sigma}(\cdot)$ onto the space orthogonal to $Z(\cdot)$.
The limiting distributions of the two statistics are
\begin{equation}\label{thm:NI:ZI}
\mathbf{T}_n
\cod \mathcal{L}_{\mathbf{T},c}=
\frac{\int_0^1J_{c,\sigma|Z}(r)\sigma(r)dB(r)+2^{-1}\{\int_0^1\sigma^2(r)dr-\sigma_u^2\}}{\int_0^1J_{c,\sigma|Z}^2(r)dr}+c
\end{equation}
and
\begin{equation}\label{thm:NI:zI}
\mathbf{t}_n
\cod  \mathcal{L}_{\mathbf{t},c}
=\frac{\int_0^1J_{c,\sigma|Z}(r)\sigma(r)dB(r)+2^{-1}\{\int_0^1\sigma^2(r)dr-\sigma_u^2\}}{\{\sigma_u^2\int_0^1J_{c,\sigma|Z}^2(r)dr\}^{1/2}}+
\frac{c\{\int_0^1J_{c,\sigma|Z}^2(r)dr\}^{1/2}}{\sigma_u},
\end{equation}
where $\sigma_u^2=\lim_{n\to\infty}n^{-1}\sum_{t=1}^nE(u_{t,n}^2)$.
\end{theorem}
When $\rho$ is far from the unity,  with $c<0$ and large in absolute value,
the limiting distributions are  far to the left compared to the limiting null distributions.
In this case, the unit root null hypothesis would be rejected with high probability.
This implies that the unit root tests based on $\mathbf{T}_n$ and $\mathbf{t}_n$ have nontrivial powers under local alternatives.
In the special case $\beta\equiv0$ and $\sigma(s)=\sigma$, i.e., when there is no deterministic trend and the error is stationary,
the two limiting distributions $\mathcal{L}_{\mathbf{T},c}$ and $\mathcal{L}_{\mathbf{t},c}$ reduce to those found in Theorem 1 of \cite{Phillips:1987b}.

Notice that when $c=0$, i.e., under the null hypothesis,  $J_{c,\sigma|Z}(r)=B_{\sigma|Z}(r)$, which implies
$\int_0^1B_{\sigma|Z}(r)\sigma(r)
\allowbreak
dB(r)=2^{-1}\{B_{\sigma|Z}^2(1)-\int_0^1\sigma^2(r)dr\}$,
by It\^{o}'s formula.\footnote{Ito's formula can be written as
$dB_\sigma(r)=\sigma(r)dB(r)$. Using It\^{o}'s formula, we derive $B_\sigma^2(r)=B_\sigma^2(0)-\int_0^r2\sigma(s)B_\sigma(s)dB(s)+2^{-1}\int_0^r2\sigma^2(s)ds$,
which leads to $\int_0^rB_{\sigma}(s)\sigma(s)dB(s)=2^{-1}\{B_{\sigma}^2(r)-\int_0^r\sigma^2(s)ds\}$.}
In this case, the limiting null distributions can be written as
\begin{equation*}\label{thm:OLS:ZI}
\mathcal{L}_{\mathbf{T}}:=\mathcal{L}_{\mathbf{T},0}=
\frac{\{B_{\sigma|Z}(1)^2-\sigma_u^2\}}{2\int_0^1B_{\sigma|Z}(r)^2dr}
~~~~~{\rm and}~~~~~ \mathcal{L}_{\mathbf{t}}:=\mathcal{L}_{\mathbf{t},0}=
\frac{\{B_{\sigma|Z}(1)^2-\sigma_u^2\}}{2\{\sigma_u^2\int_0^1B_{\sigma|Z}(r)^2dr\}^{1/2}},
\end{equation*}
where
 $B_\sigma(r)=\int_0^r\sigma(s)dB(s)$ and
$$B_{\sigma|Z}(r)=B_\sigma(r)-\left\{\int_0^1B_\sigma(s) Z(s)'ds\right\}\left\{\int_0^1 Z(s)Z(s)'ds\right\}^{-1}Z(r)$$
is the Hilbert projection of $B_\sigma(\cdot)$ onto the space orthogonal to $Z(\cdot)$.

Since the limiting null distributions contain a number of unknown parameters, one may try to directly estimate them for inferential purpose.
Consistent estimation of the limit of the average marginal variance, $\sigma_u^2$, is not difficult.
This can be done by noting that  $\sigma_u^2=\int_0^1c(s;0)ds$,
with $c(s;h)$ defined in (A4) and in the first paragraph of the technical appendix.
As a special case of Lemma A.5, 
$\sigma_u^2$ can be consistently estimated by $n^{-1}\sum_{t=1}^n \widehat{u}_{t,n}^2$, where $\widehat{u}_{t,n}=\widehat{X}_{t,n}-\widehat{\rho}_n\widehat{X}_{t-1,n}$
is the OLS residual.
However, consistent estimation of the local longrun variance $\sigma^2(s)$ is not as simple for a PLS process.
In the  case of a stationary error process, $c(s;h)$ does not depend on $s$,
and  $\sigma^2(s)=\sum_{h=-\infty}^\infty \cov(u_{t},u_{t+h})=\sigma^2$.
If it is further assumed that there is  no deterministic trend function, i.e., $\beta\equiv0$, then the limiting null distributions
$\mathcal{L}_{\mathbf{T}}$ and $\mathcal{L}_{\mathbf{t}}$
reduce to
$$\mathcal{L}_{\mathbf{T},\sigma(s)=\sigma}=\frac{\{B(1)^2-\sigma_u^2/\sigma^2\}}{2\int_0^1B(r)^2dr}~~~{\rm and}~~~
\mathcal{L}_{\mathbf{t},\sigma(s)=\sigma}=\frac{\sigma/\sigma_u\{B(1)^2-\sigma_u^2/\sigma^2\}}{2\{\int_0^1B(r)^2dr\}^{1/2}}.$$
These limiting null distributions contain only a couple of unknown parameters and coincide with those in \cite{Phillips:1987a}.
To make  inference possible in this stationary error case, as \cite{Phillips:1987a} and Phillips and Perron (1998) suggested,
the longrun variance $\sigma^2$ of the error process may be consistently estimated
using heteroscedasticity and autocorrelation consistent (HAC) estimators.
The new statistics (see page 287 of \cite{Phillips:1987a}), adjusted by using  consistent estimates of $\sigma$ and $\sigma_u$, have pivotal limiting null distributions.
However, in the piecewise locally stationary error case, the usual Phillips-Perron adjustment may not lead to
pivotal limiting null distributions.
Specifically, the HAC-based estimator of the nuisance parameter $\sigma^2(s)$, $s\in[0,1]$, is not known to be consistent in the PLS framework.
The parameter $\sigma^2(s)$ is unknown at infinitely many points, and the integral of $\sigma(s)$ over a Brownian motion needs to be estimated as well as  $\sigma^2(s)$ itself.
This makes the direct estimation of the unknown parameters in the limiting null distributions difficult.

\bigskip

\section{Bootstrap-assisted unit root test}\label{sec:boot}

To implement  the (asymptotic) level $\alpha$ test, the $\alpha$-quantiles of the limiting null distributions $\mathcal{L}_{\mathbf{T}}$ and $\mathcal{L}_{\mathbf{t}}$
need to be identified and estimated.
However, it is difficult to consistently estimate
the unknown parameters $\sigma(s)$ for all $s\in[0,1]$.
As an alternative, we shall use a bootstrap method to approximate the limiting null distributions.
When the errors are stationary, it is well known that the Phillips-Perron test or the augmented Dickey-Fuller test have size distortions in finite samples,
even though they  have been proven to work asymptotically.
Bootstrap-based methods have been proposed to improve the finite sample performance.
\cite{Psaradakis:2001}, \cite{Chang:Park:2003}, and \cite{Palm:Smeekes:Urbain:2008}
used the sieve bootstrap \citep{Kreiss:1988} assuming an infinite order AR structure for the error process.
\cite{Paparoditis:Politis:2003} applied the block bootstrap \citep{Kunsch:1989}, which randomly samples from overlapping blocks of residuals.
\cite{Swensen:2003} and \cite{Parker:Paparoditis:Politis:2006} extended the stationary bootstrap \citep{Politis:Romano:1994} to unit root testing,
where not only are  overlapping blocks  randomly chosen, but also the block size is chosen from a geometric distribution.
\cite{Cavaliere:Taylor:2009a} applied the wild bootstrap \citep{Wu:1986} for unit root M tests \citep{Perron:Ng:1996},
which are modifications of the Phillips-Perron test.

To accommodate both heteroscedasticity and temporal dependence in the error,
bootstrap-based methods have been developed by \cite{Cavaliere:Taylor:2008b} and \cite{Smeekes:Taylor:2012}.
In their papers, the error is assumed to be a linear process with heteroscedastic innovations as in equation (\ref{GLP}), and the wild bootstrap \citep{Wu:1986} was used
along with an AR sieve procedure to filter out the dependence in the error. The combination of an autoregressive sieve (or recolored) filter and wild bootstrap handles heteroscedasticity and serial correlation simultaneously, but its theoretical validity strongly depends upon the linear process assumption on the error.
We speculate that this recolored wild bootstrap (RWB) will not work for PLS errors because the naive wild bootstrap
can account for heteroscedasticity, but not for weak temporal dependence that is not completely filtered out after applying the  AR sieve.
The wild bootstrap works in the framework of (\ref{GLP}), since
 temporal dependence is removed by
the Phillips-Perron adjustment \citep{Cavaliere:Taylor:2008b} or the augmented Dickey-Fuller adjustment (i.e., the AR sieve) \citep{Cavaliere:Taylor:2009a,Smeekes:Taylor:2012}.
This type of removal is only possible under the assumption that the error is a heteroscedastic linear process,
in which case
the longrun variance $\sigma^2(s)$ can be factored into two parts;
one part is due to heteroscedasticity in  the innovations $e_t$, and the other part
is due to temporal dependence (i.e., $\sum_{j=0}^\infty c_j$) in the error, as shown recently by \cite{Rho:Shao:2015}.
As a result,  limiting null distributions after the two popular adjustments for temporal dependence
depend only on heteroscedasticity, which can be handled by the wild bootstrap.
However, for  PLS error processes, even after the Phillips-Perron or the augmented Dickey-Fuller adjustment,
the limiting null distributions are still affected by temporal dependence in the error.
Therefore, RWB is not expected to work in our setting.

To accommodate  nonstationarity and temporal dependence in the error,
we propose to adopt the so-called dependent wild bootstrap (DWB),
which was first introduced by \cite{Shao:2010b} in the context of stationary time series.
It turns out that DWB is capable of mimicking local weak dependence in the error process and provides a consistent approximation of the limiting null
distributions of $\mathbf{T}_n$ and $\mathbf{t}_n$.
Note that DWB was developed for stationary time series and its applicability was only proved for smooth function models.
\cite{Smeekes:Urbain:2014} recently proved the validity of several modified wild bootstrap methods, including DWB,  for  modulated stationary errors in a multivariate setting.
 However, as discussed in the introduction,
the modulated stationary process is somewhat restrictive due to its separable structure of temporal dependence and heteroscedasticity in its longrun variance.
Instead, our  PLS framework is considerably more
general in allowing both abrupt and smooth change in second- and higher-order properties. From a technical viewpoint, our proofs seem more involved than theirs due to the general
error framework we adopt.

In the implementation of DWB, pseudo-residuals are generated by perturbing  the original (OLS) residuals using a set $\{W_{t,n}\}_{t=1}^n$ of external variables.
The difference between DWB and the original wild bootstrap is that $\{W_{t,n}\}_{t=1}^n$ is
made to be dependent in DWB, whereas $\{W_{t,n}\}_{t=1}^n$ is assumed to be independent in the usual wild bootstrap.
The following assumptions on $\{W_{t,n}\}_{t=1}^n$ are from \cite{Shao:2010b}:
\begin{itemize}
\item[(B1)] $\{W_{t,n}\}_{t=1}^n$ is a realization from a stationary time series with $E(W_{t,n})=0$ and ${\rm var}(W_{t,n})=1$.
$\{W_{t,n}\}_{t=1}^n$ are independent of the data,  ${\rm cov}(W_{t,n},W_{t',n})=a\{(t-t')/l\}$, where $a(\cdot)$ is a kernel function
and $l=l_n$ is a bandwidth parameter that satisfies $l\asymp Cn^{\kappa}$ for some $0<\kappa<1/3$.
Assume that $W_{t,n}$ is $l$-dependent and $E(W_{1,n}^4)<\infty$.
\item[(B2)]  $a: \R\to[0,1]$ is symmetric and has compact support on [-1,1], $a(0)=1$,
$\lim_{x\to0}\{1-a(x)\}/|x|^q=k_q\neq0$ for some $q\in(0,2]$, and $\int_{-\infty}^{\infty}a(u)e^{-iux}du\geq0$ for $x\in\R$.
\end{itemize}
In practice,  $\{W_{t,n}\}_{t=1}^n$ can be sampled from a multivariate normal distribution with mean zero and  covariance function
${\rm cov}(W_{t,n},W_{t',n})=a\{(t-t')/l\}$.
There are two user-determined parameters: a kernel function $a(\cdot)$ and a bandwidth parameter $l$.
The kernel function affects the performance to a lesser degree than the bandwidth parameter $l$,
and the choice of $l$ will be discussed in Section \ref{Simulation}.
For the kernel function, some commonly used kernels, such as the Bartlett kernel, satisfy (B2).

The DWB algorithm in unit root testing is  as follows:

\begin{algorithm}\label{algorithm:DWB}{\rm [The Dependent Wild Bootstrap (DWB)]

\begin{enumerate}
\item Calculate the OLS estimate $\widehat{\beta}_n$ of
$\beta$ by fitting $y_{t,n}$ on $z_{t,n}$, and let
$\widehat{X}_{t,n}=y_{t,n}-\widehat{\beta}_n'z_{t,n}$.
\item Let $\widehat{\rho}_n$ be the OLS estimate of
$\widehat{X}_{t,n}$ on $\widehat{X}_{t-1,n}$.
Calculate the
statistics $\mathbf{T}_n=n(\widehat{\rho}_n-1)$ and
$\mathbf{t}_n=(\sum_{t=1}^n\widehat{X}_{t-1,n}^2)^{1/2}(\widehat{\rho}_n-1)/s_n$.
\item Calculate the residuals
$\widehat{u}_{t,n}=\widehat{X}_{t,n}-\widehat{\rho}_n\widehat{X}_{t-1,n}$
for all $t=1,\ldots,n$.
\item Randomly generate the $l$-dependent mean-zero stationary series $\{W_{t,n}\}_{t=1}^n$ satisfying conditions (B1)--(B2)
and generate the perturbed residuals ${u}_{t,n}^*=\widehat{u}_{t,n}W_{t,n}$.
\item
Construct the bootstrapped sample $y_{t,n}^*$ using
${u}_{t,n}^*$ as if $\rho=1$ is true:
$$(y_{t,n}^*-\widehat{\beta}_n'z_{t,n})=(y_{t-1,n}^*-\widehat{\beta}_n'z_{t-1,n})+{u}_{t,n}^*,~~~~~~t=2,\ldots,n,$$
and
$y_{1,n}^*=\widehat{\beta}_n'z_{1,n}+{u}_{1,n}^*$.
\item
Calculate $\widehat{\beta}_n^*$ by refitting $y_{t,n}^*$ on
$z_{t,n}$, and let $\widehat{X}_{t,n}^*=y_{t,n}^*-(\widehat{\beta}_n^*)'z_{t,n}$.
\item Calculate bootstrapped versions of $\widehat{\rho}_n$ and $s_n^2$, i.e.,
$\widehat{\rho}_n^*$ and ${s_n^*}^2$, based on
$\{\widehat{X}_{t,n}^*\}_{t=1}^n$, and the bootstrapped test statistics
$\mathbf{T}_n^*=n(\widehat{\rho}_n^*-1)$ and
$\mathbf{t}_n^*=\{\sum_{t=1}^n(\widehat{X}_{t-1,n}^*)^2\}^{1/2}(\widehat{\rho}_n^*-1)/s_n^*$.
\item Repeat steps 2--7 B times, and record the
bootstrapped test statistics
$\{\mathbf{T}_n^{*(1)},\ldots,\mathbf{T}_n^{*(B)}\}$ and
$\{\mathbf{t}_n^{*(1)},\ldots,
\allowbreak
\mathbf{t}_n^{*(B)}\}$. The p-values
are
$$\frac{\sum_{b=1}^B{\bf 1}\{\mathbf{T}_n^{*(b)}<\mathbf{T}_n\}}{B}~~~{\rm and}~~~~\frac{\sum_{b=1}^B{\bf 1}\{\mathbf{t}_n^{*(b)}<\mathbf{t}_n\}}{B}.$$
\end{enumerate}
}
\end{algorithm}

\begin{remark}{\rm
Notice that the null hypothesis is not enforced in step 3 of Algorithm \ref{algorithm:DWB}, i.e.,
 unrestricted residuals are used in the construction of the bootstrap samples.
Another approach constructing the bootstrap sample is discussed in \cite{Paparoditis:Politis:2003},
where the null hypothesis is imposed in step 3.
Both procedures are consistent under the null hypothesis, but as observed in \cite{Paparoditis:Politis:2003} for their residual block bootstrap,
unrestricted residuals deliver higher power. The same phenomenon was also observed  for DWB-based tests in our (unreported) simulations, so we shall not
consider the restricted residual case in detail.

}
\end{remark}

The following theorem provides the core result in the proof of the consistency of DWB in Theorem \ref{thm:NI:boot}
and may be of independent interest.
\begin{theorem}\label{keylem}
Assume (A1)--(A4), (Z1)--(Z2), and (B1)--(B2). For any $\rho=1+c/n$, $c\leq 0$, 
$$n^{-1/2}\sum_{t=1}^{\lf nr\rf} {u}_{t,n}^*\Rightarrow B_{\sigma|Z}(r)~~~~~~{\it in~probability}.$$
\end{theorem}
\noindent Note that Theorem \ref{keylem} holds not only under the null hypothesis $\rho=1$ but also under  local alternatives.
This property makes the DWB  method 
powerful because the bootstrapped distributions correctly
mimic the limiting null distributions under both the null and local alternatives.
The DWB method can still correctly approximate the limiting null distribution under local alternatives, mainly because $y_{t,n}^*$ are constructed assuming $\rho=1$ in step 5.

\begin{theorem}[Bootstrap Consistency and Power]\label{thm:NI:boot}
Assume (A1)--(A4), (Z1)--(Z2), and (B1)--(B2).
For any $c\leq 0$,
$$P(\mathbf{T}_n\leq \mathbf{T}^*_{n,\alpha}|\rho=1+c/n)\cop P(L_{T,c}\leq \mathcal{L}_{\mathbf{T}}^\alpha),$$
$$P(\mathbf{t}_n\leq \mathbf{t}^*_{n,\alpha}|\rho=1+c/n)\cop P(L_{t,c}\leq \mathcal{L}_{\mathbf{t}}^\alpha),$$
where $L_{T,c}$ and $L_{t,c}$ are random variables with distribution $\mathcal{L}_{\mathbf{T},c}$ and  $\mathcal{L}_{\mathbf{t},c}$, respectively,
which are defined in Theorem \ref{thm:NI}.
$\mathcal{L}_{\mathbf{T}}^\alpha$ and $\mathcal{L}_{\mathbf{t}}^\alpha$ are the $\alpha$-quantiles of the limiting null distributions,
$\mathcal{L}_{\mathbf{T}}$ and $\mathcal{L}_{\mathbf{t}}$, respectively.
 $\mathbf{T}^*_{n,\alpha}$ and $\mathbf{t}^*_{n,\alpha}$ are the $\alpha$-quantiles of
the  distributions of $\mathbf{T}_n^*$ and $\mathbf{t}_n^*$ conditional on the data, respectively.
\end{theorem}
\noindent Under the null hypothesis, i.e., when $c=0$, Theorem \ref{thm:NI:boot} establishes the consistency of DWB in approximating the limiting null distributions.
Since the bootstrap statistics (asymptotically) replicate the exact null distribution when $\rho=1$, the (asymptotic) size of our unit root test would be exactly the same as the level of the test.
On the other hand, if $c$ is negative and  far from 0, the probability of rejecting the null, or the asymptotic power of the test, will be close to 1.
If $c$ is not 0 but not too far from 0, Theorem \ref{thm:NI:boot} states that the probability of rejecting the null
is somewhere between the level of the test and 1.
This means that  the DWB-based unit root tests have nontrivial power under local alternatives.

\begin{remark}\label{remark:DWB}
{\rm
The  DWB method was originally developed for stationary time series (Shao, 2010).
In the construction of DWB samples, $\{W_{t,n}\}_{t=1}^n$ is generated as $l$-dependent stationary time series,
so it is natural to expect that DWB would work for stationary time series.
However, it does not seem straightforward that this simple form of bootstrap would work in the case of a locally stationary process with unknown breaks.
What Theorem \ref{keylem} suggests is that DWB is capable of capturing nonstationary behaviors,
without the need to specify any parametric forms of error structures or to know the specific form of nonstationarity such as the location of breaks.
}
\end{remark}

\section{Simulations}\label{Simulation}
In this section, the DWB method is compared
with the recolored (sieve) wild bootstrap (RWB) method, which was proposed in \citet[Section 3.3]{Cavaliere:Taylor:2009a}.
We also propose to combine the AR sieve idea in RWB with DWB and present this method as the recolored dependent wild bootstrap (RDWB).
The RDWB statistics are based on the RWB statistics using DWB to determine the critical values of the tests, so RDWB  can be considered as a generalization of RWB.

Before introducing our simulation setting and results, we first present some details about
(i)  the RDWB algorithm
and (ii) the size-corrected power calculation similar to the one   in \cite{Dominguez:Lobato:2001}.

\bigskip

First, the RDWB procedure is described below. Rewrite equation  (\ref{truemodel2}) as
\begin{equation}\label{eq:ADF}
\Delta X_{t,n}=\pi_0 X_{t-1,n}+\sum_{i=1}^k\pi_i \Delta X_{t-i,n}+u_{t,n,k},
\end{equation}
where $\Delta$ represents the difference operator.

\begin{algorithm}{\rm [The Recolored Dependent Wild Bootstrap (RDWB)]
\begin{enumerate}
\item
 Calculate the OLS estimate $\widehat{\beta}_n$ of
$\beta$ by fitting $y_{t,n}$ on $z_{t,n}$, and let
$\widehat{X}_{t,n}=y_{t,n}-\widehat{\beta}_n'z_{t,n}$.
\item Choose the number $k$ of lags  using, for example, the modified Akaike
information criterion (MAIC) in \citet[p.1529]{Ng:Perron:2001}.
That is,
$\hat{k}={\rm argmin}_{0 \leq k \leq k_{max}} {\rm MAIC}(k)$, where
${\rm MAIC}(k)={\rm ln}(\hat{\sigma}_k^2)+{2\{\tau_n(k)+k\}}/{(n-k_{max})},$
$k_{max}=\lfloor 12(n/100)^{1/4}\rfloor$,
$\tau_n(k)=(\hat{\sigma}_k^2)^{-1}\hat{\pi}_0^2\sum_{t=k_{max}+1}^n \widehat{X}_{t-1,n}^2,$
and $\hat{\sigma}_k^2=(n-k_{max})^{-1}\sum_{t=k_{max}+1}^n\hat{u}_{t,n,k}^2.$
Here, $\hat{\pi}_0$ and $\hat{u}_{t,n,k}$ are the OLS estimators and residuals from (\ref{eq:ADF}) with $k=0,\ldots,k_{max}$.
Find the OLS estimators and residuals, i.e.,  $\hat{\pi}_i$ and $\hat{u}_{t,n,\hat{k}}$ from (\ref{eq:ADF}) with $k=\hat{k}$.

\item
Let $\widehat{\rho}_n$ be the OLS estimate of
$\widehat{X}_{t,n}$ on $\widehat{X}_{t-1,n}$.
Calculate the
statistics $\mathbf{T}_n=n(\widehat{\rho}_n-1)$ and
$\mathbf{t}_n=(\sum_{t=1}^n\widehat{X}_{t-1,n}^2)^{1/2}(\widehat{\rho}_n-1)/s_n$.
\item Generate the $l$-dependent mean-zero stationary series $\{W_{t,n}\}_{t=1}^n$ satisfying conditions (B1)--(B2)
and generate the perturbed residuals ${u}_{t,n,\hat{k}}^*=\hat{u}_{t,n,\hat{k}}W_{t,n}$.

\item Construct the bootstrapped sample $y_{t,n}^*$ using
$\{{u}_{t,n}^*\}$ under the unit root null hypothesis in (\ref{eq:ADF}), i.e., $\pi_0=0$, and recolor the bootstrapped residuals:
$$\Delta(y_{t,n}^*-\widehat{\beta}_n'z_{t,n})=\sum_{i=1}^{\hat{k}}\hat{\pi}_i\Delta (y_{t-i,n}^*-\widehat{\beta}_n'z_{t-i,n})+u_{t,n,\hat{k}}^*,~~~~~~t=\hat{k},\ldots,n,$$
 and
$y_{t,n}^*=\widehat{\beta}_n'z_{t,n}+{u}_{t,n,\hat{k}}^*$ for $t=1,\ldots,\hat{k}-1$.
  \item
Calculate $\widehat{\beta}_n^*$ by refitting $y_{t,n}^*$ on
$z_{t,n}$, and let
$\widehat{X}_{t,n}^*=y_{t,n}^*-(\widehat{\beta}_n^*)'z_{t,n}$.
\item Calculate the bootstrapped versions of $\widehat{\rho}_n$ and $s_n^2$, i.e.,
$\widehat{\rho}_n^*$ and ${s_n^*}^2$, based on
$\{\widehat{X}_{t,n}^*\}_{t=1}^n$, and the bootstrapped test statistics
$\mathbf{T}_n^*=n(\widehat{\rho}_n^*-1)$ and
$\mathbf{t}_n^*=\{\sum_{t=1}^n(\widehat{X}_{t-1,n}^*)^2\}^{1/2}(\widehat{\rho}_n^*-1)/s_n^*$.
\item Repeat steps 2-7 B times, and record the
bootstrapped test statistics
$\{\mathbf{T}_n^{*(1)},\ldots,\mathbf{T}_n^{*(B)}\}$ and
$\{\mathbf{t}_n^{*(1)},\ldots,
\allowbreak
\mathbf{t}_n^{*(B)}\}$. The p-values
are
$$\frac{\sum_{b=1}^B{\bf 1}\{\mathbf{T}_n^{*(b)}<\mathbf{T}_n\}}{B}~~~{\rm and}~~~~\frac{\sum_{b=1}^B{\bf 1}\{\mathbf{t}_n^{*(b)}<\mathbf{t}_n\}}{B}.$$
\end{enumerate}
}\end{algorithm}

\begin{remark}{\rm
If $l=1$, or equivalently, if an i.i.d. sequence $W_t$ is used in RDWB step 4, then the above-described procedure coincides with RWB in \cite{Cavaliere:Taylor:2009a}.
}\end{remark}

\begin{remark}{\rm
In the above procedure, the number $\hat{k}$ of lags  is optimized for the original data  $X_{t,n}$, and the same $\hat{k}$ is used for the bootstrapped data $X_{t,n}^*$.
In general, the number of lags for the original data and that for the bootstrapped data  do not have to be the same.
For example, for the bootstrap, it can be chosen to be optimized for each bootstrapped sample; that is,
 $$k^*_b={\rm argmin}_{0 \leq k \leq k_{max}} {\rm MAIC}^*(k),$$ where ${\rm MAIC}^*(k)={\rm ln}(\hat{\sigma}_k^{2*})+\frac{2(\tau_n^*(k)+k)}{n-k_{max}},$
$\tau_n^*(k)=(\hat{\sigma}_k^{2*})^{-1}\hat{\pi}_0^{2*}\sum_{t=k_{max}+1}^n (\widehat{X}_{t-1,n}^*)^2,$
$\hat{\sigma}_k^{2*}=(n-k_{max})^{-1}\sum_{t=k_{max}+1}^n\hat{u}_{t,n,k}^{2*}.$
However, we shall keep the same $k$ for both the original and the bootstrapped data based on the finite sample findings reported in Remark 3 of \cite{Cavaliere:Taylor:2009a}.
}\end{remark}

For a fair comparison of power, the following size-corrected power procedure similar to \cite{Dominguez:Lobato:2001} is adapted.
\begin{algorithm}\label{size:adjusted}{\rm [Size-corrected Power of a Bootstrap Test]
Consider a level $\alpha$ test with $\mathbf{T}_n$ for a simple exposition.
The unit root null hypothesis is rejected if $\mathbf{T}_n<\mathcal{L}_{\mathbf{T},0,\alpha}$, where $\mathcal{L}_{\mathbf{T},0,\alpha}$ indicates the $\alpha$-quantile of $\mathcal{L}_{\mathbf{T},0}$ in Theorem \ref{thm:NI}.
\begin{enumerate}
\item Estimate the finite sample counterpart $\widehat{\mathcal{L}}_{\mathbf{T},0,\alpha}$ of $\mathcal{L}_{\mathbf{T},0,\alpha}$ based on $N$ Monte-Carlo replications.
Let $N$ be large enough so that $N\alpha$ is an integer.
That is, if $\{\mathbf{T}_n^{(1)},\mathbf{T}_n^{(2)},\ldots,\mathbf{T}_n^{(N)}\}$ indicates the set of test statistics of $N$ Monte-Carlo replications and
 $\{\mathbf{T}_n^{[1]},\mathbf{T}_n^{[2]},\ldots,\mathbf{T}_n^{[N]}\}$ indicates its ordered version from the smallest to largest,
 $\widehat{\mathcal{L}}_{\mathbf{T},0,\alpha}=\mathbf{T}_n^{[N\alpha]}$.
Note that using this infeasible critical value $\widehat{\mathcal{L}}_{\mathbf{T},0,\alpha}$, the empirical size should be similar to the nominal level $\alpha$.
\item For each Monte-Carlo replication under the null hypothesis, generate $B$ bootstrap samples and calculate the corresponding bootstrap test statistics
$\{\mathbf{T}_n^{*(b,i)}\}_{b=1}^B$,  $i=1,\ldots,N$.
Calculate the empirical size of the bootstrap test $\alpha^{(i)}$ of the $i$th Monte-Carlo replication using the infeasible critical value $\widehat{\mathcal{L}}_{\mathbf{T},0,\alpha}$, i.e.,
$$\alpha^{(i)}=B^{-1}\sum_{b=1}^B{\bf 1}\left(\mathbf{T}_n^{*(b,i)}<\widehat{\mathcal{L}}_{\mathbf{T},0,\alpha}\right).$$
 The size-corrected level $\alpha^c$ is the average of the $\alpha^{(i)}$, that is, $\alpha^c=N^{-1}\sum_{i=1}^N\alpha^{(i)}$.
\item For another set of statistics of $N$ Monte-Carlo replications under the (local) alternative, the  size-corrected power is calculated replacing $\alpha$ with  its size-corrected version, $\alpha^c$.
That is,
$$N^{-1}\sum_{i=1}^N \left(\mathbf{T}_n^{(i)}<\mathbf{T}_{n,\alpha^c}^{*(i)}\right),$$
where $\mathbf{T}_{n,\alpha^c}^{*(i)}$ is the $\alpha^c$-quantile of the bootstrapped statistics $\{\mathbf{T}_n^{*(b,i)}\}_{b=1}^B$ for the $i$th Monte-Carlo replication.
\end{enumerate}
}\end{algorithm}

\bigskip
\bigskip

The following data generating processes (DGPs) are used for comparison of DWB, RWB, and RDWB in finite samples.
For simplicity, set  $\beta\equiv 0$  so that $\widehat{X}_{t,n}=X_{t,n}$.
Consider (\ref{truemodel2}) and
$u_{t,n}$ generated from time-varying moving average (MA) and autoregressive (AR) models with lag 1,
$$({\rm MA}_{i,j})~u_{t,n}=e_{j,t,n}+\phi_i(t/n) e_{j,t-1,n},~~~~({\rm AR}_{i,j})~u_{t,n}=e_{j,t,n}+\phi_i(t/n) u_{t-1,n}$$
for $t=1,\ldots,n$, where
$e_{j,t,n}=\omega_j(t/n)\varepsilon_t$, $\varepsilon_t\stackrel{i.i.d.}{\sim}N(0,1)$.
The MA or AR coefficient $\phi_i(s)$ is possibly time-varying with the following six choices: for $s\in[0,1]$,
$$\phi_1(s)=0.8,~\phi_2(s)=-0.8,~\phi_3(s)=0.2+0.6{\bf 1}(s>0.2),$$
$$\phi_4(s)=0.2+0.6{\bf 1}(s>0.8),~\phi_5(s)=0.8-1.6s,~{\rm and}~\phi_6(s)=0.6s-0.8.$$
The function $\omega_j(s)$ governs possible heteroscedastic behavior in $u_{t,n}$ with the following five choices: for $s\in[0,1]$,
$$\omega_1(s)=0.5,~\omega_2(s)=0.1+0.5{\bf 1}(s>0.1),~\omega_3(s)=0.1+0.5{\bf 1}(s>0.9),$$
$$\omega_4(s)=0.1+0.5{\bf 1}(0.4<s<0.6),~{\rm and}~\omega_5(s)=0.5s+0.1.$$
Combinations of  $\phi_i(s)$ and $\omega_j(s)$ along with the choice of MA or AR lead to 60 DGPs that satisfy the PLS assumption in (A1)--(A4).
In particular, if $i=1$ or 2, $\phi_i(s)$ is constant over $s\in[0,1]$.
The corresponding $\{u_{t,n}\}$ processes fall into the category of linear processes with heteroscedastic error in (\ref{GLP}), making RWB consistent for any choices of $\omega_j(s)$, $j=1,\ldots,5$.
These settings are to mirror the setup of the Cavaliere-Taylor papers.
For all other settings, the asymptotic consistency of RWB is not guaranteed, whereas DWB and RDWB are expected to work asymptotically.
Sudden increases and smooth changes in MA or AR coefficients are presented in the cases with $i=3,4$ and $i=5,6$, respectively.
The variance of $e_{j,t,n}$  is a constant ($j=1$), a step function with a sudden increase in the beginning ($j=2$) and end ($j=3$) of the series, a step function with a sudden increase and decrease in the middle ($j=4$), or a smoothly increasing sequence ($j=5$).

The sample sizes $n=100$ and $400$ are considered.
The number of Monte-Carlo replications is 2000, and  the number of bootstrap replications is $B=1000$ for all bootstrap methods.
 For  local alternatives, $c=0,  -5, -10, -15, -20, -25, -30$ are considered.
In particular, for DWB and RDWB, in each replication,
pseudoseries $(W_{1,n},\ldots,W_{n,n})'$ are generated from i.i.d. $N(\mathbf{0}_n,\Sigma)$,
where $\Sigma$ is an $n$ by $n$ matrix with its $(i,j)$th element being $a\{(i-j)/l\}$.
Here the Bartlett kernel is used, i.e., $a(s)=(1-|s|){\bf 1}(|s|\leq 1)$.
For DWB and RDWB, the bandwidth parameter $l$ is chosen as $l=\lfloor 6(n/100)^{1/4}\rfloor$. That is,
$l=6$ if  $n=100$, and $l=8$ if $n=400$.
In Section B of the supplementary material, (i) full details on the effect of different choices of $l$ for selected DGPs are presented
and (ii)  a data-driven choice of $l$, the minimum volatility method, is proposed.
It seems that the empirical sizes are not overly sensitive to the choice of $l$, as long as $l$ is not too small,
and the finite sample size comparison with the MV method in Section B  of supplementary material supports the above deterministic choice of $l$.

\bigskip

Tables \ref{sizetable:MA} and \ref{sizetable:AR} present the empirical sizes of the three methods when the nominal size is 5\%.
When the model is stationary with positive coefficient $\phi(s)=\phi_1(s)=0.8$, i.e., (${\rm MA/AR}_{1,j}$) for $j=1,\ldots,5$, all three bootstrap methods produce reasonably accurate  sizes, except that the DWB method tends to under-reject for the AR models.
This under-rejecting behavior of  DWB is observed consistently for most  AR models.
This might be due to the fact that the DWB method mimics the time dependence in the original data in a  manner similar to MA models,
 so that it does not produce as accurate sizes for AR models as for MA models.
The RDWB method nicely compensates this shortcoming by applying an AR-based prewhitening.
The prewhitening effect is  most noticeable when
the model is stationary with negative coefficient $\phi(s)=\phi_2(s)=-0.8$, i.e.,  (${\rm MA/AR}_{2,j}$) for $j=1,\ldots,5$. For these models with negative autocorrelation,
the size-distortion of the DWB method is very large at both sample sizes with slight less distortion for larger sample size.
This suggests that although the DWB method should work asymptotically for the negative coefficient case, this convergence could be too slow to be useful in practice.
On the other hand, after applying  the AR-based prewhitening, similar to RWB, finite sample sizes are brought closer to the nominal level.

A careful examination of RWB shows that it has a fairly accurate size, especially when it is theoretically supported ($i=1,2$).
However, for some DGPs with changing MA or AR coefficients ($i=3,4,5,6$),  RWB does not seem to be consistent.
In particular, in the MA models, the sizes of RWB tend to further deviate from the nominal level as the sample size $n$ increases
 when there is a sudden increase in the variance in innovations at the latter part of the series ($j=3$) with changing variance
(see $({\rm MA}_{3,3})$, $({\rm MA}_{4,3})$, $({\rm MA}_{5,3})$,  and $({\rm MA}_{6,3})$) or when both MA coefficient and variance of innovations change smoothly (see $({\rm MA}_{5,5})$).
In the AR models,  RWB tends to have heavier size distortion as $n$ increases when the AR coefficient changes drastically from negative to positive ($i=5$; see $({\rm AR}_{5,1})$,  $({\rm AR}_{5,2})$,  $({\rm AR}_{5,3})$, and $({\rm AR}_{5,5})$) or  when the AR coefficient is negative and changes smoothly and the variance in innovations suddenly increases at the latter part of the series (see $({\rm AR}_{6,3})$).
This size distortion might be an indication that  the AR prewhitening (RWB) alone does not work in theory, and the dependence in the error is not completely filtered out.
By contrast, RDWB tends to have  more accurate sizes for these models, although size distortion due to inaccurate prewhitening is still apparent
to a lesser degree.
On the other hand, as long as the MA or AR coefficients are nonnegative, DWB without prewhitening is always 
demonstrated to have more accurate size as $n$ increases.
In particular, for MA models with changing MA coefficient ($i=3,4,5$), DWB tends to produce the best size with the most consistent behavior among the three bootstrap methods.

Overall, the size for RDWB seems to be the most reliable among the three bootstrap methods if the underlying DGP is not known.
In some unreported simulations, we have observed the following: (i) the large size distortion associated with
the DWB method for negative autocorrelation models, (${\rm MA/AR}_{2,1}$), can be reduced to below the nominal  $5\%$ level if we use restricted residuals, at the  price of power loss; (ii) a comparison with residual block bootstrap in \cite{Paparoditis:Politis:2003} shows that the size for the residual block bootstrap can be quite distorted
for some DGPs, e.g., (${\rm MA}_{5,5}$). This  indicates the inability of residual block bootstrap to consistently approximate the limiting null distribution when the error process is PLS.

\begin{figure}[h!]\centering
\resizebox{1\textwidth}{!}{\includegraphics{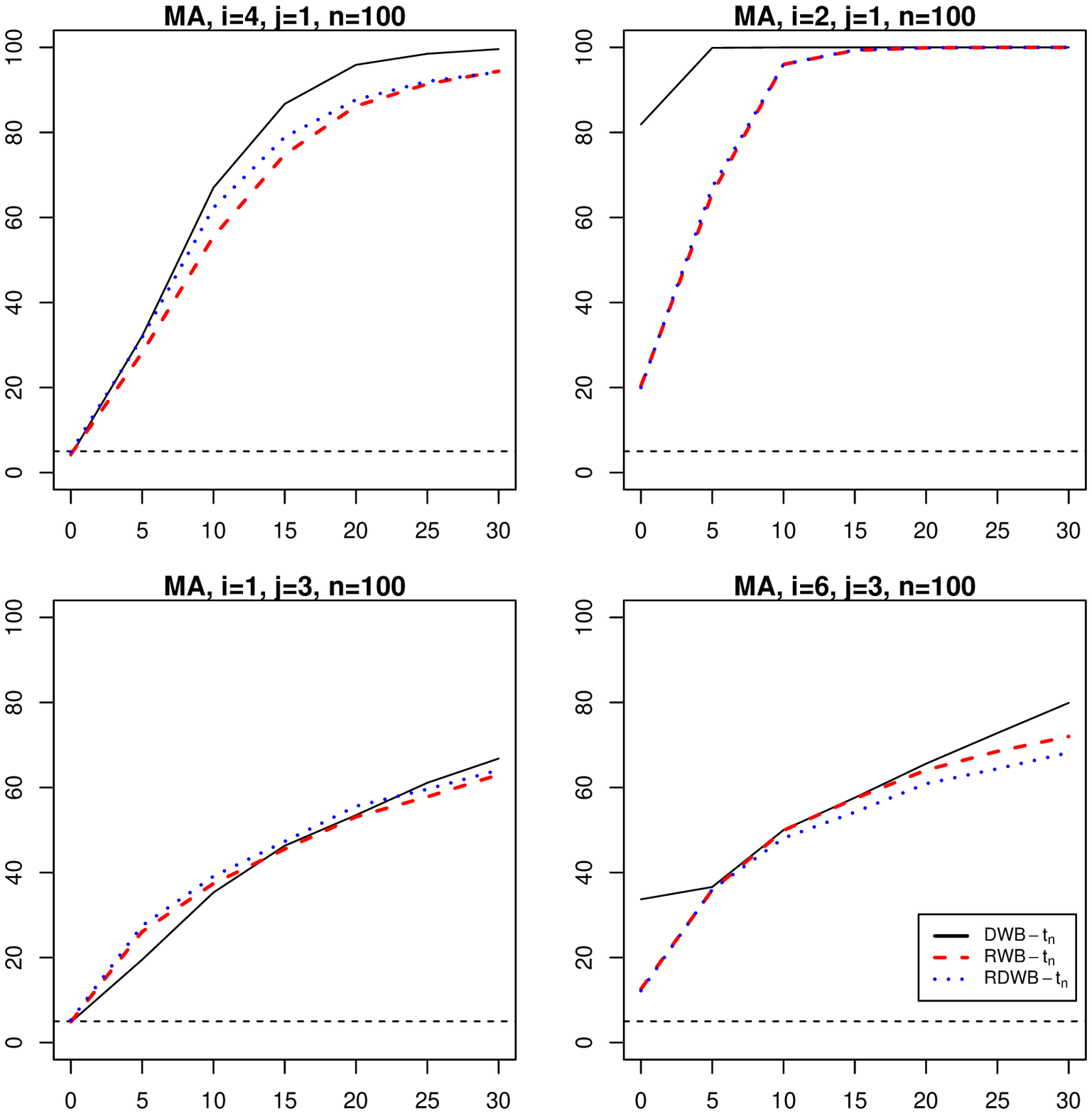}}
\caption{Rejection frequencies (\%) versus $-c$, where $\rho=1+c/n$ for DWB, RWB, and RDWB unit root tests when the error $u_{t,n}$ is generated following the MA processes with
$\phi_i(s)$ and $\omega_j(s)$ for selected $(i,j)$s.
The plots present empirical sizes when $c=0$ and size-adjusted powers when $c\neq0$. 2000 Monte-Carlo replications and 1000 bootstrap replications are used.
The sample size is $n=100$ and the nominal level is 5\%.}
\label{MA50_n100-paper}
\end{figure}

\begin{figure}[h!]\centering
\resizebox{1\textwidth}{!}{\includegraphics{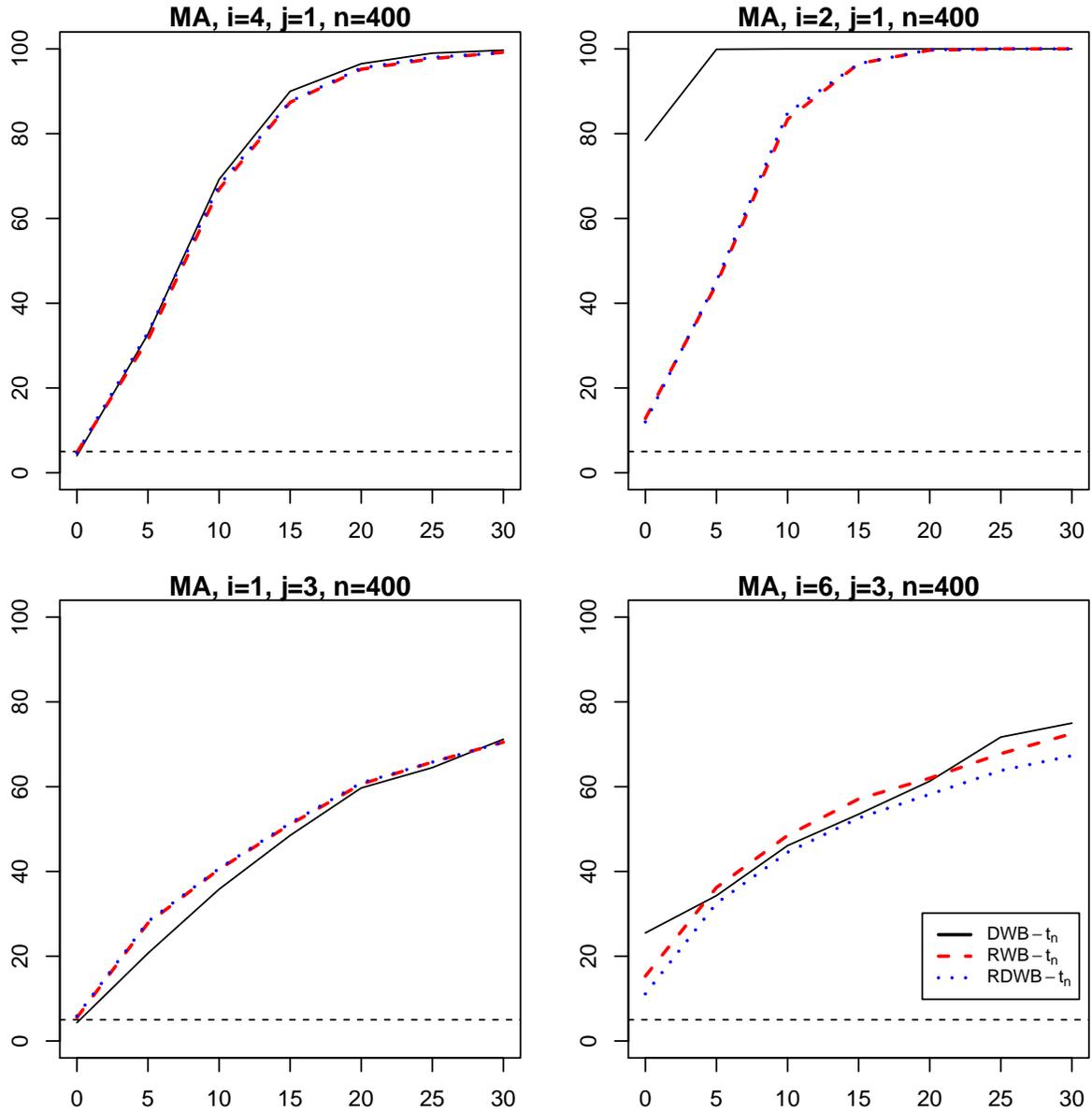}}
\caption{Rejection frequencies (\%) versus $-c$, where $\rho=1+c/n$ for DWB, RWB, and RDWB unit root tests when the error $u_{t,n}$ is generated following the MA processes with
$\phi_i(s)$ and $\omega_j(s)$ for selected $(i,j)$s.
The plots present empirical sizes when $c=0$ and size-adjusted powers when $c\neq0$. 2000 Monte-Carlo replications and 1000 bootstrap replications are used.
The sample size is $n=400$ and the nominal level is 5\%.}
\label{MA50_n400-paper}
\end{figure}

Figures \ref{MA50_n100-paper} and \ref{MA50_n400-paper} present the power curves of $\mathbf{t}_n$ for DWB, RWB, and RDWB for selected DGPs with $n=100$ and $400$, respectively.
The size-adjusted power curves in the first panel, (${\rm MA}_{4,1}$), are representative for most of the cases where all three bootstrap methods
have reasonably accurate sizes, where (i)  DWB tends to have the best power, (ii) RDWB tends to have slightly better power than  RWB when $n=100$, and (iii)  RWB and RDWB are fairly comparable in terms of sizes and powers in general.
Most MA or AR models with positive MA or AR coefficient for at least part of a series ($\phi_i(s)$ with $i=1,3,4,5$) and constant, early break, or smooth change in error variance ($\omega_j(s)$ with $j=1,2,5$)
tend to have a similar shape.
The second panel, (${\rm MA}_{2,1}$), represents the size-adjusted curves when the MA or AR coefficients are negative at all time points ($i=2,6$) so that the finite sample size of  DWB is highly distorted.
Even though DWB has the best power, it is not recommended due to its big size distortion in this case.
It seems that  RWB and RDWB do not have much difference in terms of size-adjusted power.

The last two panels focus on the comparison between  RWB and RDWB.
The third panel, (${\rm MA}_{1,3}$), is representative when there is a jump in $\omega(s)$ at the end of the series ($j=3$) and when both RWB and RDWB
have reasonable sizes. Models with $i=1,2,6$ and $j=3,4$ tend to have a similar pattern if  DWB is ignored due to its high size distortion when $i=2,6$.
In this case, RWB and RDWB have similar powers, as RWB tends to have slightly better power when $i = 5$ or 6, whereas RDWB tends to have slightly higher power when $i = 1$.
The last panel, (${\rm MA}_{6,3}$), is representative for the case when  RWB is not consistent.
(${\rm AR}_{i,j}$) or (${\rm MA}_{i,j}$) with $i=3,4,5$ and $j=3,4$ fall into this category.
In this case,  RDWB seems to present the most reasonable size and power. Even though RWB appears to have the best power,
RWB does not seem to be consistent due to the considerable increase in its finite sample size as $n$ increases for some models.
Complete power curves for all DGPs are presented in the supplementary material as Figures C.1--C.4.
It is worth noting that  $\mathbf{t}_n$ tends to produce more accurate sizes with little power loss (or slightly better power) than $\mathbf{T}_n$.

In summary,  RDWB, the combination of  RWB and DWB, appears to work well in finite samples. It tends to produce reasonably high powers and fairly accurate sizes in all models under examination.
In the situation when DWB or RWB have a large size distortion, the size accuracy of RDWB is well maintained and its power appears quite reasonable in all cases.
One downside associated with RDWB is that it requires two tuning parameters: the truncation lag in the AR sieve and the bandwidth parameter in DWB.
In this paper, we  choose the number of lags for  RWB and RDWB using the MAIC method.
As for the bandwidth parameter, it seems that  DWB and RDWB are not sensitive to the choice of the bandwidth parameter
and the proposed deterministic choice
seems to perform reasonably well  in finite samples.
Given that the DGP is unknown in practice, we shall  recommend the use of RDWB.

\section{Conclusion}\label{Conclusion}
In this paper, we present a new bootstrap-based unit root testing procedure that is robust to changing second- and higher-order properties in the error process.
The error  is modeled as a piecewise locally stationary (PLS) process, which is general enough to include time-varying nonlinear processes as well as heteroscedastic
linear processes as special cases. In particular, the PLS process does not impose a separable structure on its longrun variance as do heteroscedastic linear processes and modulated stationary processes,
which have been adopted in the literature to model heteroscedasticity and weak dependence of the error.
Under the PLS framework, the limiting null distributions of two popular test statistics are derived
and the dependent wild bootstrap (DWB) method is used to approximate these non-pivotal distributions.
The functional central limiting theorem has been established for the standardized partial sum process of the DWB residuals, and bootstrap consistency is justified under local alternatives. The DWB-based unit root test has asymptotically nontrivial local power. The DWB method was originally proposed for stationary time series. By showing its consistency in the PLS setting, we broaden its applicability and its use in the locally stationary context is worth further exploration.
For finite sample simulations, we propose a recolored DWB (RDWB), combining the AR sieve idea used in the RWB test with DWB to improve the performance of the DWB-based test. In many cases, the RDWB method tends to provide the most accurate sizes and reasonably good power, compared to the use of DWB or RWB alone. In practice, with little knowledge of the error structure, the RDWB-based test seems preferable due to its robustness for a large class of nonstationary error processes.

\bigskip

\centerline {\bf\large\sc Acknowledgements}
This research was partially supported by NSF grant DMS-1104545. We are grateful to the co-editor and the three referees for their constructive comments and suggestions that led to a substantial improvement of the paper.
In particular, we are most grateful to Peter C. B. Phillips, who has gone beyond the call of duty for an editor in
carefully correcting our English.
We also thank Fabrizio Zanello, Mark Gockenbach, Benjamin Ong, and Meghan Campbell for proofreading.
Superior, a high performance computing cluster at Michigan Technological University, was used in obtaining results presented in this publication.

\bibliography{References}{}

\begin{thebibliography}{}

\bibitem[\protect\citeauthoryear{Adak}{Adak}{1998}]{Adak:1998}
Adak, S. (1998).
\newblock Time-dependent spectral analysis of nonstationary time series.
\newblock {\em Journal of the American Statistical Association\/}~{\em
  93\/}(444), 1488--1501.

\bibitem[\protect\citeauthoryear{Andrews}{Andrews}{1984}]{Andrews:1984}
Andrews, D. W.~K. (1984).
\newblock Non-strong mixing autoregressive processes.
\newblock {\em Journal of Applied Probability\/}~{\em 21\/}(4), 930--934.

\bibitem[\protect\citeauthoryear{Andrews}{Andrews}{1991}]{Andrews:1991}
Andrews, D. W.~K. (1991).
\newblock {Heteroskedasticity and autocorrelation consistent covariance matrix
  estimation}.
\newblock {\em Econometrica\/}~{\em 59\/}(3), 817--858.

\bibitem[\protect\citeauthoryear{Billingsley}{Billingsley}{1968}]{Billingsley:1968}
Billingsley, P. (1968).
\newblock {\em Convergence of Probability Measures}.
\newblock New York: Wiley.

\bibitem[\protect\citeauthoryear{Busetti and Taylor}{Busetti and
  Taylor}{2003}]{Busetti:Taylor:2003}
Busetti, F. and A.~M.~R. Taylor (2003).
\newblock Variance shifts, structural breaks, and stationarity tests.
\newblock {\em Journal of Business and Economic Statistics\/}~{\em 21\/}(4),
  510--531.

\bibitem[\protect\citeauthoryear{Cavaliere and Taylor}{Cavaliere and
  Taylor}{2007}]{Cavaliere:Taylor:2007}
Cavaliere, G. and A.~M.~R. Taylor (2007).
\newblock {Testing for unit roots in time series models with non-stationary
  volatility}.
\newblock {\em Journal of Econometrics\/}~{\em 140\/}(2), 919--947.

\bibitem[\protect\citeauthoryear{Cavaliere and Taylor}{Cavaliere and
  Taylor}{2008a}]{Cavaliere:Taylor:2008b}
Cavaliere, G. and A.~M.~R. Taylor (2008a).
\newblock {Bootstrap unit root tests for time series with nonstationary
  volatility}.
\newblock {\em Econometric Theory\/}~{\em 24\/}(1), 43--71.

\bibitem[\protect\citeauthoryear{Cavaliere and Taylor}{Cavaliere and
  Taylor}{2008b}]{Cavaliere:Taylor:2008a}
Cavaliere, G. and A.~M.~R. Taylor (2008b).
\newblock Time-transformed unit root tests for models with non-stationary
  volatility.
\newblock {\em Journal of Time Series Analysis\/}~{\em 29\/}(2), 300--330.

\bibitem[\protect\citeauthoryear{Cavaliere and Taylor}{Cavaliere and
  Taylor}{2009}]{Cavaliere:Taylor:2009a}
Cavaliere, G. and A.~M.~R. Taylor (2009).
\newblock {Bootstrap M unit root tests}.
\newblock {\em Econometric Reviews\/}~{\em 28\/}(5), 393--421.

\bibitem[\protect\citeauthoryear{Chang and Park}{Chang and
  Park}{2002}]{Chang:Park:2002}
Chang, Y. and J.~Y. Park (2002).
\newblock On the asymptotics of adf tests for unit roots.
\newblock {\em Econometric Reviews\/}~{\em 21}, 431--447.

\bibitem[\protect\citeauthoryear{Chang and Park}{Chang and
  Park}{2003}]{Chang:Park:2003}
Chang, Y. and J.~Y. Park (2003).
\newblock A sieve bootstrap for the test of a unit root.
\newblock {\em Journal of Time Series Analysis\/}~{\em 24\/}(4), 379--400.

\bibitem[\protect\citeauthoryear{Dahlhaus}{Dahlhaus}{1997}]{Dahlhaus:1997}
Dahlhaus, R. (1997).
\newblock Fitting time series models to nonstationary processes.
\newblock {\em The Annals of Statistics\/}~{\em 25\/}(1), 1--37.

\bibitem[\protect\citeauthoryear{Dahlhaus and Subba~Rao}{Dahlhaus and
  Subba~Rao}{2006}]{Dahlhaus:Rao:2006}
Dahlhaus, R. and S.~Subba~Rao (2006).
\newblock Statistical inference for time-varying arch processes.
\newblock {\em The Annals of Statistics\/}~{\em 34\/}(3), 1075--1114.

\bibitem[\protect\citeauthoryear{Dickey and Fuller}{Dickey and
  Fuller}{1979}]{Dickey:Fuller:1979}
Dickey, D.~A. and W.~A. Fuller (1979).
\newblock Distribution of the estimators for autoregressive time series with a
  unit root.
\newblock {\em Journal of the American Statistical Association\/}~{\em
  74\/}(366), 427--431.

\bibitem[\protect\citeauthoryear{Dickey and Fuller}{Dickey and
  Fuller}{1981}]{Dickey:Fuller:1981}
Dickey, D.~A. and W.~A. Fuller (1981).
\newblock Likelihood ratio statistics for autoregressive time series with a
  unit root.
\newblock {\em Econometrica\/}~{\em 49\/}(4), 1057--1072.

\bibitem[\protect\citeauthoryear{Dom\'{i}nguez and Lobato}{Dom\'{i}nguez and
  Lobato}{2001}]{Dominguez:Lobato:2001}
Dom\'{i}nguez, M.~A. and I.~N. Lobato (2001).
\newblock Size corrected power for bootstrap tests.
\newblock Working Papers 102, Centro de Investigacion Economica, ITAM.

\bibitem[\protect\citeauthoryear{Draghicescu, Guillas, and Wu}{Draghicescu
  et~al.}{2009}]{Draghicescu:Guillas:Wu:2009}
Draghicescu, D., S.~Guillas, and W.~B. Wu (2009).
\newblock Quantile curve estimation and visualization for nonstationary time
  series.
\newblock {\em Journal of Computational and Graphical Statistics\/}~{\em
  18\/}(1), 1--20.

\bibitem[\protect\citeauthoryear{Elliott, Rothenberg, and Stock}{Elliott
  et~al.}{1996}]{Elliott_etal:1996}
Elliott, G., T.~J. Rothenberg, and J.~H. Stock (1996).
\newblock Efficient tests for an autoregressive unit root.
\newblock {\em Econometrica\/}~{\em 64\/}(4), 813--836.

\bibitem[\protect\citeauthoryear{Fryzlewicz, Sapatinas, and
  Subba~Rao}{Fryzlewicz et~al.}{2008}]{Fryzlewicz_etal:2008}
Fryzlewicz, P., T.~Sapatinas, and S.~Subba~Rao (2008).
\newblock Normalized least-squares estimation in time-varying arch models.
\newblock {\em The Annals of Statistics\/}~{\em 36\/}(2), 742--786.

\bibitem[\protect\citeauthoryear{Fryzlewicz and Subba~Rao}{Fryzlewicz and
  Subba~Rao}{2011}]{Fryzlewicz:SubbaRao:2011}
Fryzlewicz, P. and S.~Subba~Rao (2011).
\newblock Mixing properties of arch and time-varying arch processes.
\newblock {\em Bernoulli\/}~{\em 17\/}(1), 320--346.

\bibitem[\protect\citeauthoryear{Giurcanu and Spokoiny}{Giurcanu and
  Spokoiny}{2004}]{Giurcanu:Spokoiny:2004}
Giurcanu, M. and V.~Spokoiny (2004).
\newblock Confidence estimation of the covariance function of stationary and
  locally stationary processes.
\newblock {\em Statistics and Decisions\/}~{\em 22\/}(4), 283--300.

\bibitem[\protect\citeauthoryear{Kim and Nelson}{Kim and
  Nelson}{1999}]{Kim:Nelson:1999}
Kim, C.-J. and C.~R. Nelson (1999).
\newblock {Has the U.S. economy become more stable? A Bayesian approach based
  on a Markov-switching model of the business cycle}.
\newblock {\em The Review of Economics and Statistics\/}~{\em 81\/}(4),
  608--616.

\bibitem[\protect\citeauthoryear{Kreiss}{Kreiss}{1988}]{Kreiss:1988}
Kreiss, J.-P. (1988).
\newblock Asymptotic statistical inference for a class of stochastic processes.
\newblock Habilitationsschrift, Universit\"{a}t Hamburg.

\bibitem[\protect\citeauthoryear{K\"{u}nsch}{K\"{u}nsch}{1989}]{Kunsch:1989}
K\"{u}nsch, H.~R. (1989).
\newblock The jackknife and the bootstrap for general stationary observations.
\newblock {\em The Annals of Statistics\/}~{\em 17\/}(3), 1217--1241.

\bibitem[\protect\citeauthoryear{Mallat, Papanicolaou, and Zhang}{Mallat
  et~al.}{1998}]{Mallat:Papanicolaou:Zhang:1998}
Mallat, S., G.~Papanicolaou, and Z.~Zhang (1998).
\newblock Adaptive covariance estimation of locally stationary processes.
\newblock {\em The Annals of Statistics\/}~{\em 26\/}(1), 1--47.

\bibitem[\protect\citeauthoryear{McConnell and Perez-Quiros}{McConnell and
  Perez-Quiros}{2000}]{McConnell:Perez-Quiros:2000}
McConnell, M.~M. and G.~Perez-Quiros (2000).
\newblock Output fluctuations in the united states: What has changed since the
  early 1980's?
\newblock {\em American Economic Review\/}~{\em 90\/}(5), 1464--1476.

\bibitem[\protect\citeauthoryear{M\"{u}ller and Elliott}{M\"{u}ller and
  Elliott}{2003}]{Muller:Elliott:2003}
M\"{u}ller, U.~K. and G.~Elliott (2003).
\newblock Tests for unit roots and the initial condition.
\newblock {\em Econometrica\/}~{\em 71\/}(4), 1269--1286.

\bibitem[\protect\citeauthoryear{Newey and West}{Newey and
  West}{1987}]{Newey:West:1987}
Newey, W. and K.~D. West (1987).
\newblock A simple, positive semi-definite, heteroskedasticity and
  autocorrelation consistent covariance matrix.
\newblock {\em Econometrica\/}~{\em 55\/}(3), 703--708.

\bibitem[\protect\citeauthoryear{Ng and Perron}{Ng and
  Perron}{2001}]{Ng:Perron:2001}
Ng, S. and P.~Perron (2001).
\newblock Lag length selection and the construction of unit root tests with
  good size and power.
\newblock {\em Econometrica\/}~{\em 69\/}(6), 1519--1554.

\bibitem[\protect\citeauthoryear{Palm, Smeekes, and Urbain}{Palm
  et~al.}{2008}]{Palm:Smeekes:Urbain:2008}
Palm, F.~C., S.~Smeekes, and J.-P. Urbain (2008).
\newblock Bootstrap unit-root tests: Comparison and extensions.
\newblock {\em Journal of Time Series Analysis\/}~{\em 29\/}(2), 371--401.

\bibitem[\protect\citeauthoryear{Paparoditis and Politis}{Paparoditis and
  Politis}{2002}]{Paparoditis:Politis:2002}
Paparoditis, E. and D.~N. Politis (2002).
\newblock Local block bootstrap.
\newblock {\em Comptes Rendus Mathematique\/}~{\em 335\/}(11), 959--962.

\bibitem[\protect\citeauthoryear{Paparoditis and Politis}{Paparoditis and
  Politis}{2003}]{Paparoditis:Politis:2003}
Paparoditis, E. and D.~N. Politis (2003).
\newblock Residual-based block bootstrap for unit root testing.
\newblock {\em Econometrica\/}~{\em 71\/}(3), 813--855.

\bibitem[\protect\citeauthoryear{Paparoditis and Politis}{Paparoditis and
  Politis}{2005}]{Paparoditis:Politis:2005}
Paparoditis, E. and D.~N. Politis (2005).
\newblock Bootstrapping unit root tests for autoregressive time series.
\newblock {\em Journal of the American Statistical Association\/}~{\em
  100\/}(470), 545--553.

\bibitem[\protect\citeauthoryear{Parker, Paparoditis, and Politis}{Parker
  et~al.}{2006}]{Parker:Paparoditis:Politis:2006}
Parker, C., E.~Paparoditis, and D.~N. Politis (2006).
\newblock Unit root testing via the stationary bootstrap.
\newblock {\em Journal of Econometrics\/}~{\em 133\/}(2), 601--638.

\bibitem[\protect\citeauthoryear{Perron and Ng}{Perron and
  Ng}{1996}]{Perron:Ng:1996}
Perron, P. and S.~Ng (1996).
\newblock Useful modifications to some unit root tests with dependent errors
  and their local asymptotic properties.
\newblock {\em Review of Economic Studies\/}~{\em 63\/}(3), 435--463.

\bibitem[\protect\citeauthoryear{Phillips}{Phillips}{1987a}]{Phillips:1987a}
Phillips, P. C.~B. (1987a).
\newblock Time series regression with a unit root.
\newblock {\em Econometrica\/}~{\em 55\/}(2), 277--301.

\bibitem[\protect\citeauthoryear{Phillips}{Phillips}{1987b}]{Phillips:1987b}
Phillips, P. C.~B. (1987b).
\newblock Towards a unified asymptotic theory for autoregression.
\newblock {\em Biometrika\/}~{\em 74\/}(3), 535--547.

\bibitem[\protect\citeauthoryear{Phillips and Perron}{Phillips and
  Perron}{1988}]{Phillips:Perron:1988}
Phillips, P. C.~B. and P.~Perron (1988).
\newblock Testing for a unit root in time series regression.
\newblock {\em Biometrika\/}~{\em 75\/}(2).

\bibitem[\protect\citeauthoryear{Phillips and Solo}{Phillips and
  Solo}{1992}]{Phillips:Solo:1992}
Phillips, P. C.~B. and V.~Solo (1992).
\newblock Asymptotics for linear processes.
\newblock {\em The Annals of Statistics\/}~{\em 20\/}(2), 971--1001.

\bibitem[\protect\citeauthoryear{Phillips and Xiao}{Phillips and
  Xiao}{1998}]{Phillips:Xiao:1998}
Phillips, P. C.~B. and Z.~Xiao (1998).
\newblock A primer on unit root testing.
\newblock {\em Journal of Economic Surveys\/}~{\em 12\/}(5), 423--469.

\bibitem[\protect\citeauthoryear{Politis and Romano}{Politis and
  Romano}{1994}]{Politis:Romano:1994}
Politis, D.~N. and J.~P. Romano (1994).
\newblock The stationary bootstrap.
\newblock {\em Journal of the American Statistical Association\/}~{\em
  89\/}(428), 1303--1313.

\bibitem[\protect\citeauthoryear{Priestley}{Priestley}{1965}]{Priestley:1965}
Priestley, M.~B. (1965).
\newblock Evolutionary spectra and non-stationary processes.
\newblock {\em Journal of the Royal Statistical Society: Series B\/}~{\em
  27\/}(2), 204--237.

\bibitem[\protect\citeauthoryear{Psaradakis}{Psaradakis}{2001}]{Psaradakis:2001}
Psaradakis, Z. (2001).
\newblock Bootstrap tests for an autoregressive unit root in the presence of
  weakly dependent errors.
\newblock {\em Journal of Time Series Analysis\/}~{\em 22\/}(5), 577--594.

\bibitem[\protect\citeauthoryear{Rho and Shao}{Rho and
  Shao}{2015}]{Rho:Shao:2015}
Rho, Y. and X.~Shao (2015).
\newblock Inference for time series regression models with weakly dependent and
  heteroscedastic errors.
\newblock {\em Journal of Business \& Economic Statistics\/}~{\em 33\/}(3),
  444--457.

\bibitem[\protect\citeauthoryear{Said and Dickey}{Said and
  Dickey}{1984}]{Said:Dickey:1984}
Said, S.~E. and D.~A. Dickey (1984).
\newblock {Testing for unit roots in autoregressive-moving average models of
  unknown order}.
\newblock {\em Biometrika\/}~{\em 71\/}(3), 599--607.

\bibitem[\protect\citeauthoryear{Sensier and van Dijk}{Sensier and van
  Dijk}{2004}]{Sensier:vanDijk:2004}
Sensier, M. and D.~van Dijk (2004).
\newblock {Testing for volatility changes in U.S. macroeconomic time series}.
\newblock {\em The Review of Economics and Statistics\/}~{\em 86\/}(3),
  833--839.

\bibitem[\protect\citeauthoryear{Shao}{Shao}{2010}]{Shao:2010b}
Shao, X. (2010).
\newblock The dependent wild bootstrap.
\newblock {\em Journal of the American Statistical Association\/}~{\em
  105\/}(489), 218--235.

\bibitem[\protect\citeauthoryear{Shao and Wu}{Shao and Wu}{2007}]{Shao:Wu:2007}
Shao, X. and W.~B. Wu (2007).
\newblock Asymptotic spectral theory for nonlinear time series.
\newblock {\em The Annals of Statistics\/}~{\em 35\/}(4), 1773--1801.

\bibitem[\protect\citeauthoryear{Smeekes}{Smeekes}{2013}]{Smeekes:2013}
Smeekes, S. (2013).
\newblock Detrending bootstrap unit root tests.
\newblock {\em Econometric Reviews\/}~{\em 32\/}(8), 869--891.

\bibitem[\protect\citeauthoryear{Smeekes and Taylor}{Smeekes and
  Taylor}{2012}]{Smeekes:Taylor:2012}
Smeekes, S. and A.~M.~R. Taylor (2012).
\newblock Bootstrap union tests for unit roots in the presence of nonstationary
  volatility.
\newblock {\em Econometric Theory\/}~{\em 28\/}(2), 422--456.

\bibitem[\protect\citeauthoryear{Smeekes and Urbain}{Smeekes and
  Urbain}{2014}]{Smeekes:Urbain:2014}
Smeekes, S. and J.-P. Urbain (2014).
\newblock {A multivariate invariance principle for modified wild bootstrap
  methods with an application to unit root testing}.
\newblock Technical report.

\bibitem[\protect\citeauthoryear{St{\u{a}}ric{\u{a}} and
  Granger}{St{\u{a}}ric{\u{a}} and Granger}{2005}]{Starica:Granger:2005}
St{\u{a}}ric{\u{a}}, C. and C.~Granger (2005).
\newblock Nonstationarities in stock returns.
\newblock {\em Review of Economics and Statistics\/}~{\em 87\/}(3), 503--522.

\bibitem[\protect\citeauthoryear{Stock and Watson}{Stock and
  Watson}{1999}]{Stock:Watson:1999}
Stock, J. and M.~Watson (1999).
\newblock A comparison of linear and nonlinear univariate models for
  forecasting macroeconomic time series.
\newblock In R.~Engle and H.~White (Eds.), {\em Cointegration, Causality and
  Forecasting: A Festschrift for Clive W.J. Granger}, pp.\  1--44. Oxford:
  Oxford University Press.

\bibitem[\protect\citeauthoryear{Swensen}{Swensen}{2003}]{Swensen:2003}
Swensen, A.~R. (2003).
\newblock Bootstrapping unit root tests for integrated processes.
\newblock {\em Journal of Time Series Analysis\/}~{\em 24\/}(1), 99--126.

\bibitem[\protect\citeauthoryear{Wu}{Wu}{1986}]{Wu:1986}
Wu, C. F.~J. (1986).
\newblock Jackknife, bootstrap and other resampling methods in regression
  analysis (with discussion).
\newblock {\em The Annals of Statistics\/}~{\em 14\/}(4), 1261--1350.

\bibitem[\protect\citeauthoryear{Wu}{Wu}{2005}]{Wu:2005}
Wu, W.~B. (2005).
\newblock Nonlinear system theory: Another look at dependence.
\newblock {\em Proceedings of the National Academy of Sciences of the United
  States of America\/}~{\em 102\/}(40), 14150--14154.

\bibitem[\protect\citeauthoryear{Wu and Zhou}{Wu and Zhou}{2011}]{Wu:Zhou:2011}
Wu, W.~B. and Z.~Zhou (2011).
\newblock {Gaussian approximations for non-stationary multiple time series.}
\newblock {\em Statistica Sininca\/}~{\em 21\/}(3), 1397--1413.

\bibitem[\protect\citeauthoryear{Zhou}{Zhou}{2013}]{Zhou:2013}
Zhou, Z. (2013).
\newblock Heteroscedasticity and autocorrelation robust structural change
  detection.
\newblock {\em Journal of the American Statistical Association\/}~{\em
  108\/}(502), 726--740.

\bibitem[\protect\citeauthoryear{Zhou and Wu}{Zhou and Wu}{2009}]{Zhou:Wu:2009}
Zhou, Z. and W.~B. Wu (2009).
\newblock Local linear quantile estimation for nonstationary time series.
\newblock {\em The Annals of Statistics\/}~{\em 37\/}(5B), 2696--2729.

\end{thebibliography}


\begin{thebibliography}{}

\bibitem[\protect\citeauthoryear{Chang and Park}{Chang and
  Park}{2003}]{Chang:Park:2003}
Chang, Y. and J.~Y. Park (2003).
\newblock A sieve bootstrap for the test of a unit root.
\newblock {\em Journal of Time Series Analysis\/}~{\em 24\/}(4), 379--400.

\bibitem[\protect\citeauthoryear{Kurtz}{Kurtz}{2001}]{Kurtz:2001}
Kurtz, T.~G. (2001).
\newblock {\em {Lectures on Stochastic Analysis}}.
\newblock University of Wisconsin-Madison: available at
  http://www.math.wisc.edu/~kurtz/735/main735.pdf.

\bibitem[\protect\citeauthoryear{Paparoditis and Politis}{Paparoditis and
  Politis}{2003}]{Paparoditis:Politis:2003}
Paparoditis, E. and D.~N. Politis (2003).
\newblock Residual-based block bootstrap for unit root testing.
\newblock {\em Econometrica\/}~{\em 71\/}(3), 813--855.

\bibitem[\protect\citeauthoryear{Phillips}{Phillips}{1987}]{Phillips:1987b}
Phillips, P. C.~B. (1987).
\newblock Towards a unified asymptotic theory for autoregression.
\newblock {\em Biometrika\/}~{\em 74\/}(3), 535--547.

\bibitem[\protect\citeauthoryear{Politis, Romano, and Wolf}{Politis
  et~al.}{1999}]{Politis:Romano:Wolf:1999}
Politis, D.~N., J.~P. Romano, and M.~Wolf (1999).
\newblock {\em Subsampling}.
\newblock New York: Springer.

\bibitem[\protect\citeauthoryear{Rosenblatt}{Rosenblatt}{1985}]{Rosenblatt:1985}
Rosenblatt, M. (1985).
\newblock {\em Stationary Squences and Random Fields}.
\newblock Birkh{\~u}ser.

\bibitem[\protect\citeauthoryear{Shao and Yu}{Shao and Yu}{1996}]{Shao:Yu:1996}
Shao, Q.-M. and H.~Yu (1996).
\newblock Weak convergence for weighted empirical processes of dependent
  sequences.
\newblock {\em The Annals of Probability\/}~{\em 24\/}(4), 2098--2127.

\bibitem[\protect\citeauthoryear{Shao}{Shao}{2010}]{Shao:2010b}
Shao, X. (2010).
\newblock The dependent wild bootstrap.
\newblock {\em Journal of the American Statistical Association\/}~{\em
  105\/}(489), 218--235.

\bibitem[\protect\citeauthoryear{Shao and Wu}{Shao and Wu}{2007}]{Shao:Wu:2007}
Shao, X. and W.~B. Wu (2007).
\newblock Asymptotic spectral theory for nonlinear time series.
\newblock {\em The Annals of Statistics\/}~{\em 35\/}(4), 1773--1801.

\bibitem[\protect\citeauthoryear{Wu}{Wu}{2005}]{Wu:2005}
Wu, W.~B. (2005).
\newblock Nonlinear system theory: Another look at dependence.
\newblock {\em Proceedings of the National Academy of Sciences of the United
  States of America\/}~{\em 102\/}(40), 14150--14154.

\bibitem[\protect\citeauthoryear{Wu and Shao}{Wu and Shao}{2004}]{Wu:Shao:2004}
Wu, W.~B. and X.~Shao (2004).
\newblock Limit theorems for iterated random functions.
\newblock {\em Journal of Applied Probability\/}~{\em 41\/}(2), 425--436.

\bibitem[\protect\citeauthoryear{Zhou}{Zhou}{2013}]{Zhou:2013}
Zhou, Z. (2013).
\newblock Heteroscedasticity and autocorrelation robust structural change
  detection.
\newblock {\em Journal of the American Statistical Association\/}~{\em
  108\/}(502), 726--740.

\end{thebibliography}
\bibliographystyle{chicago}

\clearpage
\newpage

\clearpage
\newpage
\begin{table}[ht]
\caption{Empirical sizes for DWB, RWB, and RDWB for MA models with $\phi_i(s)$ and $\omega_j(s)$ based on 2000 Monte-Carlo replications and 1000 Bootstrap replications under $\rho=1$.
The nominal level is 5\%.}\small
\label{sizetable:MA}
\centering
\begin{tabular}{rr|rr|rr|rr||rr|rr|rr}
  \hline
 & &\multicolumn{6}{|c||}{$n=100$} &\multicolumn{6}{|c}{$n=400$} \\\hline
  & & \multicolumn{2}{|c|}{DWB}&\multicolumn{2}{|c|}{RWB} &\multicolumn{2}{|c||}{RDWB}  & \multicolumn{2}{|c|}{DWB}&\multicolumn{2}{|c|}{RWB} &\multicolumn{2}{|c}{RDWB}  \\ \hline
$i$&$j$& $T_n$ & $t_n$ & $T_n$ & $t_n$  & $T_n$ & $t_n$ & $T_n$ & $t_n$& $T_n$ & $t_n$ & $T_n$ & $t_n$\\
  \hline
\multirow{5}{*}{1} &1 & 2.9 & 3.1 & 4.8 & 5.1 & 4.7 & 4.7 & 3.9 & 3.8 & 4.9 & 4.5 & 4.5 & 4.4 \\ 
  &2 & 3.2 & 3.2 & 4.5 & 4.5 & 5.0 & 4.9 & 3.1 & 3.3 & 4.4 & 4.5 & 4.5 & 4.5 \\ 
  &3 & 4.3 & 4.8 & 4.2 & 4.9 & 4.9 & 5.3 & 4.5 & 4.3 & 5.3 & 5.5 & 5.9 & 5.9 \\ 
  &4 & 4.3 & 4.3 & 3.5 & 3.6 & 5.3 & 5.3 & 4.3 & 4.5 & 5.5 & 5.7 & 5.9 & 6.0 \\ 
  &5 & 2.8 & 2.9 & 3.9 & 4.0 & 4.1 & 4.0 & 3.5 & 3.6 & 4.5 & 5.0 & 4.9 & 5.0 \\ \hline
  \multirow{5}{*}{2} &1 & 82.2 & 81.9 & 20.5 & 20.4 & 19.8 & 20.0 & 78.5 & 78.4 & 12.8 & 12.8 & 12.2 & 12.0 \\ 
  &2 & 87.2 & 86.8 & 22.9 & 23.1 & 22.3 & 22.5 & 80.3 & 80.2 & 11.1 & 11.0 & 10.6 & 10.7 \\ 
  &3 & 95.5 & 94.5 & 21.6 & 20.4 & 21.4 & 20.6 & 96.2 & 95.8 & 13.0 & 13.0 & 12.4 & 12.4 \\ 
  &4 & 80.9 & 80.7 & 17.4 & 17.4 & 18.4 & 18.3 & 73.5 & 73.3 & 10.8 & 10.8 & 11.1 & 11.2 \\ 
  &5 & 91.8 & 91.2 & 23.9 & 23.9 & 23.4 & 23.5 & 87.3 & 87.1 & 13.6 & 13.6 & 12.8 & 12.8 \\ \hline
  \multirow{5}{*}{3} &1 & 3.5 & 3.7 & 5.5 & 5.1 & 5.0 & 5.0 & 3.5 & 3.5 & 4.9 & 4.8 & 4.7 & 4.5 \\ 
  &2 & 4.0 & 3.8 & 5.1 & 5.0 & 5.2 & 5.0 & 3.7 & 3.6 & 5.3 & 5.5 & 4.8 & 5.1 \\ 
  &3 & 5.2 & 5.4 & 5.3 & 5.9 & 6.1 & 6.3 & 4.7 & 4.8 & 7.8 & 7.8 & 7.4 & 7.1 \\ 
  &4 & 4.7 & 5.0 & 5.0 & 5.3 & 5.5 & 5.8 & 4.2 & 4.0 & 6.8 & 6.8 & 6.2 & 6.3 \\ 
  &5 & 4.2 & 4.3 & 6.2 & 5.9 & 6.0 & 5.8 & 4.1 & 3.8 & 6.3 & 5.8 & 5.2 & 5.2 \\ \hline
  \multirow{5}{*}{4} & 1& 4.0 & 4.0 & 4.4 & 4.3 & 5.0 & 4.9 & 3.6 & 4.0 & 4.9 & 4.8 & 4.6 & 4.7 \\ 
  &2 & 4.0 & 4.0 & 5.1 & 4.7 & 5.2 & 5.0 & 3.2 & 3.4 & 4.2 & 4.0 & 4.0 & 4.2 \\ 
  &3 & 6.0 & 6.7 & 4.9 & 6.4 & 5.3 & 6.6 & 4.9 & 5.2 & 9.5 & 9.0 & 7.7 & 7.5 \\ 
  &4 & 5.7 & 5.7 & 2.8 & 2.8 & 4.8 & 4.9 & 5.5 & 5.5 & 5.5 & 5.2 & 6.2 & 6.2 \\ 
  &5 & 3.5 & 3.1 & 4.1 & 3.5 & 4.7 & 4.2 & 3.6 & 3.7 & 5.4 & 5.2 & 4.9 & 4.7 \\ \hline
  \multirow{5}{*}{5} &1 & 7.4 & 7.6 & 6.6 & 6.5 & 6.2 & 6.4 & 6.2 & 5.9 & 8.3 & 7.9 & 5.7 & 5.2 \\ 
  &2 & 5.9 & 6.1 & 5.1 & 4.9 & 5.1 & 5.2 & 6.7 & 6.9 & 7.3 & 7.3 & 6.2 & 6.5 \\ 
  &3 & 8.5 & 8.0 & 7.8 & 10.3 & 7.3 & 9.2 & 5.9 & 6.5 & 16.4 & 15.2 & 10.1 & 10.0 \\ 
  &4 & 8.7 & 8.6 & 4.1 & 4.3 & 5.5 & 5.5 & 7.4 & 7.5 & 6.6 & 6.3 & 5.7 & 5.9 \\ 
  &5 & 5.1 & 5.1 & 6.4 & 6.2 & 5.3 & 5.5 & 4.3 & 4.5 & 9.2 & 8.8 & 5.1 & 5.1 \\ \hline
  \multirow{5}{*}{6} & 1& 33.6 & 33.1 & 11.3 & 11.1 & 10.0 & 9.5 & 25.4 & 24.9 & 10.5 & 10.5 & 8.0 & 8.1 \\ 
  &2 & 30.7 & 30.8 & 10.7 & 10.2 & 9.8 & 9.9 & 23.4 & 23.2 & 8.2 & 8.1 & 6.8 & 7.0 \\ 
  &3 & 35.4 & 33.7 & 12.8 & 12.6 & 12.1 & 12.2 & 25.9 & 25.5 & 16.2 & 15.3 & 11.6 & 11.1 \\ 
  &4 & 33.1 & 33.1 & 7.9 & 7.8 & 8.6 & 8.6 & 26.5 & 26.2 & 6.9 & 6.8 & 6.7 & 6.7 \\ 
  &5 & 27.3 & 26.7 & 11.4 & 10.5 & 9.8 & 9.7 & 20.0 & 19.7 & 9.2 & 8.8 & 6.7 & 6.7 \\ 
   \hline
\end{tabular}
\end{table}

\clearpage
\newpage
\begin{table}[ht]
\caption{Empirical sizes for DWB, RWB, and RDWB for AR models with $\phi_i(s)$ and $\omega_j(s)$ based on 2000 Monte-Carlo replications and 1000 Bootstrap replications under $\rho=1$.
The nominal level is 5\%.}\small
\label{sizetable:AR}
\centering
\begin{tabular}{rr|rr|rr|rr||rr|rr|rr}
  \hline
 & &\multicolumn{6}{|c||}{$n=100$} &\multicolumn{6}{|c}{$n=400$} \\\hline
  & & \multicolumn{2}{|c|}{DWB}&\multicolumn{2}{|c|}{RWB} &\multicolumn{2}{|c||}{RDWB}  & \multicolumn{2}{|c|}{DWB}&\multicolumn{2}{|c|}{RWB} &\multicolumn{2}{|c}{RDWB}  \\ \hline
$i$&$j$& $T_n$ & $t_n$ & $T_n$ & $t_n$  & $T_n$ & $t_n$ & $T_n$ & $t_n$& $T_n$ & $t_n$ & $T_n$ & $t_n$\\
  \hline
\multirow{5}{*}{1} &1 & 0.2 & 0.5 & 3.8 & 4.2 & 4.0 & 4.2 & 0.8 & 0.8 & 3.5 & 3.8 & 3.7 & 3.9 \\ 
&  2 & 0.8 & 0.6 & 4.0 & 4.3 & 4.6 & 4.7 & 0.4 & 0.6 & 3.8 & 3.7 & 4.0 & 4.1 \\ 
 & 3 & 0.5 & 0.6 & 5.5 & 6.9 & 5.9 & 7.2 & 0.7 & 1.1 & 3.8 & 4.6 & 4.3 & 5.1 \\ 
  &4 & 0.3 & 0.4 & 1.8 & 1.9 & 2.8 & 2.9 & 1.0 & 1.1 & 5.1 & 5.0 & 5.6 & 5.6 \\ 
 & 5 & 0.4 & 0.4 & 3.2 & 3.3 & 3.5 & 3.8 & 0.5 & 0.7 & 4.0 & 3.9 & 4.1 & 4.0 \\ \hline
  \multirow{5}{*}{2} &1 & 45.1 & 45.0 & 7.8 & 7.6 & 7.5 & 7.7 & 32.9 & 33.0 & 5.3 & 5.3 & 5.7 & 5.5 \\ 
  &2 & 45.1 & 45.1 & 6.2 & 6.0 & 7.3 & 7.2 & 32.9 & 33.0 & 5.2 & 5.2 & 5.5 & 5.5 \\ 
  &3 & 62.4 & 61.5 & 10.4 & 9.6 & 11.0 & 10.6 & 53.2 & 52.4 & 9.7 & 9.3 & 9.8 & 9.8 \\ 
  &4 & 44.1 & 44.2 & 5.5 & 5.5 & 6.6 & 6.7 & 34.8 & 34.8 & 6.4 & 6.4 & 6.8 & 6.8 \\ 
  &5 & 51.7 & 51.5 & 7.6 & 7.4 & 9.0 & 9.0 & 41.4 & 41.2 & 5.5 & 5.7 & 6.0 & 6.0 \\ \hline
  \multirow{5}{*}{3} &1 & 0.5 & 0.6 & 4.8 & 5.0 & 4.0 & 4.3 & 0.9 & 1.0 & 5.8 & 5.8 & 4.8 & 4.9 \\ 
  &2 & 0.4 & 0.5 & 4.5 & 4.5 & 4.1 & 4.1 & 0.7 & 0.7 & 4.1 & 3.9 & 3.5 & 3.6 \\ 
  &3 & 0.7 & 0.8 & 7.3 & 9.8 & 6.8 & 8.9 & 1.1 & 1.1 & 5.2 & 5.9 & 4.9 & 5.7 \\ 
  &4 & 0.2 & 0.4 & 3.2 & 3.2 & 3.3 & 3.5 & 0.9 & 1.0 & 6.2 & 5.8 & 5.5 & 5.3 \\ 
  &5 & 0.5 & 0.9 & 3.8 & 4.2 & 3.5 & 4.0 & 0.9 & 0.8 & 5.1 & 5.0 & 4.9 & 4.7 \\ \hline
  \multirow{5}{*}{4} &1 & 1.7 & 1.7 & 3.5 & 3.5 & 3.6 & 3.7 & 2.1 & 2.0 & 6.5 & 5.7 & 5.1 & 4.8 \\ 
  &2 & 1.7 & 1.7 & 4.0 & 4.1 & 3.9 & 3.9 & 1.8 & 1.8 & 6.3 & 6.3 & 5.4 & 5.2 \\ 
  &3 & 1.5 & 1.5 & 6.1 & 11.9 & 6.0 & 11.3 & 1.8 & 2.1 & 9.4 & 10.5 & 7.4 & 8.6 \\ 
  &4 & 3.4 & 3.3 & 2.1 & 2.0 & 3.6 & 3.6 & 3.7 & 3.8 & 3.9 & 3.8 & 5.4 & 5.1 \\ 
  &5 & 1.1 & 1.0 & 4.3 & 4.3 & 3.6 & 3.8 & 1.2 & 1.4 & 8.8 & 8.0 & 5.4 & 5.1 \\ \hline
  \multirow{5}{*}{5} &1 & 3.1 & 3.4 & 6.6 & 6.3 & 5.2 & 5.3 & 2.8 & 3.0 & 9.3 & 8.3 & 5.5 & 5.0 \\ 
  &2 & 3.1 & 2.9 & 4.2 & 4.2 & 3.9 & 3.7 & 3.5 & 3.8 & 9.2 & 8.7 & 6.0 & 6.3 \\ 
  &3 & 3.2 & 3.1 & 7.5 & 15.2 & 6.9 & 14.1 & 1.8 & 2.0 & 13.5 & 14.0 & 9.2 & 10.5 \\ 
  &4 & 6.6 & 6.3 & 3.1 & 3.0 & 4.5 & 4.2 & 5.8 & 5.7 & 5.8 & 5.4 & 5.8 & 5.5 \\ 
  &5 & 2.1 & 2.2 & 6.4 & 6.3 & 4.2 & 4.3 & 2.3 & 2.2 & 11.2 & 10.3 & 6.2 & 6.4 \\ \hline
  \multirow{5}{*}{6} & 1& 20.8 & 20.6 & 5.8 & 5.6 & 6.3 & 6.2 & 15.2 & 15.3 & 5.5 & 5.5 & 5.5 & 5.5 \\ 
  &2& 18.4 & 18.6 & 6.0 & 5.9 & 6.5 & 6.5 & 13.2 & 13.3 & 5.2 & 5.0 & 4.8 & 4.8 \\ 
  &3 & 26.1 & 25.4 & 9.4 & 9.3 & 9.3 & 9.3 & 17.3 & 16.8 & 10.9 & 10.1 & 8.8 & 8.6 \\ 
  &4 & 23.4 & 23.6 & 4.7 & 4.8 & 6.4 & 6.4 & 16.2 & 16.2 & 5.5 & 5.5 & 5.9 & 5.8 \\ 
 &5 & 18.9 & 18.2 & 5.7 & 5.7 & 6.2 & 6.5 & 12.7 & 12.2 & 5.7 & 5.4 & 5.1 & 5.0 \\ 
   \hline
\end{tabular}
\end{table}

\clearpage
\newpage

\end{document}


\appendix
\begin{center}
{\bf\large Supplementary Material for ``Bootstrap-Assisted Unit Root Testing With Piecewise Locally Stationary Errors"}
\end{center}
\centerline{\textsc{Yeonwoo Rho$^1$ and Xiaofeng Shao$^2$}}
\centerline {\it  Michigan Technological University$^1$}
\centerline {\it  University of Illinois at Urbana-Champaign$^2$}
\bigskip
\centerline{\today}
\bigskip
\bigskip

This supplementary material contains  all  technical proofs for results (Section \ref{append:tech}),   implementation details for  the DWB and RDWB methods (Section \ref{append:MV}), and full power curves for all the models (Section \ref{append:power}).

\section{Technical Appendix}\label{append:tech}

The symbols $O_p(1)$ and $o_p(1)$ signify being bounded in probability and convergence to zero in probability, respectively.
Denote by $P^*,~E^*,$ and $\var^*$ the probability, expectation, and variance, respectively, conditional on data $\mathcal{X}_n=(X_{1,n},\ldots,X_{n,n})$.
For notational simplicity, the dependence of  $X_{t,n}$, $u_{t,n}$, and $W_{t,n}$ on $n$ are often suppressed, and these quantities are written as $X_t$, $u_t$, and $W_t$, respectively.
For a sequence of random variables $\{Y_n\}$, $Y_n=o_{p}^*(1)$ in probability is used if for any $\epsilon>0$,
$P^*\{|Y_n|>\epsilon\}\to0$ {\rm in~probability},
as defined in \citet[p.386]{Chang:Park:2003}.
We define $S_t=S_{t,n}=\sum_{i=1}^t u_{i,n}$.
The positive constant $C$ is generic and may vary from place to place.
The symbol $\mathcal{I}_j$ is used in different places to indicate different objects.
For notational simplicity, we often write $G(s,\mathcal{F}_t):=G_{\zeta_s}(s,\mathcal{F}_t)$ and $c(s;h):=c_{\zeta_s}(s;h)$,
omitting the subscript $\zeta_s$, where $\zeta_s=j$ such that $s\in[b_j,b_{j+1})$ and $\zeta_s=\tau$ if $s=1$.
Let $\gamma_h(r)=\int_0^rc(s;h)ds$.
Notice that by definition, $\gamma_0(1)=\sigma_u^2$, and these symbols are interchangeably used in the proofs.

Recall that
$\mathcal{F}_t=(\ldots,\varepsilon_{t-1},\varepsilon_{t})$ with
$\varepsilon_t$ i.i.d. (0,1), and $\{\varepsilon_t'\}$ is an  i.i.d.
copy of $\{\varepsilon_t\}$. Following \cite{Wu:2005}, for $I\subset\Z$,
define $\mathcal{F}_{t,I}$ be the same as $\mathcal{F}_t$ except
that $\varepsilon_j$ is replaced by $\varepsilon_j$ for $j\in
I$. In particular, for $i\leq t$,
$\mathcal{F}_{t,\{i\}}=(\ldots,\varepsilon_{i-1},\varepsilon_{i}',\varepsilon_{i+1},\ldots,\varepsilon_t)$.
Denote by $\mathcal{F}_{t,i}^*=\mathcal{F}_{t,\{k\in\Z:k\leq i\}}$.

To keep the proofs concise,  the case with no deterministic trend functions, i.e., $\beta\equiv0$, is presented.
The statements in Theorems 2.1, 3.1, and 3.2 hold by replacing $B_{\sigma}(r)$ with $B_{\sigma|Z}(r)$ and $X_t$ with $\widehat{X}_t$.
The following four lemmas prove some basic properties of $\{u_t\}$ and $\{X_t\}$ that are useful in the subsequent proofs.

\begin{lemma}\label{lemma1-1}
Assume (A1)-(A4). Fix $j\in\{0,1,\ldots,\tau\}$.
\begin{itemize}
\item[(i)] For any $t,t'\in[b_j n,b_{j+1}n)$, $\big|\cov(u_{t},u_{t'})-c_j(t/n;|t-t'|)\big|\leq C(|t-t'|/n)$.
\item[(ii)] For any $s\neq s'\in[b_j,b_{j+1}]$, $\big|c_j(s;h)-c_j(s';h)\big|\leq  C|s-s'|$ uniformly over $h\in\field{N}$.
\item[(iii)] For any $\rho>0$, $\sup_{s\in[b_j,b_{j+1}]}\sum_{h=0}^\infty \big|h^{\rho}c_j(s,h)\big|\leq C\sum_{h=0}^\infty h^{\rho}\chi^h<\infty$.
\item[(iv)]  $\displaystyle\sup_{b_j\leq s\neq s'<b_{j+1}}\displaystyle\frac{|\sigma(s)-\sigma(s')|}{|s-s'|(-\log|s-s'|+1)}\leq C$.
\end{itemize}
In addition, if $j=\tau$, (i) and (iv) also hold for all $t,t'\in[b_{\tau} n,n]$ or for supremum over $\{b_\tau\leq s\neq s'\leq 1\}$.
\end{lemma}

\begin{proof}[Proof of Lemma \ref{lemma1-1}]
(i)
For all $t,t'\in[b_j n,b_{j+1}n)$,
$$ \cov(u_{t},u_{t'}) = c_j(t/n;|t-t'|) - 
\cov\big\{ G_j(t/n,\mathcal{F}_t), G_j(t/n,\mathcal{F}_{t'})-G_j(t'/n,\mathcal{F}_{t'})\big\}.$$
From the Cauchy-Schwartz inequality and (A1),
$\big|\cov\big\{G_j(t/n,\mathcal{F}_t),G_j(j/n,\mathcal{F}_{t'})-G_j(t'/n,\mathcal{F}_{t'})\big\}\big|
 \leq ||G_j(t/n,\mathcal{F}_t)||_2$ $ ||G_j(t/n,\mathcal{F}_{t'})-G_j(t'/n,\mathcal{F}_{t'})||_2 \leq C(|t-t'|/n),$
which completes the proof.
If $j=\tau$, the same argument holds for all $t,t'\in[b_{\tau} n,n]$.

(ii)
It follows from the triangular inequality, Cauchy-Schwartz inequality, and (A1) that for any $s\neq s'\in[b_j,b_{j+1}]$,
$|c_j(s;h)-c_j(s';h)|
 \leq ||G_j(s,\mathcal{F}_0)||_2||G_j(s,\mathcal{F}_h)-G_j(s',\mathcal{F}_h)||_2+||G_j(s',\mathcal{F}_h)||_2||G_j(s,\mathcal{F}_0)-G_j(s',\mathcal{F}_0)\}||_2
\leq C|s-s'|$
holds uniformly over $h\in\field{N}$.

    (iii)
This is a straightforward consequence of Lemma A.1 in \cite{Shao:Wu:2007}, Theorem 1 in \cite{Wu:2005}, and the assumption (A3).

(iv)
It follows from (A3) that $|c_j(s;h)-c_j(s';h)|\leq 2C\chi^h$ for all $h\in\field{N}$ and $s,s'\in[b_j,b_{j+1})$.
Let $m$ be the smallest positive integer such that $\chi^m\leq|s-s'|$.
Then using (ii),
$|\sigma(s)-\sigma(s')|\leq \sum_{h=-\infty}^\infty|c_j(s;h)-c_j(s';h)|\leq C(\sum_{|h|\leq m-1}|s-s'|+\sum_{|h|\geq m}\chi^h)
\leq C\{m|s-s'|+\chi^m(1-\chi)^{-1}\}\leq C|s-s'|(-\log\chi)^{-1}(-\log|s-s'|)+C(1-\chi)^{-1}|s-s'|\leq C|s-s'|(-\log|s-s'|+1)$.
Notice that constant $C$'s do not depend on $s$ or $s'$. Thus the proof is complete.
If $j=\tau$, the same argument holds for $s,s'\in[b_\tau,1]$.
\end{proof}

\begin{lemma}\label{lem:uiuih}
Under the conditions (A2)-(A3), for any $i=1,\ldots,n-h$,
$h=0,\ldots,n-i$,
$$|E(u_iu_{i+h})|\leq C\chi^h,$$
where $C$ is a constant that does not depend on $h$, $i$, or $n$ and $\chi$ is from (A3).
\end{lemma}

\begin{proof}[Proof of Lemma \ref{lem:uiuih}]
By definition,  $\mathcal{F}_i$ and $\mathcal{F}^*_{i+h,i}$ are independent.
Therefore,  $E\{G(i/n,\mathcal{F}_i)G((i+h)/n,\mathcal{F}^*_{i+h,i})\}=0$, and 
$$\begin{array}{lll}
E(u_iu_{i+h})&=&
E\left[G(i/n,\mathcal{F}_i)\{G((i+h)/n,\mathcal{F}_{i+h})-G((i+h)/n,\mathcal{F}_{i+h,\{i\}})\}\right]
\\ &&+ E\left[G(i/n,\mathcal{F}_i)\{G((i+h)/n,\mathcal{F}_{i+h,\{i\}})-G((i+h)/n,\mathcal{F}^*_{i+h,i})\}\right]
\end{array}$$
Then, by the
Cauchy-Schwartz inequality,
$$\begin{array}{lll}
|E(u_iu_{i+h})|
 &\leq&|| G(i/n,\mathcal{F}_i)||_2||G((i+h)/n,\mathcal{F}_{i+h})-G((i+h)/n,\mathcal{F}_{i+h,\{i\}})||_2
\\ &&+||G(i/n,\mathcal{F}_i)||_2||G((i+h)/n,\mathcal{F}_{i+h,\{i\}})-G((i+h)/n,\mathcal{F}^*_{i+h,i})||_2
\end{array}$$
By (A2),  $|| G(i/n,\mathcal{F}_i)||_2<C<\infty$, and
by (A3)
$||G((i+h)/n,\mathcal{F}_{i+h})-G((i+h)/n,\mathcal{F}_{i+h,\{i\}})||_2<||G((i+h)/n,\mathcal{F}_{i+h})-G((i+h)/n,\mathcal{F}_{i+h,\{i\}})||_4\leq
C\chi^h$. Thus the first term is bounded by $C\chi^h$, where $C$
does not depend on $h$, $i$, or $n$.

Now the proof is complete if the following statement is shown:
$$||G((i+h)/n,\mathcal{F}_{i+h,\{i\}})-G((i+h)/n,\mathcal{F}^*_{i+h,i})||_4\leq C\chi^h.$$
Define $\mathcal{F}^*_{i+h,\{i\},m}=\mathcal{F}_{i+h,A}$, where
$A=\{k\in\Z:k\leq i-m-1\}\bigcup\{i\}$. In particular, if $m=0$,
$\mathcal{F}^*_{i+h,\{i\},0}=\mathcal{F}^*_{i+h,i}$. Then
$||G((i+h)/n,\mathcal{F}_{i+h,\{i\}})-G((i+h)/n,\mathcal{F}^*_{i+h,i})||_4
=||\sum_{m=0}^\infty G((i+h)/n,\mathcal{F}^*_{i+h,\{i\},m})-G((i+h)/n,\mathcal{F}^*_{i+h,\{i\},m+1})||_4
\leq\sum_{m=0}^\infty||G((i+h)/n,\mathcal{F}^*_{i+h,\{i\},m})-G((i+h)/n,\mathcal{F}^*_{i+h,\{i\},m+1})||_4
\leq C\sum_{m=0}^\infty \chi^{h+m+1}
\allowbreak
=C\chi^{h+1}/(1-\chi) \leq C\chi^h,
$
where the last $C$ does not depend on $h$, $i$, or $n$. Thus the proof
is complete.
\end{proof}

Let ${\rm cum}(Y_0,Y_1,Y_2,Y_3)$ denote the fourth-order cumulant.
When $E(Y_i)=0,~i=0,1,2,3$,  the following relation (see page 36
in \cite{Rosenblatt:1985}, for example) is often used in the subsequent proofs:
\begin{equation}\label{cumrel}
\cov(Y_0Y_1,Y_2Y_3)=E(Y_0Y_2)E(Y_1Y_3)+E(Y_0Y_3)E(Y_1Y_2)+{\rm
cum}(Y_0,Y_1,Y_2,Y_3).
\end{equation}

\begin{lemma}\label{cumulant}
Assume (A1)-(A4). Then
$$\sup_{1\leq t_1\leq \ldots \leq t_4\leq n}|{\rm cum}(u_{t_1},u_{t_2},u_{t_3},u_{t_4})|\leq C\chi^{(t_4-t_1)/3},$$
with $\chi$ as in (A3).
\end{lemma}

\begin{proof}[Proof of Lemma \ref{cumulant}]
Let  $\mathcal{F}_{t}'=\mathcal{F}_{t,0}^*$ if $t>0$,
and $\mathcal{F}_{t}'=\mathcal{F}_t$  if $t\leq0$.
Define $\mathcal{F}'_{t,m}=\mathcal{F}_{t,\{k\in\Z:-m\leq k\leq0\}}$ for $m\geq 0$ and $t>0$.
The argument is similar to the proof of Proposition 2 in \cite{Wu:Shao:2004}.
Let $1\leq t_1\leq
\ldots \leq t_4\leq n$, and  $m_k=t_{k+1}-t_k$ for
$k\in\{1,2,3\}$. Since for a fixed $s\in[0,1]$, the process
$\{G(s,\mathcal{F}_t)\}_t$ is stationary, 
$$\begin{array}{lll}
&& {\rm cum}(u_{t_1},u_{t_2},u_{t_3},u_{t_4})
\\ &=& {\rm cum}\{G(t_1/n,\mathcal{F}_{t_1-t_k}),G(t_2/n,\mathcal{F}_{t_2-t_k})-G(t_2/n,\mathcal{F}'_{t_2-t_k}),
G(t_3/n,\mathcal{F}_{t_3-t_k}),G(t_4/n,\mathcal{F}_{t_4-t_k})\}
\\ && +{\rm cum}\{G(t_1/n,\mathcal{F}_{t_1-t_k}),G(t_2/n,\mathcal{F}'_{t_2-t_k}),
G(t_3/n,\mathcal{F}_{t_3-t_k})-G(t_3/n,\mathcal{F}'_{t_3-t_k}),G(t_4/n,\mathcal{F}_{t_4-t_k})\}
\\ && +{\rm cum}\{G(t_1/n,\mathcal{F}_{t_1-t_k}),G(t_2/n,\mathcal{F}'_{t_2-t_k}),
G(t_3/n,\mathcal{F}'_{t_3-t_k}),G(t_4/n,\mathcal{F}_{t_4-t_k})-G(t_4/n,\mathcal{F}'_{t_4-t_k})\}
\\ && +{\rm cum}\{G(t_1/n,\mathcal{F}_{t_1-t_k}),G(t_2/n,\mathcal{F}'_{t_2-t_k}),
G(t_3/n,\mathcal{F}'_{t_3-t_k}),G(t_4/n,\mathcal{F}'_{t_4-t_k})\}
\\ & := & \mathcal{I}_1+\mathcal{I}_2+\mathcal{I}_3+\mathcal{I}_4
\end{array}$$
due to the additive property of cumulants
[the property (iii) on page 35 in \cite{Rosenblatt:1985}].
First we claim that $\mathcal{I}_{4}=0$.
If $k=1$, $\mathcal{F}_{t_1-t_k}=\mathcal{F}_0$ is independent of
$\mathcal{F}'_{t_2-t_k}$,  $\mathcal{F}'_{t_3-t_k}$, $\mathcal{F}'_{t_4-t_k}$,
so $\mathcal{I}_4=0$ using the property (ii) on page 35 in \cite{Rosenblatt:1985}.
If $k=2$, then $\mathcal{F}_{t_2-t_k}'=\mathcal{F}_{0}'=\mathcal{F}_{0}$ by definition,
and $\mathcal{F}_{t_1-t_k}$ and $\mathcal{F}_{0}$ are independent of
 $\mathcal{F}'_{t_3-t_k}$, $\mathcal{F}'_{t_4-t_k}$, which leads to $\mathcal{I}_4=0$.
Similarly, if $k=3$,
$\mathcal{F}_{t_1-t_k}$, $\mathcal{F}_{t_2-t_k}'=\mathcal{F}_{t_2-t_k}$, and $\mathcal{F}_{0}$ are independent of
 $\mathcal{F}'_{t_4-t_k}$.
 Thus $\mathcal{I}_4=0$ for all $k=1,2,3$.
Also, notice that since $\mathcal{F}_t'=\mathcal{F}_t$ if $t\leq 0$,
it can be shown that $\mathcal{I}_1=0$ if $k=2$ and $\mathcal{I}_1=\mathcal{I}_2=0$ if $k=3$.
Thus the proof is done if the following statement is proved  for each $k=1,2,3$:
\begin{equation}\label{cum}
\max_{k\leq i\leq 3}|\mathcal{I}_{i}|\leq C\chi^{m_k}.
\end{equation}
Once (\ref{cum}) is shown, it follows that
for each $k=1,2,3$,
$|{\rm cum}(u_{t_1},u_{t_2},u_{t_3},u_{t_4})|\leq C\chi^{m_k}$.
Taking the minimum over $k$ for both sides yields
$|{\rm cum}(u_{t_1},u_{t_2},u_{t_3},u_{t_4})|
\leq
C\min_{k=1,2,3}\chi^{m_k} =C\chi^{\max_{k=1,2,3}m_k}
\allowbreak\leq  C\chi^{(t_4-t_1)/3}$, since
$t_4-t_1=\sum_{j=2}^4(t_j-t_{j-1})\leq
3\max_{j=2,3,4}(t_j-t_{j-1})=3\max_{k=1,2,3}m_k$.

The subsequent arguments prove (\ref{cum}). For each $k=1,2,3$, fix any $j=k+1,\ldots,4$.
Let
$Y_0=G(t_j/n,\mathcal{F}_{t_j-t_k})-G(t_j/n,\mathcal{F}'_{t_j-t_k})$
and $Y_1$, $Y_2$, and $Y_3$ be the other variables in
$\mathcal{I}_{j-1}$ so that we can write $\mathcal{I}_{j-1}={\rm
cum}(Y_0,Y_1,Y_2,Y_3)$.
Since
$
||Y_0||_4\leq
||\{G(t_j/n,\mathcal{F}_{t_j-t_k})-G(t_j/n,\mathcal{F}'_{t_j-t_k,0})\}||_4
+\sum_{m=0}^\infty ||\{G(t_j/n,\mathcal{F}'_{t_j-t_k,m})-G(t_j/n,\mathcal{F}'_{t_j-t_k,m+1})\}||_4
\allowbreak
\leq C\{\chi^{t_j-t_k}+\sum_{m=0}^\infty \chi^{t_j-t_k+m+1}\}\leq C\chi^{t_j-t_k}
$
holds by the triangular inequality and (A3), it follows that
\begin{equation}\label{Y0}
||Y_0||_4\leq C\chi^{t_j-t_k},
\end{equation}
where $C$ is a constant that does not depend on $t_j$, $j$, or $n$.
Observe that due to (\ref{cumrel}),
$\mathcal{I}_{j-1}=E(Y_0Y_1Y_2Y_3)-E(Y_0Y_1)E(Y_2Y_3)-E(Y_0Y_2)E(Y_1Y_3)-E(Y_0Y_3)E(Y_1Y_2)$.
By H\"{o}lder's inequality, (\ref{Y0}), and (A2), it follows that  $|E(Y_0Y_1Y_2Y_3)|\leq ||Y_0||_4||Y_1Y_2Y_3||_{4/3}\leq
C\chi^{t_j-t_k}$ and $|E(Y_0Y_i)|\leq ||Y_0||_2||Y_i||_{2}\leq
C\chi^{t_j-t_k}$.
Thus $|\mathcal{I}_{j-1}|\leq
C\chi^{t_j-t_k}\leq C\chi^{m_k}$, and (\ref{cum}) is proved.
\end{proof}

\begin{lemma}\label{lem:xt2}
Assume (A1)-(A4).
Under the local alternatives $\rho=1+c/n$, $c\leq 0$,
$$\sup_{1\leq t\leq n}\{{E(X_t^2)}/{t}\}\leq C~~~~{\rm and}~~~~\sup_{1\leq t\leq n}\{{E(X_t^4)}/{t^2}\}\leq C,$$
where
$C$ is a positive constant that does not depend on $n$.
\end{lemma}

\begin{proof}[Proof of Lemma \ref{lem:xt2}]
In this proof, all $C$'s  indicate a constant that do not depend on $t$ or $n$.
Suppose $1\leq i_1\leq i_2\leq i_3\leq i_4\leq t$ for some $t=1,\ldots,n$.
By (\ref{cumrel}) and Lemmas \ref{lem:uiuih} and \ref{cumulant},
$$\begin{array}{lll}
E(u_{i_1}u_{i_2}u_{i_3}u_{i_4})&=&E(u_{i_1}u_{i_2})E(u_{i_3}u_{i_4})+E(u_{i_1}u_{i_3})E(u_{i_2}u_{i_4})+E(u_{i_1}u_{i_4})E(u_{i_2}u_{i_3})
\\&&+{\rm cum}(u_{i_1},u_{i_2},u_{i_3},u_{i_4})
\\ &\leq& C(\chi^{i_2-i_1}\chi^{i_4-i_3}+\chi^{i_3-i_1}\chi^{i_4-i_2}+\chi^{i_4-i_1}\chi^{i_3-i_2}
+\chi^{(i_4-i_1)/3}).
\end{array}$$
It follows that
$ E(X_t^4)=E(\sum_{i=1}^t\rho^{t-i}u_i)^4=24\sum_{1\leq i_1\leq i_2\leq i_3\leq i_4\leq t}\rho^{4t-i_1-i_2-i_3-i_4}E(u_{i_1}u_{i_2}u_{i_3}u_{i_4})\leq Ct^2,$
where the last inequality holds by observing the following four simple facts:
\begin{itemize}
\item[i.]
$\sum_{1\leq i_1\leq i_2\leq i_3\leq i_4\leq t}\chi^{(i_4-i_1)/3}=\sum_{h=0}^{t-1}(t-h)(h+1)^2\chi^{h/3}\leq Ct.$
\item[ii.]$\{\sum_{i_1,i_2}\chi^{|i_1-i_2|}\}\{\sum_{i_3,i_4}\chi^{|i_3-i_4|}\}\leq (Ct)^2.$
\item[iii.]$\chi~\mbox{is strictly positive}.$
\item[iv.]$\rho^{t-i}=(1+c/n)^{t-i}\leq 1~\mbox{for any}~ i\leq t.$
\end{itemize}
Similarly,
$E(X_t^2)=E(\sum_{i=1}^t\rho^{t-i}u_i)^2=2\sum_{1\leq i_1\leq i_2\leq t}\rho^{2t-i_1-i_2}E(u_{i_1}u_{i_2})\leq C\sum_{h=0}^{t-1}(t-h)\chi^h\leq Ct.$
\end{proof}

The following lemmas contain key results needed in the proof of Theorem 2.1 
and they may be of independent interest.
\begin{lemma}\label{lem:ut}
Assume (A1)-(A4).
\begin{itemize}
\item[(i)] $n^{-1/2}S_{\lf nr\rf}=n^{-1/2}\sum_{i=1}^{\lf nr\rf}u_{i}\Rightarrow B_\sigma(r)=\int_0^r\sigma(s)dB(s)$.
\item[(ii)] For a fixed $r\in(0,1]$ and a fixed integer $h\geq0$,  $|n^{-1}\sum_{i=1}^{\lf nr\rf\wedge(n-h)}u_{i}u_{i+h}-\gamma_h(r)|=o_p(1)$.
Recall that $\gamma_h(r)=\int_0^rc_{\zeta_s}(s;h)ds$, where $\zeta_s=j$ such that $s\in[b_j,b_{j+1})$.
\end{itemize}
\end{lemma}

\begin{proof}[Proof of Lemma \ref{lem:ut}]
(i) Define a step function $\sigma_n(s)=\sigma(t/n)$ for $s\in[t/n,(t+1)/n)$ and $t=0,1,\ldots,n$, with $\sigma_n(1)=\sigma(1)$.
Let $\check{B}_{n,\sigma}(r)=\int_0^{\lf nr\rf/n}\sigma_n(s)dB(s)$ and $\tilde{B}_{n,\sigma}(r)=\int_0^r\sigma_n(s)dB(s)$.
Recall that $B_\sigma(r)=\int_0^r\sigma(s)dB(s)$.

By the triangle inequality,
$\sup_{r\in[0,1]}|\check{B}_{n,\sigma}(r)-B_\sigma(r)|\leq\sup_{r\in[0,1]}|\check{B}_{n,\sigma}(r)-\tilde{B}_{n,\sigma}(r)|+
\sup_{r\in[0,1]}|\tilde{B}_{n,\sigma}(r)-B_\sigma(r)|
=:\mathcal{I}_1+\mathcal{I}_2$.
It follows that $\mathcal{I}_1=o_p(1)$ because $\sup_{r\in[0,1]}|r-\lf nr\rf/n|\leq 1/n$ and
$\sup_{r\in[0,1]}|\int_{\lf nr\rf/n}^r\sigma_n(s)dB(s)|
\allowbreak
\leq C\sup_{t=1,\ldots,n}|B(t/n)-B((t-1)/n)|=o_p(1)$.
Notice that by Lemma \ref{lemma1-1} (iv),
$\sup_{r\in[0,1]}|\sigma_n(r)-\sigma(r)|
=\sup_{0\leq j\leq \tau}\sup_{b_j\leq s < b_{j+1}}|\sigma_n(s)-\sigma(s)|
=\sup_{0\leq j\leq \tau}\sup_{b_j\leq s < b_{j+1}}|\sigma(\lf ns\rf/n)-\sigma(s)|
\leq (\tau+1) C|\lf ns\rf/n-s|(-\log|\lf ns\rf/n-s|+1)
=O(n^{-1}\log n)=o(1)$.
Thus $\mathcal{I}_2=o_p(1)$ holds by \citet[Proposition 5.19]{Kurtz:2001}.
It follows that
\begin{equation}\label{lem:ut:i1}
\sup_{r\in[0,1]}|\check{B}_{n,\sigma}(r)-B_\sigma(r)|=o_p(1).
\end{equation}
From Proposition 5 in \cite{Zhou:2013}, on a richer probability space, there exist i.i.d.
standard normal random variables $V_1,\ldots,V_n$  such that
\begin{equation}\label{lem:ut:i2}
\sup_{r\in[0,1]}|n^{-1/2}S_{\lf nr\rf}-\hat{B}_{n,\sigma}(r)|=o_p(1),
\end{equation}
where $\hat{B}_{n,\sigma}(r)= n^{-1/2}\sum_{i=1}^{\lf nr\rf}\sigma(i/n)V_i$.
Since
$\{\hat{B}_{n,\sigma}(r)\}_{r\in[0,1]}\eqd\{\sum_{t=1}^{\lf nr\rf}\sigma(t/n)[B(t/n)-B\{(t-1)/n\}]\}_{r\in[0,1]}
\allowbreak \eqd\{\check{B}_{n,\sigma}(r)\}_{r\in[0,1]},$
\begin{equation}\label{lem:ut:i3}\hat{B}_{n,\sigma}(r)\Rightarrow B_\sigma(r)
\end{equation}
by (\ref{lem:ut:i1}).
Then (i) follows from (\ref{lem:ut:i2}) and (\ref{lem:ut:i3}).

\bigskip
(ii)
Define $Y_i=Y_{i,n}=u_{i}u_{i+h}-E(u_{i}u_{i+h})$. We claim that $|n^{-1}\sum_{i=1}^{\lf nr\rf\wedge(n-h)}Y_{i,n}|=o_p(1).$
Observe that by (\ref{cumrel}), for $i\geq i'$,
$E(Y_iY_{i'})=\cov(u_{i}u_{i+h},u_{i'}u_{i'+h})=E(u_{i}u_{i'})E(u_{i+h}u_{i'+h})+E(u_{i}u_{i'+h})E(u_{i+h}u_{i'})+{\rm cum}(u_{i},u_{i+h},u_{i'},u_{i'+h})
\leq C\chi^{2|i-i'|}+C\chi^{|i-i'-h|+|i+h-i'|}+C\chi^{|i+h-i'|/3}\leq C\chi^{|i-i'|/3}$,
where the first inequality is  due to Lemmas \ref{lem:uiuih} and \ref{cumulant}.
Then, by Chebyshev's inequality, for any $\delta>0$,
$P(|\sum_{i=1}^{\lf nr\rf\wedge(n-h)}Y_{i,n}|>n\delta)\leq (n\delta)^{-2}E(\sum_{i=1}^{\lf nr\rf\wedge(n-h)}Y_{i,n})^2
\leq (n\delta)^{-2}\sum_{i,i'=1}^{{\lf nr\rf\wedge(n-h)}}E(Y_{i,n}Y_{i',n})\leq C (n\delta)^{-2}\sum_{i,i'=1}^{{\lf nr\rf\wedge(n-h)}}\chi^{|i-i'|/3}
\leq (n\delta)^{-2}Cn
=o(1)$.
Therefore,
$|n^{-1}\sum_{i=1}^{n-h}\{u_{i}u_{i+h}-E(u_{i}u_{i+h})\}|=o_p(1).$

Now it remains to show that $|n^{-1}\sum_{i=1}^{\lf nr\rf\wedge(n-h)}E(u_{i}u_{i+h})-\gamma_h(r)|=o(1)$.
For $r\in(0,1]$, let $\mathcal{B}_{r}=\{i: i/n<b_j<(i+h)/n~\mbox{for some}~b_j~\mbox{and}~1\leq i\leq\lf nr\rf\wedge(n-h)\}$ and
$\tau_r$ be the number of break points in $(0,r)$.
Since $n^{-1}\sum_{i=1}^{\lf nr\rf\wedge(n-h)}E(u_{i}u_{i+h})=n^{-1}\sum_{i\not\in\mathcal{B}_{r}}E(u_{i}u_{i+h})
+n^{-1}\sum_{i\in\mathcal{B}_{r}}E(u_{i}u_{i+h})=\mathcal{I}_{1,r}+\mathcal{I}_{2,r}$,
it suffices to show that
\begin{equation}\label{lem:ut:ii}
\sup_{r\in(0,1]}|\mathcal{I}_{1,r}-\gamma_h(r)|=o(1)~~~~~{\rm and}~~~~~\sup_{r\in(0,1]}|\mathcal{I}_{2,r}|=o(1).
\end{equation}
For $\mathcal{I}_{1,r}$, it follows from Lemma \ref{lemma1-1} (i) that
$|\mathcal{I}_{1,r}-\gamma_h(r)|\leq n^{-1}\sum_{i\not\in\mathcal{B}_{r}}|E(u_{i}u_{i+h})-c_{\zeta_{i/n}}(i/n;h)|+ n^{-1}\sum_{i\in\mathcal{B}_{r}}|c_{\zeta_{i/n}}(i/n;h)|
\leq Ch/n$ holds for a constant $C$ that does not depend on $r$.
For $\mathcal{I}_{2,r}$,
$\sup_{r\in[0,1)}|\mathcal{I}_{2,r}|\leq \sup_{r\in[0,1)} C\tau_rh/n\leq C\tau h/n=O(h/n)$.
Thus (\ref{lem:ut:ii}) holds, and the proof is complete.
\end{proof}

\begin{lemma}\label{lem:xt}
Assume (A1)-(A4). Let $S_t=\sum_{i=1}^tu_{i,n}$. The following statements hold jointly.
\begin{itemize}
\item[(i)] For any $r\in(0,1]$, $n^{-2}\sum_{t=1}^{\lf nr\rf}S_{t-1}^2\cod\int_0^rB_\sigma^2(s)ds$.
\item[(ii)] For any $r\in(0,1]$,  $n^{-1}\sum_{t=1}^{\lf nr\rf}S_{t-1}u_{t}\cod 2^{-1}\left\{B_\sigma^2(r)-\gamma_0(r)\right\}$.
\item[(iii)] For any $r\in(0,1]$,  $n^{-1}\sum_{t=1}^{\lf nr\rf\wedge (n-h)}S_{t-1}u_{t+h}\cod 2^{-1}\left\{B_\sigma^2(r)-\gamma_0(r)\right\}-\sum_{k=1}^h\gamma_k(r)$
for any fixed integer $h\geq1$.
\item[(iv)] For any $r\in(0,1]$, $n^{-3/2}\sum_{t=1}^{\lf nr\rf}S_{t-1}\cod\int_0^r B_\sigma(s)ds$.
\end{itemize}
\end{lemma}

\begin{proof}[Proof of Lemma \ref{lem:xt}]
The proof can be done by standard arguments using the identity $2S_{t-1}u_{t}=S_{t}^2-S_{t-1}^2-u_{t}^2$,
the continuous mapping theorem, and Lemma \ref{lem:ut}.
\end{proof}

\begin{lemma}\label{lem:sn}
Assume (A1)-(A4). Under the local alternatives $\rho=1+c/n$, $c\leq 0$,
\begin{equation}\label{thm:NI:ii}
n^{-2}\sum_{t=1}^{\lf nr\rf}X_{t-1}^2\Rightarrow\int_0^rJ_{c,\sigma}^2(s)ds,
\end{equation}
\begin{equation}\label{thm:NI:iii}
n^{-1}\sum_{t=1}^{n}X_{t-1}u_t\cod \int_0^1J_{c,\sigma}(r)\sigma(r)dB(r)+2^{-1}\left\{\int_0^1\sigma^2(r)dr-\sigma_u^2\right\},
\end{equation}
and
\begin{equation}\label{sn:sigmau}
s_n^2=(n-2)^{-1}\sum_{t=1}^n(X_t-\widehat{\rho}_nX_{t-1})^2\cop\gamma_0(1)=\sigma_u^2~~~~~~{\rm for}~c=0.
\end{equation}
\end{lemma}

\begin{proof}[Proof of Lemma \ref{lem:sn}]
First observe that $e^{c/n}=1+c/n+O(n^{-2})$
so that $\rho_n=e^{c/n}+O(n^{-2})$.
Then
$X_t$ is asymptotically equivalent to $\sum_{j=1}^t e^{(t-j)c/n}u_j$, i,e,
$X_t=\sum_{j=1}^t\rho_n^{t-j}u_j=\sum_{j=1}^t e^{(t-j)c/n}u_j +O_p(n^{-3/2})$.
Following the argument in \cite{Phillips:1987b}, page 539, and using Lemma \ref{lem:ut} (i), it can be shown that
$n^{-1/2} \sum_{j=1}^{\lf nr\rf} e^{(t-j)c/n}u_j \Rightarrow J_{c,\sigma}(r),$
which implies
\begin{equation}\label{lem:NI:ut}
n^{-1/2}X_{\lf nr\rf}\Rightarrow J_{c,\sigma}(r).
\end{equation}
Then (\ref{thm:NI:ii}) follows from the continuous mapping theorem.
For (\ref{thm:NI:iii}),  squaring both sides of (3) yields 
$X_t^2=(1+cn^{-1})^2X_{t-1}^2+u_t^2+2(1+cn^{-1})X_{t-1}u_t$
so that
$\sum_{t=1}^nX_{t}^2=(1+2cn^{-1})\sum_{t=1}^nX_{t-1}^2+\sum_{t=1}^nu_t^2+2\sum_{t=1}^nX_{t-1}u_t+O_p(1)$.
Thus
$$\begin{array}{lll}
\textstyle 2n^{-1}\sum_{t=1}^nX_{t-1}u_t&=&n^{-1}X_n^2-2cn^{-2}\sum_{t=1}^nX_{t-1}^2-n^{-1}\sum_{t=1}^nu_t^2+O_p(n^{-1})
\\[1mm]&\cod&J_{c,\sigma}^2(1)-2c\int_0^1J_{c,\sigma}^2(r)dr-\sigma_u^2
\\[1mm] &=& 2\int_0^1J_{c,\sigma}(r)\sigma(r)dB(r)+\{\int_0^1\sigma^2(r)dr-\sigma_u^2\},
\end{array}$$
which implies (\ref{thm:NI:iii}).
Here, the last equality is due to
$$\textstyle J_{c,\sigma}^2(1)=\int_0^1\sigma^2(r)dr+2c\int_0^1J_{c,\sigma}^2(r)dr+2\int_0^1J_{c,\sigma}(r)\sigma(r)dB(r),$$
which follows from It\^{o}'s formula.\footnote{Recall that $J_{c,\sigma}(r)$ is defined as $dJ_{c,\sigma}(r)=cJ_{c,\sigma}(r)dr+\sigma(r)dB(r)$.
Using It\^{o}'s formula, we can derive
$J_{c,\sigma}^2(r)=J_{c,\sigma}^2(0)+\int_0^r2cJ_{c,\sigma}^2(s)ds+\int_0^r2\sigma(s)J_{c,\sigma}(s)dB(s)+\int_0^r\sigma^2(s)ds$,
which leads to the desired result.}

For (\ref{sn:sigmau}), notice that $s_n^2=(n-2)^{-1}\sum_{t=1}^n(X_t-\widehat{\rho}_nX_{t-1})^2
=(n-2)^{-1}\sum_{t=1}^n u_t^2+(n-2)^{-1}(\widehat{\rho}_n-\rho )^2\sum_{t=1}^nX_{t-1}^2
+2(n-2)^{-1}(\rho-\widehat{\rho}_n)\sum_{t=1}^nX_{t-1}u_t
:=\mathcal{I}_1+\mathcal{I}_2+\mathcal{I}_3$.
Here $\mathcal{I}_1\cop\gamma_0(1)$ by Lemma \ref{lem:ut} (ii).
For $\mathcal{I}_2$, under the null hypothesis $\rho=1$, $S_t=X_t=\sum_{i=1}^tu_i$.
By Lemma \ref{lem:xt} (i) and (ii), $\widehat{\rho}_n-\rho =O_p(n^{-1})$ and
$\sum_{t=1}^nX_{t-1}^2=O_p(n^2)$ so that $\mathcal{I}_2=O_p(n^{-1})$.
Under the null, $\sum_{t=1}^nX_{t-1}u_t=O_p(n^{3/2})$ by the Cauchy-Schwartz inequality,
which leads to $\mathcal{I}_3=O_p(n^{-1/2})$.
Thus the proof is complete.

\end{proof}

\begin{proof}[Proof of Theorem 2.1] 
The proof is straightforward using the continuous mapping theorem, Lemma \ref{lem:sn}, and Slutsky's theorem.
\end{proof}

We now prove bootstrap consistency.
The proof can be done using the large-block small-block argument as presented in the proof of Theorem 3.1 in \cite{Shao:2010b}.
Let $L_n=\lf (n/l_n)^{1/2}\rf $ be the length of a large-block and $l_n$ be that of a small-block.
Note that $L_n\to\infty$ and  $l_n=o(L_n)$.
Our goal is to assign points $t\in\{1,2,\ldots,\lf nr\rf\}$ to alternating large and small blocks.
Let $K_n=K_{n,r}=\lf \lf nr\rf(L_n+l_n)^{-1}\rf$ be the number of the large (small) blocks.
Define the $k$th large-block $\mathcal{L}_k=\{j\in\N:(k-1)(L_n+l_n)+1\leq j \leq k(l_n+L_n)-l_n\}$ for $1\leq k \leq K_n$,
and the $k$th small-block $\mathcal{S}_k=\{j\in\N:k(L_n+l_n)-l_n+1\leq j \leq k(l_n+L_n)\}$ for $1\leq k\leq K_n-1$
and $\mathcal{S}_{K_n}=\{j\in\N:K_n(L_n+l_n)-l_n+1\leq j\leq \lf nr \rf \}$.

Let $U_k=\sum_{j\in\mathcal{L}_k}  W_ju_{j}$ and $V_k=\sum_{j\in\mathcal{S}_k}  W_ju_{j}$, $k=1,...,K_n$.
Define $\mathcal{B}_L=\{k:\mathcal{L}_k$ contains a break point $b_j$ for some $j=0,\ldots,\tau\}$
and  $\mathcal{B}_S=\{k:\mathcal{S}_k$ contains a break point $b_j$ for some $j=0,\ldots,\tau\}$.
Notice that there are only finitely many (less than $\tau$) elements in $\mathcal{B}_L$ and $\mathcal{B}_S$.

\begin{lemma}\label{lemma2-1}
Assume (A1)-(A4) and (B1)-(B2). Then
\begin{equation}\label{lemma:sigma}
\sup_{r\in[0,1]}\left|n^{-1}\sum_{k=1}^{K_n}\sum_{j,j'\in\mathcal{L}_k} \cov(u_{j},u_{j'})a\{(j-j')/l_n\}-\int_0^r\sigma^2(s)ds\right|=o(1).
\end{equation}
\end{lemma}

\begin{proof}[Proof of Lemma \ref{lemma2-1}]
Suppose $k\not\in\mathcal{B}_L$.
We shall first show that
\begin{equation}\label{lemma2-1:nBl}
\sup_{r\in[0,1]}\left|n^{-1}\sum_{k\not\in\mathcal{B}_L}\sum_{j,j'\in\mathcal{L}_k}\cov(u_{j},u_{j'})a\{(j-j')/l_n\}
-\int_0^r\sigma^2(s)ds\right|=o(1).
\end{equation}
Recall that $\zeta_s=j$ such that $s\in[b_j,b_{j+1})$ and $\zeta_1=\tau$,
and $c(s;h)=c_{\zeta_s}(s;h)$.
Since $a(\cdot)=0$ outside of its support [-1,1], by Lemma \ref{lemma1-1} (i) and (ii), it follows that
$L_n^{-1}\sum_{j,j'\in\mathcal{L}_k}\cov(u_{j},u_{j'})a\{(j-j')/l_n\}
 =c(k/K_n;0)+O(L_n/n)+2\sum_{h=1}^{l_n}(1-h/L_n)a(h/l_n)\{c(k/K_n;h)+O(L_n/n)\}
 =\sigma^2(k/K_n)-2\sum_{h=1}^{\infty}  d_{h}c(k/K_n;h)+O(l_nL_n/n),$
where $d_{h}=1-(1-h/L_n)a(h/l_n)$ if $0\leq h\leq l_n$ and 1 if $h>l_n$.
By (B2) and Lemma \ref{lemma1-1} (iii),
$\sum_{h=1}^{\infty} d_{h} c(k/K_n;h)\leq Cl_n^{-q}\{k_q+o(1)\}\sum_{h=1}^{\infty} h^q c(k/K_n;h)
+C\overline{a}L_n^{-1}\sum_{h=1}^{\infty}h c(k/K_n;h)
\allowbreak\leq C(l_n^{-q}+L_n^{-1}) =o(1),$
where $\overline{a}=\sup_{s\in[-1,1]}a(s)$ and $C$ is a constant that  does not depend on $k$ or $r$.

Therefore,
\begin{equation}\label{lemma:sigma0}
\sup_{k\not\in\mathcal{B}_L}\left|L_n^{-1}\sum_{j,j'\in\mathcal{L}_k}\cov(u_{j},u_{j'})a\{(j-j')/l_n\}-\sigma^2(k/K_n)\right|
\leq C\{l_n^{-q}+L_n^{-1}\}=o(1)
\end{equation}
so that $\sup_{r\in[0,1]}|n^{-1}\sum_{k\not\in\mathcal{B}_L}\sum_{j,j'\in\mathcal{L}_k}\cov(u_{j},u_{j'})a\{(j-j')/l_n\}
-n^{-1}\sum_{k\not\in\mathcal{B}_L} \sigma^2(k/K_n)L_n|=o(1).
$
Since  $\sup_{r\in[0,1]}|n^{-1}\sum_{k\not\in\mathcal{B}_L}\sigma^2(k/K_n)L_n-
\sum_{k\not\in\mathcal{B}_L}\int_{(k-1)/K_n}^{k/K_n}\sigma^2(s)ds|=o(1)$ by Lemma \ref{lemma1-1} (iv)
and $\sup_{r\in[0,1]}|
\allowbreak
\sum_{k\not\in\mathcal{B}_L} \int_{(k-1)/K_n}^{k/K_n}\sigma^2(s)ds-\int_0^r\sigma^2(s)ds|=o(1)$,  (\ref{lemma2-1:nBl}) is proved.
If $k\in\mathcal{B}_L$,
(A2) implies that
\begin{equation}\label{lemma2-1:Bl}
\left|n^{-1}\sum_{k\in\mathcal{B}_L}\sum_{j,j'\in\mathcal{L}_k}\cov(u_{j},u_{j'})a\{(j-j')/l_n\}\right|=O(n^{-1}\tau L_n^2)=o(1).
\end{equation}
Thus (\ref{lemma:sigma}) follows from  (\ref{lemma2-1:nBl}) and (\ref{lemma2-1:Bl}).
\end{proof}

\begin{lemma}\label{lemma2}
Assume (A1)-(A4) and (B1)-(B2). For a fixed constant $r\in(0,1]$,
\begin{equation}\label{thm2:toprove}
n^{-1/2}\sum_{t=1}^{\lfloor nr\rfloor}W_tu_{t}\cod N\bigg(0,\int_0^r\sigma^2(s)ds\bigg) ~~~~~\mbox{in probability.}
\end{equation}
\end{lemma}

\begin{proof}[Proof of Lemma \ref{lemma2}]
The left-hand side of (\ref{thm2:toprove}) can be decomposed into large- and small-block parts as
$n^{-1/2}\sum_{t=1}^{\lf nr\rf} W_tu_{t}=n^{-1/2}\sum_{k=1}^{K_n}U_k+n^{-1/2}\sum_{k=1}^{K_n}V_k.$
Note that $E^*(U_k)=0$ for all $k=1,...,K_n$ and since $W_t$'s are $l_n$-dependent, $U_1,...,U_{K_n}$ are independent random variables conditional on $\mathcal{X}_n$.
The same property holds for $V_1,...,V_{K_n}$.

First it will be shown that the large-block part converges to the limit in (\ref{thm2:toprove}), i.e.,
\begin{equation}\label{thm2:toprove1}
n^{-1/2}\sum_{k=1}^{K_n}U_k~\cod~N\bigg(0,\int_0^r\sigma^2(s)ds\bigg) ~~~~~\mbox{in probability.}
\end{equation}
Using the same argument as in the equation (A.3) in \cite{Shao:2010b} and H$\ddot{\rm o}$lder's inequality, it follows that
\begin{equation}\label{uk}
\sum_{k=1}^{K_n}E^*|U_k|^{4}\leq C
l_n^{2}L_n \sum_{k=1}^{K_n}\sum_{j\in\mathcal{L}_k}|u_{j}|^{4}.
\end{equation}
The argument in \cite{Shao:2010b} applies here because everything is conditional on $\mathcal{X}_n$, and the property of $W_t$ remains the same.
From (A2), $E|u_{j}|^4\leq C$ for $j=1,\ldots,n$, so that $\sum_{k=1}^{K_n}\sum_{j\in\mathcal{L}_k}|u_{j}|^{4}\leq\sum_{j=1}^n|u_{j}|^4=O_p(n)$.
It follows that $\sum_{k=1}^{K_n}E^*|U_k|^{4}=O_p(l_n^2L_nn)=O_p\{(nl_n)^{3/2}\}$.
Since for any $\epsilon>0$,
$ E^*\{U_k^2{\bf 1}(|U_k|>{n}^{1/2}\epsilon)\}\leq {(n^{1/2}\epsilon)}^{-2} E^*\{|U_k|^{4}{\bf 1}(|U_k|>{n}^{1/2}\epsilon)\}
\leq  n^{-1}\epsilon^{-2}E^*|U_k|^{4}$
holds for all $k$, it follows that $n^{-1}\sum_{k=1}^{K_n}E^*\{U_k^2{\bf 1}(|U_k|>n^{1/2}\epsilon)\}=O_p\{{(l_n^3/n)}^{1/2}\}=o_p(1)$.
Then (\ref{thm2:toprove1}) follows from Lemma \ref{lemma2-1}.

Next it will be shown that the contribution from small-blocks $n^{-1/2}\sum_{k=1}^{K_n}V_k$ is negligible, i.e.,
\begin{equation}\label{thm2:toprove2}
n^{-1/2}\sum_{k=1}^{K_n}V_k=o_{p}^*(1).
\end{equation}
For $k\not\in\mathcal{B}_S$, by Lemma \ref{lemma1-1} (i) and (iii),
$E\{E^*(V_k^2)\}=E\big[\sum_{j,j'\in\mathcal{S}_k}u_{j}u_{j'}a\{(j-j')/l_n\}\big] =\sum_{j,j'\in\mathcal{S}_k}\cov(u_{j},u_{j'})a\{(j-j')/l_n\}
\leq l_n\sum_{h=0}^{l_n-1}\{c(k/K_n;h)+C(l_n/n)\}a(h/l_n)\leq Cl_n.$
For $k=K_n$, using a similar argument,
$E\{E^*(V_{K_n}^2)\}\leq CL_n.$
For $k\in\mathcal{B}_S$ and $k\neq K_n$, $E\{E^*(V_{K_n}^2)\}\leq C\tau l_n^2.$
Since $\tau<\infty$, it follows that $\sum_{k=1}^{K_n}  E\{E^*(V_k^2)\}\leq C(K_nl_n+l_n^2+L_n)=o(n)$.
Then (\ref{thm2:toprove2}) follows from the Markov inequality, independence of $V_k$'s, and linearity of expectation.
The proof is completed in view of (\ref{thm2:toprove1}) and (\ref{thm2:toprove2}).
\end{proof}

The following two lemmas are used in the proof of Theorem 3.1, 

\begin{lemma}\label{lemma3}
Assume (A1)-(A4)  and (B1)-(B2).
Then for
$0< r_1 <r_2\leq1$ and $n\geq n_0$ for some positive integer $n_0$, conditional on the data $\mathcal{X}_n$,
\begin{equation}\label{tightness}
 E^*\bigg|n^{-1/2}\sum_{t=\lf nr_1 \rf+1}^{\lf nr_2 \rf}W_tu_{t}\bigg|^{4}
\leq \overline{C}(\mathcal{X}_n)\big\{(r_2-r_1)^2+n^{-p_1}(r_2-r_1)\big\},
\end{equation}
for some $p_1>0$,  $\overline{C}(\mathcal{X}_n)$ that does not depend on $r_1$ or $r_2$, and $\overline{C}(\mathcal{X}_n)=O_p(1)$.
Furthermore,
\begin{equation}\label{thm3:toprove3}
n^{-1/2}\sum_{t=1}^{\lf nr\rf}W_tu_{t}\Rightarrow B_\sigma(r)~~~~~\mbox{in~probability.}
\end{equation}
\end{lemma}

\begin{proof}[Proof of Lemma \ref{lemma3}]
First (\ref{tightness}) will be proved using
 the large-block small-block argument.
Recall that $U_k=\sum_{j\in\mathcal{L}_k}  W_ju_{j}$ and $V_k=\sum_{j\in\mathcal{S}_k}  W_ju_{j}$ for $k=1,...,K_n$, $L_n=\lf(n/l_n)^{1/2}\rf$,
and $K_{n,r}=O(\lf \lf nr\rf (L_n+l_n)^{-1}\rf)$.
Let $K_1=K_{n,r_1}$ and $K_2=K_{n,r_2}$ for convenience.
Define $p_2=(1-3\kappa)/2>0$ and $p_3=\kappa q$, where $\kappa$ and $q$ are from (B1) and (B2), respectively.
Define $p_1=\min(p_2,p_3)$.
By the Cr-inequality,
$$E^*\bigg|\sum_{t=\lf nr_1 \rf+1}^{\lf nr_2 \rf}W_tu_{t}\bigg|^{4}=E^*\bigg|\sum_{k=K_{1}+1}^{K_{2}}U_k+\sum_{k=K_{1}+1}^{K_{2}}V_k\bigg|^{4}
\leq 2^3\left(E^*\bigg|\sum_{k=K_{1}+1}^{K_{2}}U_k\bigg|^{4}+E^*\bigg|\sum_{k=K_{1}+1}^{K_{2}}V_k\bigg|^{4}\right).$$
Since $U_k$ and $V_k$ are independent conditional on the data and have mean 0,
$$E^*\bigg|\sum_{k=K_{1}+1}^{K_{2}}U_k\bigg|^{4}=\sum_{k=K_{1}+1}^{K_{2}}E^*(U_k^4)+\sum_{k\neq k'}E^*(U_k^2U_{k'}^2)
\leq \sum_{k=K_{1}+1}^{K_{2}}E^*(U_k^4)+\Bigg\{\sum_{k=K_{1}+1}^{K_{2}}E^*(U_k^2)\Bigg\}^2,$$
and similarly for  $V_k$.

For the large-block part, from  (\ref{uk}) and (A2),
\begin{equation}\label{eq1}
n^{-2}\sum_{k=K_{1}+1}^{K_{2}}E^*(U_k^4)\leq n^{-2}Cl_n^2L_n\sum_{k=K_{1}+1}^{K_2}\sum_{j\in\mathcal{L}_k}|u_{j}|^{4}\leq C_1(\mathcal{X}_n)n^{-p_2}(r_2-r_1),
\end{equation}
where $C_1(\mathcal{X}_n)=O_p(1)$.
By (\ref{lemma:sigma}), (\ref{lemma:sigma0}), and (\ref{lemma2-1:Bl}),
for any $0\leq r_1 <r_2\leq 1$,
$E\big|n^{-1} \sum_{k=K_{1}+1}^{K_{2}}E^*(U_k^2)-\int_{r_1}^{r_2}\sigma^2(s)ds\big|\leq C\{l_n^{-q}+{L_n^{-1}}\}
\leq C(n^{-p_3}+n^{-p_2})\leq Cn^{-p_1}$.
Note that the constant $C$ does not depend on $r_1$ or $r_2$.
Therefore,
\begin{equation}\label{eq2}
n^{-2} \left\{\sum_{k=K_{1}+1}^{K_{2}}E^*(U_k^2)\right\}^2\leq C_{2}(\mathcal{X}_n)(r_2-r_1)^2+C_{3}(\mathcal{X}_n)n^{-p_1}(r_2-r_1),
\end{equation}
where $c=\{\sup_{s\in[0,1]}\sigma^2(s)\}^2<\infty$ is a constant and $C_{2}(\mathcal{X}_n)$ and $C_{3}(\mathcal{X}_n)$ are both $O_p(1)$.

For the small block part, note that $K_2-K_1\leq Cn(r_2-r_1)/L_n=C(r_2-r_1) {(nl_n)}^{1/2}$ by the definition of $K_1$, $K_2$, and $L_n$, and
$E^* (V_k^4)=O_p(l_n^4)$ by (A2) and (B1).
Therefore,
\begin{equation}\label{eq3}
n^{-2}\sum_{k=K_{1}+1}^{K_{2}}E^*(V_k^4)=O_p\{n^{-2}l_n^4(K_2-K_1)\}=C_4(\mathcal{X}_n)(l_n^3/n)n^{-p_2}(r_2-r_1),
\end{equation}
where $C_4(\mathcal{X}_n)=O_p(1)$.
Also, it has been shown that $n^{-1}\sum_{k=K_{1}+1}^{K_{2}}E^*(V_k^2)=O_p\{(K_2-K_1)l_n/n\}=O_p(1)n^{-p_2}(r_2-r_1),$ which implies that
\begin{equation}\label{eq4}
\left\{n^{-1}\sum_{k=K_{1}+1}^{K_{2}}E^*(V_k^2)\right\}^2=C_5(\mathcal{X}_n)n^{-2p_2}(r_2-r_1)^2,
\end{equation}
where $C_5(\mathcal{X}_n)=O_p(1)$.
It is worth noting that $C_j(\mathcal{X}_n)$, $j=1,\ldots,5$ in (\ref{eq1}), (\ref{eq2}), (\ref{eq3}), and (\ref{eq4}), does not depend on $r_1$ or $r_2$.
Therefore an upper bound for the left-hand side of (\ref{tightness}) is
$$2^3\left[\big\{C_2(\mathcal{X}_n)+C_5(\mathcal{X}_n)n^{-2p_2}\big\}(r_2-r_1)^2+\big\{C_1(\mathcal{X}_n)+C_3(\mathcal{X}_n)+C_4(\mathcal{X}_n)(l_n^3/n)\big\}n^{-p_1}(r_2-r_1)\right],$$
so that (\ref{tightness}) holds for large enough $n$ with
$\overline{C}(\mathcal{X}_n)=2^3\max\{C_2(\mathcal{X}_n),C_1(\mathcal{X}_n)+C_3(\mathcal{X}_n)\}+1$.

For (\ref{thm3:toprove3}), the finite-dimensional convergence,
\begin{equation*}
\left(n^{-1/2}\sum_{t=1}^{\lf nr_1\rf}W_tu_{t},\ldots,n^{-1/2}\sum_{t=1}^{\lf nr_k\rf}W_tu_{t}\right)
\cod \left\{\int_0^{r_1}\sigma(s)dB(s),\ldots,\int_0^{r_k}\sigma(s)dB(s)\right\}
\end{equation*}
in probability for any $k\in\field{N}$ and $r_1,\ldots,r_k$, follows from a similar argument presented in Lemma \ref{lemma2} and the Cram\'{e}r-Wold device.
The tightness follows from (\ref{tightness}) and the argument of Theorem 2.1 in \cite{Shao:Yu:1996}.
This completes the proof for (\ref{thm3:toprove3}).
\end{proof}

\begin{lemma}\label{lem1:new}
Under the conditions (A1)-(A4) and (B1)-(B2),
$$n^{-1/2}\sum_{t=1}^{\lf nr \rf}X_{t-1}W_t(\rho-\widehat{\rho}_n)\Rightarrow 0~~~~~~\mbox{in probability}$$
under the local alternatives $\rho=1+c/n$, $c\leq 0$.
\end{lemma}

\begin{proof}[Proof of Lemma \ref{lem1:new}]
The proof follows once the following two statements are established:
\begin{equation}\label{lem1:new:eq1}
\left|n^{-1/2}(\rho-\widehat{\rho}_n)\sum_{t=1}^{\lf nr\rf}X_{t-1}W_t\right|=o_p^*(1)~~~~~~\mbox{for any}~r\in[0,1]
\end{equation}
and
\begin{equation}\label{lem1:new:eq2}
E^*\left|n^{-1/2}(\rho-\widehat{\rho}_n)\sum_{t=\lf nr_1\rf+1}^{\lf nr_2\rf}X_{t-1}W_t\right|^4\leq \overline{C}(\mathcal{X}_n)\{(r_2-r_1)^2+n^{-p_1}(r_2-r_1)\},
\end{equation}
where $p_1>0$, $\overline{C}(\mathcal{X}_n)$ is a constant that does not depend on $r_1$ or $r_2$ such that $\overline{C}(\mathcal{X}_n)=O_p(1)$.
Note that $n(\widehat{\rho}_n-\rho)=(n^{-1}\sum_{t=1}^nX_{t-1}u_t)/(n^{-2}\sum_{t=1}^nX_{t-1}^2)=O_p(1)$ under the local alternatives by Lemma \ref{lem:sn} and the continuous mapping theorem.

Equation (\ref{lem1:new:eq1}) holds trivially if $r=0$.
For any fixed $r\in(0,1]$,
by Chebyshev's inequality,
$P^*(|\sum_{t=1}^{\lf nr\rf}X_{t-1}W_t|>\lambda)\leq \lambda^{-2} E^*|\sum_{t=1}^{ \lf nr\rf}X_{t-1}
\allowbreak W_t|^2=C\lambda^{-2} \sum_{t=1}^{\lf nr\rf}\sum_{h=0}^{l_n}X_{t-1}X_{t+h-1}a(h/l_n)$
for any $\lambda>0$.
Observe that $E|X_{t-1}X_{t+h-1}|\leq ||X_{t-1}||_2||X_{t+h-1}||_2\leq C(t+h)$ by the Cauchy-Schwarz inequality and Lemma \ref{lem:xt2}.
For any $\delta>0$, by letting $\lambda=n^{3/2}\delta$, it follows that
$E\{P^*(|n^{-3/2}\allowbreak\sum_{t=1}^{\lf nr\rf}X_{t-1}W_t|> \delta)\}\leq Cn^{-3}\delta^{-2}\sum_{t=1}^n\sum_{h=0}^{l_n}(t+h)
\leq Cn^{-3}\delta^{-2}(n^2l_n)=O(n^{-1}l_n)=o(1)$.
Thus (\ref{lem1:new:eq1}) is established.

Equation (\ref{lem1:new:eq2}) can be shown using the large- and small- block argument.
Define indices for large and small blocks $\mathcal{S}_k$ and $\mathcal{L}_k$ as before.
Decompose $\sum_{t=1}^{\lf nr\rf}X_{t-1}W_t=\sum_{k=1}^{K_{n,r}}\mathbf{U}_k+\sum_{k=1}^{K_{n,r}}\mathbf{V}_k$ into large and small blocks.
Recall that $K_{n,r}=\lf \lf nr\rf (L_n+l_n)^{-1}\rf$ is the number of large and small blocks,  $L_n=\lf (n/l)^{1/2}\rf$ is the length of the large block,
and $l_n\asymp Cn^{\kappa}$ with $\kappa\in(0,1/3)$.
Let $K_1=K_{n,r_1}$ and $K_2=K_{n,r_2}$.

Following the same argument used in the proof of (\ref{tightness}),  the upper bounds of
$\sum_{k=K_1+1}^{K_2}E^*(\mathbf{U}_k^4)$, $\sum_{k=K_1+1}^{K_2}E^*(\mathbf{U}_k^2)$,
$\sum_{k=K_1+1}^{K_2}E^*(\mathbf{V}_k^4)$, and $\sum_{k=K_1+1}^{K_2}E^*(\mathbf{V}_k^2)$ shall be examined.
In the subsequent argument, $C(\mathcal{X}_n)$, $C_1(\mathcal{X}_n)$, $C_2(\mathcal{X}_n)$, $C_3(\mathcal{X}_n)$, and $C_4(\mathcal{X}_n)$ are all $O_p(1)$
and do not depend on $r_2$ or $r_1$.
In particular, $C(\mathcal{X}_n)$ may have different values in different places.

Following the same argument as in (22) or (A.3) in \cite{Shao:2010b},
 $\sum_{k=K_1+1}^{K_2}E^*(\mathbf{U}_k^4)\leq Cl_n^2L_n\sum_{k=K_1+1}^{K_2}\sum_{j\in\mathcal{L}_k}|X_{j-1}|^4
\leq C(\mathcal{X}_n)l_n^2L_n\sum_{j=\lf nr_1\rf+1}^{\lf nr_2\rf} j^2
\allowbreak
\leq C(\mathcal{X}_n)
l_n^2L_n (\lf nr_2\rf^3-\lf nr_1\rf^3)
\leq C(\mathcal{X}_n)l_n^2L_n n^3 (r_2-r_1)$,
where the second inequality is due to Lemma \ref{lem:xt2}.
Since $l_n^2L_n n^{-3}=l^{3/2}n^{-5/2}\asymp Cn^{-(3\kappa+5)/2}$, letting $p_1=(3\kappa+5)/2$,
it follows that
\begin{equation}\label{null:U4}
n^{-6}\sum_{k=K_1+1}^{K_2}E^*(\mathbf{U}_k^4)\leq C_1(\mathcal{X}_n) n^{-p_1}(r_2-r_1).
\end{equation}
By Lemma \ref{lem:xt2},
$E\{E^*(\mathbf{U}_k^2)\}=E\{E^*(\sum_{t\in\mathcal{L}_k} X_{t-1}W_t)^2\}
\leq \sum_{t\in\mathcal{L}_k}\allowbreak\sum_{h=-l}^l
|E(X_{t-1}X_{t-1+h})|a(h/l)
\leq C\sum_{t\in\mathcal{L}_k}\sum_{h=-l}^lt$ so that
$\sum_{k=K_1+1}^{K_2}E\{E^*(\mathbf{U}_k^2)\}\leq Cl_n(\lf nr_2\rf^2-\lf nr_1\rf^2)\leq Cl_nn^2(r_2-r_1)$
and
\begin{equation}\label{null:U2}
n^{-6}\left\{\sum_{k=K_1+1}^{K_2}E^*(\mathbf{U}_k^2)\right\}^2\leq l_n^2n^{-2}C_2(\mathcal{X}_n)(r_2-r_1)^2.
\end{equation}
The same arguments work for small blocks, replacing $\mathbf{U}_k$ in (\ref{null:U4}) and (\ref{null:U2}) with $\mathbf{V}_k$, which complete the proof of (\ref{lem1:new:eq2}).
\end{proof}

We are now ready to prove Theorems 3.1 and 3.2. 
\begin{proof}[Proof of Theorem 3.1] 
Observe that
$n^{-1/2}\sum_{t=1}^{\lf nr \rf}{u}_{t}^*
=n^{-1/2}
\sum_{t=1}^{\lf nr \rf}\widehat{u}_{t}W_t
=n^{-1/2}\sum_{t=1}^{\lf nr \rf}(X_{t}-\widehat{\rho}_nX_{t-1})
\allowbreak
W_t
=n^{-1/2}\sum_{t=1}^{\lf nr \rf}(\rho X_{t-1}+u_{t}-\widehat{\rho}_nX_{t-1})W_t
=\left\{n^{-1/2}\sum_{t=1}^{\lf nr \rf}X_{t-1}W_t\right\}(\rho-\widehat{\rho}_n)+n^{-1/2}\sum_{t=1}^{\lf nr \rf}W_tu_{t}
\allowbreak =: \mathcal{I}_{1,r}+\mathcal{I}_{2,r}$.
Noting that $\mathcal{I}_{1,r}\Rightarrow 0$ in probability by Lemma \ref{lem1:new} and
$\mathcal{I}_{2,r}\Rightarrow B_\sigma(r)$ in probability by Lemma \ref{lemma3},
the proof is complete.
\end{proof}

\begin{proof}[Proof of Theorem 3.2] 
We claim that under the local alternatives,
\begin{equation}\label{thm:boot:const:claim0}
n^{-1}\sum_{t=1}^n \{(u_{t}^*)^2-E^*(u_{t}^*)^2\}=o_p^*(1)~~~{\rm and}
\end{equation}
\begin{equation}\label{thm:boot:const:claim}
n^{-1}\sum_{t=1}^n \{E^*(u_{t}^*)^2-u_t^2\}=o_p(1).
\end{equation}
Once (\ref{thm:boot:const:claim0}) and (\ref{thm:boot:const:claim}) are established,
it follows that $n^{-1}\sum_{t=1}^n \{(u_{t}^*)^2-u_t^2\}=o_p^*(1)$.
Then using a similar argument as in the proof of Lemma \ref{lem:xt} (i) and (ii),
Theorem 3.2 
follows from an application of  the continuous mapping theorem, Theorem 3.1, 
and the fact that
$n^{-1}\sum_{t=1}^n {u}_{t}^2\cop\sigma_u^2$, which is due to Lemma \ref{lem:ut} (ii) and the argument in the proof of Theorem 5.1 in \cite{Paparoditis:Politis:2003}.

To prove (\ref{thm:boot:const:claim}),
write
$n^{-1}\sum_{t=1}^n \{E^*(u_{t}^*)^2-u_t^2\}=n^{-1}\sum_{t=1}^n (\widehat{u}_t^2-u_t^2)
=n^{-1}\sum_{t=1}^n [\{u_t
\allowbreak
+(\rho-\widehat{\rho}_n)X_{t-1}\}^2-u_t^2]
=(\rho-\widehat{\rho}_n)^2n^{-1}\sum X_{t-1}^2
+2(\rho-\widehat{\rho}_n)n^{-1}\sum X_{t-1}u_t
=:\mathcal{I}_1+\mathcal{I}_2$.
Lemma \ref{lem:sn} implies that  $\mathcal{I}_k=O_p(n^{-1})$ for all $k=1,2$ under the local alternatives.

Now we shall prove (\ref{thm:boot:const:claim0}).
Observe that
$\sum_{t=1}^n \{(u_{t}^*)^2-E^*(u_{t}^*)^2\}=\sum_{t=1}^n\widehat{u}_t^2(W_t^2-1)$.
For any $\delta>0$,
$
P^*\{|\sum_{t=1}^n\widehat{u}_t^2(W_t^2-1)|>n\delta\}
\leq (n\delta)^{-2}E^*\{\sum_{t=1}^n\widehat{u}_t^2(W_t^2-1)\}^2
\leq (n\delta)^{-2}C \{\sum_{t=1}^n\sum_{h=0}^{l_n}\widehat{u}_t^2\widehat{u}_{t+h}^2\},$
and it remains to show $\sum_{t=1}^n\sum_{h=0}^{l_n}\widehat{u}_t^2\widehat{u}_{t+h}^2=o_p(n^2)$.
Since $\widehat{u}_t=u_t+(\rho-\widehat{\rho}_n)X_{t-1}$, 
$$\begin{array}{lll}
\sum_{t=1}^n\sum_{h=0}^{l_n}\widehat{u}_t^2\widehat{u}_{t+h}^2&=&
\sum_{t=1}^n\sum_{h=0}^{l_n}u_t^2u_{t+h}^2
\\&&+2(\rho-\widehat{\rho}_n)\sum_{t=1}^n\sum_{h=0}^{l_n}\{u_t^2u_{t+h}X_{t+h-1}+u_{t+h}^2u_tX_{t-1}\}
\\&&+(\rho-\widehat{\rho}_n)^2\sum_{t=1}^n\sum_{h=0}^{l_n}\{u_t^2X_{t+h-1}^2+u_{t+h}^2X_{t-1}^2+4u_tu_{t+h}X_{t-1}X_{t+h-1}\}
\\&&+2(\rho-\widehat{\rho}_n)^3\sum_{t=1}^n\sum_{h=0}^{l_n}\{u_{t+h}X_{t-1}^2X_{t+h-1}+u_tX_{t+h-1}^2X_{t-1}\}
\\&&+(\rho-\widehat{\rho}_n)^4\sum_{t=1}^n\sum_{h=0}^{l_n}X_{t-1}^2X_{t+h-1}^2
\\ &=:& \mathcal{I}_1+\mathcal{I}_2+\mathcal{I}_3+\mathcal{I}_4+\mathcal{I}_5.
\end{array}$$
We claim that $\mathcal{I}_j=o_p(n^2)$ for all $j=1,\ldots,5$.

For $\mathcal{I}_1$, since
$\sup_{t_1,t_2} E|u_{t_1}^2u_{t_2}^2|\leq C$,
$\mathcal{I}_1=O_p(nl_n)=o_p(n^2)$.

For $\mathcal{I}_2$, write $\mathcal{I}_2=:\mathcal{I}_{2,1}+\mathcal{I}_{2,2}$.
Observe that
$$
\begin{array}{ll}
\mathcal{I}_{2,1}
&=\sum_{t=1}^nu_t^2\left(\sum_{h=0}^{l_n}X_{t+h-1}u_{t+h}\right)
\leq \left\{\sum_{t=1}^nu_t^4\right\}^{1/2}\left\{\sum_{t=1}^n\left(\sum_{h=0}^{l_n}X_{t+h-1}u_{t+h}\right)^2\right\}^{1/2}
\\&~~~~~~~~~~~~~~~~~~~~~~~~~~= \left\{O_p(n)\right\}^{1/2}\left\{\sum_{t=1}^n\sum_{h=0}^{l_n}\sum_{h'=0}^{l_n}X_{t+h-1}u_{t+h}X_{t+h'-1}u_{t+h'}\right\}^{1/2}
\end{array}
$$
by H\"{o}lder's inequality.
Since $E|X_{t+h-1}u_{t+h}X_{t+h'-1}u_{t+h'}|\leq ||X_{t+h-1}u_{t+h}||_2||X_{t+h'-1}u_{t+h'}||_2
\leq (||X_{t+h-1}^2||_2
\allowbreak
||u_{t+h}^2||_2||X_{t+h'-1}^2||_2||u_{t+h'}^2||_2)^{1/2}\leq C\{(t+h-1)(t+h'-1)\}^{1/2}$
by H\"{o}lder's inequality, (A2), and Lemma
\ref{lem:xt2},
it follows that
$E(|\sum_{t=1}^n\sum_{h=0}^{l_n}\sum_{h'=0}^{l_n} X_{t+h-1}u_{t+h}X_{t+h'-1}u_{t+h'}|)
\leq C\sum_{t=1}^n\{(t+l_n)^{3/2}-t^{3/2}\}^2=O(l_n^2n^2)$.
Thus $\mathcal{I}_{2,1}=O_p\{n^{1/2}(l_n^2n^2)^{1/2}\}
=O_p(n^{3/2}l_n)=o_p(n^2)$.
Similarly, it can be shown that $\mathcal{I}_{2,2}=o_p(n^2)$,
which leads to $\mathcal{I}_2=o_p(n^{2})$.
The proof for $\mathcal{I}_3$, $\mathcal{I}_4$, and $\mathcal{I}_5$
can be  done
using H\"{o}lder's inequality, (A2), and Lemma
\ref{lem:xt2}
for all summands.
Specifically,
for $\mathcal{I}_3=:\mathcal{I}_{3,1}+\mathcal{I}_{3,2}+\mathcal{I}_{3,3}$,
observe that for any $t_1,t_2,t_3,t_4\in\{1,\ldots,n\}$,
$$\textstyle E|u_{t_1}u_{t_2}X_{t_3}X_{t_4}|\leq ||u_{t_1}u_{t_2}||_2||X_{t_3}X_{t_4}||_2
\leq \{E(u_{t_1}^4)E(u_{t_2}^4)E(X_{t_3}^4)E(X_{t_4}^4)\}^{1/4}
\leq Cn.$$
Thus $\mathcal{I}_3=O_p(n^{-2}nnl_n)=o_p(n)$.
For $\mathcal{I}_4=:\mathcal{I}_{4,1}+\mathcal{I}_{4,2}$,
observe that for any $t_1,t_2,t_3\in\{1,\ldots,n\}$,
$$\begin{array}{lll}
E|u_{t_1}X_{t_2}^2X_{t_3}| &\leq & ||u_{t_1}||_4||X_{t_2}^2X_{t_3}||_{4/3}
\leq C\{E(X_{t_2}^{8/3}X_{t_3}^{4/3})\}^{3/4} \leq C(||X_{t_2}^{8/3}||_{3/2}||X_{t_3}^{4/3}||_3)^{3/4}
\\&\leq& C\{(EX_{t_2}^4)^{2/3}(EX_{t_3}^4)^{1/3}\}^{3/4}
=C(EX_{t_2}^4)^{1/2}(EX_{t_3}^4)^{1/4}
\leq Ct_2t_3\leq Cn^{2}.
\end{array}$$
Thus
$\mathcal{I}_4=O_p(n^{-3}n^{2}nl_n)=o_p(n)$.
For $\mathcal{I}_5$,
notice that
$E(X_{t-1}^2X_{t+h-1}^2)\leq
||X_{t-1}^2||_2||X_{t+h-1}^2||_2
=\{E(X_{t-1}^4)E(X_{t+h-1}^4)\}^{1/2}
\leq C(t-1)(t+h-1)
\leq Cn^2$.
Thus $\mathcal{I}_5=O_p(n^{-4}n^2nl_n)=o_p(n^2)$,
which completes the proof.
\end{proof}

\section{The Choice of $l$ and the Minimum Volatility Method}\label{append:MV}
In this section, we shall investigate the effect of the choice of $l$ on the finite sample behavior of the DWB and RDWB methods.
A data-driven approach, the minimum volatility (MV) method, is first proposed.
The deterministic choice of $l$, as suggested in Section 4 of the paper, is compared to the MV method. 

The idea behind the MV method is similar in spirit to that in \cite{Politis:Romano:Wolf:1999}.
The rationale behind the MV method is that the approximation of the limiting distribution should be stable if the bandwidth parameter $l$ is in an appropriate range.
We shall propose the following MV algorithm in the context of finding the optimal bandwidth parameter for the DWB method.

\begin{algorithm} {\rm [The Minimum Volatility (MV) Method]
\begin{enumerate}
\item Choose some candidates $l_1,\ldots,l_k$.
\item For each $l_i$ ($i=1,\ldots,k$), generate the bootstrap sample $y_{t,n}^{*(i)}$ ($t=1,\ldots,n$) and calculate $\mathbf{T}_n^{(1,i)}$
\item Repeat $B$ times so that we have $(\mathbf{T}_n^{*(1,i)},\ldots,\mathbf{T}_n^{*(B,i)})$ for each $l_i$.
\item Let $D_i$ be the empirical distribution function of $(\mathbf{T}_n^{*(1,i)},\ldots,\mathbf{T}_n^{*(B,i)})$, i.e.,
$D_i(x)=B^{-1}\sum_{b=1}^B{\bf 1}(\mathbf{T}_n^{*(b,i)}
\allowbreak
\leq x)$.
For $i=1,\ldots,k-1$, calculate the Kolmogorov-Smirnov distance between $D_{i}$ and $D_{i+1}$,
$H_i=\sup_{x\in\R}|D_i(x)-D_{i+1}(x)|.$
\item The optimal $l$ is $l_{\widehat{i}}$, where $\widehat{i}={\rm argmin}_{i=1,\ldots,k-1} H_i$.
\end{enumerate}
}
\end{algorithm}
The MV procedure  above is described for the $\mathbf{T}_n$ statistics and DWB for simplicity.
The same method can be applied to $\mathbf{t}_n$ and  RDWB as well.
Note that the MV choice of $l$ depends on the data $\{X_{t,n}\}$.
Tables \ref{table:MA} and \ref{table:AR}  present the details of how the choice of $l$ affects the empirical size,
along with the  average of the chosen $l$ for selected DGPs.
Here, the candidates are $l=1,\ldots,\lfloor 12(n/100)^{1/4}\rfloor$.
Thus, the maximum value of $l$ that is considered equals 13 if $n = 100$, and 17 if $n = 400$.
Although the MV method may not necessarily choose  a theoretically optimal $l$, it seems to provide a reasonable practical guidance  as long as the range of the candidates for $l$
is appropriate.

On the other hand, the MV method is computationally costly, with the computational time proportional to the number of candidate bandwidths we include and the number of bootstrap replications. 
Tables \ref{table:MA} and \ref{table:AR}  indicate that the empirical rejection rates for  DWB and RDWB are not too sensitive to the choice of $l$, as long as $l$ is not too small. We propose to use the middle value, $l=\lfloor 6(n/100)^{1/4}\rfloor$, as a computationally efficient practical alternative.
Table \ref{table:RR1} further compares this deterministic choice with the MV method for RDWB, which is recommended in the paper for its finite sample performance.
It seems that the two choices of $l$ are comparable in almost all DGPs for  RDWB.
This behavior is observed not just for the size but also for the power in our unreported simulations.
Therefore, we shall recommend  RDWB with the aforementioned deterministic choice of $l$.

\begin{table}[ht]
\centering\footnotesize\caption{\footnotesize Empirical sizes for DWB and RDWB with various choices of $l$ for selected DGPs, matching with the four panels in Figures 1 and 2. The table represents MA models with 2000 Monte-Carlo replications, 1000 Bootstrap replications,  and $\rho=1$.
 The $l_{MV}$ rows indicate the average of optimal $l$ chosen by the MV method.
 The MV rows indicates the empirical sizes with DWB and RDWB using the MB method.
The nominal level is 5\%.}\label{table:MA}

\begin{tabular}{r@{\hskip0.5pt}r|r@{\hskip6pt}r|r@{\hskip6pt}r||r@{\hskip6pt}r|r@{\hskip6pt}r||r@{\hskip6pt}r|r@{\hskip6pt}r||r@{\hskip6pt}r|r@{\hskip6pt}r}
  \hline
&  &\multicolumn{4}{|c||}{$({\rm MA}_{4,1})$}  & \multicolumn{4}{|c||}{$({\rm MA}_{2,1})$}&\multicolumn{4}{|c||}{$({\rm MA}_{1,3})$}  & \multicolumn{4}{|c}{$({\rm MA}_{6,3})$}\\\hline
& &\multicolumn{2}{|c|}{DWB}  & \multicolumn{2}{|c||}{RDWB}&\multicolumn{2}{|c|}{DWB}  & \multicolumn{2}{|c||}{RDWB}&\multicolumn{2}{|c|}{DWB}  & \multicolumn{2}{|c||}{RDWB}&\multicolumn{2}{|c|}{DWB}  & \multicolumn{2}{|c}{RDWB}\\ \hline
 $n$&$l$& $\mathbf{T}_n$ & $\mathbf{t}_n$ & $\mathbf{T}_n$ & $\mathbf{t}_n$& $\mathbf{T}_n$ & $\mathbf{t}_n$ & $\mathbf{T}_n$ & $\mathbf{t}_n$& $\mathbf{T}_n$ & $\mathbf{t}_n$ & $\mathbf{T}_n$ & $\mathbf{t}_n$& $\mathbf{T}_n$ & $\mathbf{t}_n$& $\mathbf{T}_n$ & $\mathbf{t}_n$\\
  \hline
   \multirow{15}{*}{100} &1 & 1.5 & 1.5 & 4.4 & 4.3 & 93.5 & 93.2 & 20.5 & 20.4 & 0.9 & 0.9 & 4.2 & 4.9 & 43.9 & 40.5 & 12.8 & 12.6 \\ 
 & 2 & 3.3 & 3.2 & 5.1 & 4.7 & 84.9 & 84.7 & 20.6 & 20.9 & 2.9 & 3.2 & 4.8 & 4.9 & 34.0 & 32.5 & 12.3 & 12.0 \\ 
  &3 & 3.5 & 3.6 & 4.3 & 4.4 & 81.2 & 81.0 & 19.9 & 20.2 & 4.0 & 4.2 & 4.4 & 5.0 & 33.6 & 31.6 & 12.2 & 11.8 \\ 
  &4 & 4.0 & 3.7 & 4.9 & 4.5 & 81.0 & 80.8 & 20.0 & 20.0 & 4.5 & 4.7 & 4.6 & 5.2 & 33.9 & 32.0 & 11.9 & 11.8 \\ 
  &5 & 3.9 & 3.9 & 4.7 & 4.5 & 81.0 & 81.2 & 20.0 & 20.1 & 4.4 & 4.5 & 4.8 & 5.3 & 34.5 & 33.4 & 12.2 & 12.0 \\ 
  &6 & 4.0 & 4.0 & 5.0 & 4.9 & 82.2 & 81.9 & 19.8 & 20.0 & 4.3 & 4.8 & 4.9 & 5.3 & 35.4 & 33.7 & 12.1 & 12.2 \\ 
  &7 & 3.8 & 4.0 & 4.9 & 5.0 & 82.7 & 82.9 & 19.1 & 19.5 & 4.6 & 4.8 & 5.1 & 5.4 & 36.6 & 34.8 & 12.4 & 12.5 \\ 
  &8 & 3.8 & 4.0 & 4.9 & 4.8 & 83.9 & 83.8 & 19.1 & 19.4 & 4.5 & 4.5 & 5.3 & 5.5 & 37.6 & 35.6 & 12.5 & 12.4 \\ 
  &9 & 3.9 & 4.0 & 5.0 & 5.1 & 84.9 & 85.0 & 19.2 & 19.4 & 4.0 & 4.5 & 5.5 & 5.8 & 38.7 & 36.3 & 12.8 & 13.0 \\ 
  &10 & 3.9 & 4.2 & 5.1 & 5.0 & 85.2 & 85.0 & 19.2 & 19.2 & 4.2 & 4.4 & 5.5 & 5.8 & 39.8 & 38.0 & 13.0 & 13.2 \\ 
  &11 & 3.8 & 4.0 & 5.1 & 5.5 & 86.1 & 86.1 & 19.5 & 19.8 & 4.2 & 4.2 & 5.6 & 5.8 & 40.9 & 38.5 & 13.1 & 13.2 \\ 
  &12 & 3.8 & 3.8 & 5.5 & 5.2 & 86.9 & 87.0 & 19.1 & 19.2 & 4.3 & 4.3 & 5.6 & 5.9 & 41.8 & 39.6 & 13.2 & 13.3 \\ 
  &13 & 3.6 & 4.2 & 5.5 & 5.2 & 87.3 & 87.4 & 20.0 & 19.9 & 4.0 & 4.2 & 5.8 & 5.7 & 42.6 & 40.8 & 13.4 & 13.2 \\ \hline
 & MV & 4.0 & 4.2 & 5.0 & 4.8 & 83.5 & 83.7 & 19.8 & 19.9 & 4.4 & 4.6 & 5.1 & 5.6 & 37.6 & 36.0 & 12.3 & 12.2 \\ \hline
  &$l_{MV}$ & 7.3 & 7.3 & 6.9 & 6.5 & 7.5 & 7.4 & 6.4 & 6.3 & 8.0 & 7.8 & 7.2 & 7.0 & 8.1 & 7.9 & 7.3 & 7.2 \\ \hline\hline
   \multirow{19}{*}{400}&  1 & 0.8 & 1.0 & 4.9 & 4.8 & 97.2 & 97.0 & 12.8 & 12.8 & 0.5 & 0.6 & 5.3 & 5.5 & 44.9 & 41.9 & 16.2 & 15.3 \\ 
  &2& 2.4 & 2.6 & 4.7 & 4.9 & 89.0 & 88.7 & 13.1 & 13.1 & 2.5 & 2.7 & 5.2 & 5.4 & 31.2 & 29.7 & 14.4 & 13.5 \\ 
  &3 & 3.1 & 3.5 & 4.5 & 4.6 & 82.8 & 82.2 & 12.7 & 12.7 & 3.2 & 3.4 & 5.3 & 5.3 & 26.6 & 25.8 & 13.2 & 12.4 \\ 
  &4 & 3.8 & 3.8 & 4.5 & 4.8 & 79.7 & 79.6 & 12.8 & 12.8 & 3.6 & 3.7 & 5.5 & 5.2 & 25.4 & 24.6 & 12.1 & 11.8 \\ 
  &5 & 3.8 & 3.9 & 4.7 & 4.7 & 78.0 & 78.2 & 12.1 & 12.2 & 3.8 & 4.2 & 5.4 & 5.7 & 25.0 & 24.2 & 11.9 & 11.5 \\ 
  &6 & 3.8 & 4.0 & 4.2 & 4.4 & 77.8 & 77.9 & 12.3 & 12.4 & 4.0 & 4.0 & 5.5 & 5.9 & 25.2 & 24.8 & 11.5 & 10.9 \\ 
  &7 & 3.7 & 4.0 & 4.6 & 4.7 & 78.0 & 77.9 & 12.4 & 12.2 & 4.0 & 4.5 & 5.8 & 5.8 & 25.7 & 24.7 & 11.6 & 11.2 \\ 
  &8 & 3.6 & 4.0 & 4.6 & 4.7 & 78.5 & 78.4 & 12.2 & 12.0 & 4.5 & 4.3 & 5.9 & 5.9 & 25.9 & 25.5 & 11.6 & 11.1 \\ 
  &9 & 3.8 & 4.1 & 4.6 & 4.7 & 78.7 & 78.6 & 11.8 & 11.8 & 4.5 & 4.5 & 5.9 & 5.9 & 26.8 & 26.2 & 11.3 & 11.0 \\ 
  &10 & 3.8 & 4.0 & 4.7 & 4.8 & 78.8 & 79.0 & 11.7 & 11.7 & 4.3 & 4.7 & 6.3 & 6.2 & 27.0 & 26.5 & 11.9 & 11.5 \\ 
  &11 & 4.0 & 4.0 & 4.7 & 4.6 & 79.6 & 79.5 & 11.6 & 11.6 & 4.3 & 4.9 & 6.2 & 6.3 & 27.8 & 27.3 & 12.3 & 11.7 \\ 
  &12 & 3.8 & 3.9 & 4.6 & 4.7 & 80.5 & 80.2 & 11.5 & 11.5 & 4.5 & 4.6 & 6.3 & 6.4 & 28.2 & 27.8 & 12.0 & 11.3 \\ 
  &13 & 3.9 & 4.3 & 4.7 & 4.8 & 81.0 & 80.9 & 11.3 & 11.5 & 4.8 & 4.7 & 6.4 & 6.3 & 28.8 & 28.2 & 12.2 & 11.7 \\ 
  &14 & 3.7 & 4.0 & 4.4 & 4.8 & 81.3 & 81.3 & 11.2 & 11.3 & 4.6 & 5.0 & 6.7 & 6.3 & 30.0 & 29.3 & 12.2 & 11.8 \\ 
  &15 & 3.8 & 4.1 & 4.8 & 4.9 & 82.1 & 82.2 & 11.3 & 11.3 & 4.6 & 4.9 & 6.7 & 6.6 & 30.3 & 29.1 & 12.6 & 11.9 \\ 
  &16 & 3.8 & 4.3 & 4.8 & 4.8 & 82.7 & 82.7 & 11.0 & 11.0 & 4.5 & 4.8 & 6.7 & 6.8 & 30.7 & 30.2 & 12.8 & 12.2 \\ 
  &17 & 3.6 & 4.0 & 4.8 & 4.5 & 83.2 & 83.0 & 11.0 & 10.8 & 4.4 & 4.8 & 6.8 & 6.9 & 31.8 & 30.6 & 12.5 & 12.2 \\ \hline
  &MV & 3.5 & 3.7 & 4.5 & 4.6 & 79.5 & 79.2 & 11.8 & 11.8 & 4.2 & 4.3 & 6.2 & 5.9 & 27.1 & 26.8 & 12.0 & 11.8 \\ \hline
  & $l_{MV}$ & 9.1 & 9.1 & 8.5 & 8.2 & 9.7 & 9.6 & 8.4 & 8.2 & 9.6 & 9.6 & 8.8 & 8.6 & 10.0 & 9.8 & 9.2 & 9.0 \\    \hline
\end{tabular}
\end{table}

\begin{table}[ht]
\centering\footnotesize\caption{\footnotesize Empirical sizes for DWB and RDWB with various choices of $l$ for selected DGPs, matching with the four panels in Figures 1 and 2. 
The table represents AR models with 2000 Monte-Carlo replications, 1000 Bootstrap replications,  and $\rho=1$.
 The $l_{MV}$ rows indicate the average of optimal $l$ chosen by the MV method.
 The MV rows indicates the empirical sizes with DWB and RDWB using the MB method.
The nominal level is 5\%.}\label{table:AR}

\begin{tabular}{r@{\hskip0.5pt}r|r@{\hskip6pt}r|r@{\hskip6pt}r||r@{\hskip6pt}r|r@{\hskip6pt}r||r@{\hskip6pt}r|r@{\hskip6pt}r||r@{\hskip6pt}r|r@{\hskip6pt}r}
  \hline
&  &\multicolumn{4}{|c||}{$({\rm AR}_{4,1})$}  & \multicolumn{4}{|c||}{$({\rm AR}_{2,1})$}&\multicolumn{4}{|c||}{$({\rm AR}_{1,3})$}  & \multicolumn{4}{|c}{$({\rm AR}_{6,3})$}\\\hline
& &\multicolumn{2}{|c|}{DWB}  & \multicolumn{2}{|c||}{RDWB}&\multicolumn{2}{|c|}{DWB}  & \multicolumn{2}{|c||}{RDWB}&\multicolumn{2}{|c|}{DWB}  & \multicolumn{2}{|c||}{RDWB}&\multicolumn{2}{|c|}{DWB}  & \multicolumn{2}{|c}{RDWB}\\ \hline
 $n$&$l$& $\mathbf{T}_n$ & $\mathbf{t}_n$ & $\mathbf{T}_n$ & $\mathbf{t}_n$& $\mathbf{T}_n$ & $\mathbf{t}_n$ & $\mathbf{T}_n$ & $\mathbf{t}_n$& $\mathbf{T}_n$ & $\mathbf{t}_n$ & $\mathbf{T}_n$ & $\mathbf{t}_n$& $\mathbf{T}_n$ & $\mathbf{t}_n$& $\mathbf{T}_n$ & $\mathbf{t}_n$\\
  \hline
   \multirow{15}{*}{100} &1 & 0.2 & 0.2 & 3.5 & 3.5 & 63.6 & 62.4 & 7.8 & 7.6 & 0.0 & 0.0 & 5.5 & 6.9 & 31.6 & 29.7 & 9.4 & 9.3 \\ 
&  2 & 0.6 & 0.5 & 3.5 & 3.5 & 34.9 & 35.0 & 7.7 & 7.7 & 0.0 & 0.2 & 5.7 & 7.0 & 20.7 & 19.4 & 8.7 & 8.6 \\ 
 & 3 & 1.1 & 1.1 & 3.6 & 3.5 & 43.8 & 43.9 & 7.6 & 7.6 & 0.2 & 0.3 & 5.5 & 7.0 & 23.1 & 22.1 & 9.0 & 8.9 \\ 
  &4 & 1.5 & 1.5 & 3.5 & 3.4 & 40.2 & 40.1 & 7.3 & 7.4 & 0.2 & 0.4 & 5.6 & 7.1 & 23.4 & 22.4 & 9.2 & 9.0 \\ 
  &5 & 1.5 & 1.5 & 3.6 & 3.5 & 44.6 & 44.5 & 7.6 & 7.7 & 0.5 & 0.6 & 5.8 & 7.0 & 24.8 & 24.1 & 9.2 & 9.3 \\ 
  &6 & 1.7 & 1.7 & 3.6 & 3.7 & 45.1 & 45.0 & 7.5 & 7.7 & 0.5 & 0.6 & 5.9 & 7.2 & 26.1 & 25.4 & 9.3 & 9.3 \\ 
  &7 & 1.7 & 1.8 & 3.5 & 3.8 & 47.0 & 47.2 & 7.6 & 7.9 & 0.5 & 0.7 & 5.9 & 7.4 & 27.5 & 26.0 & 9.9 & 9.6 \\ 
  &8 & 1.8 & 1.9 & 3.8 & 4.0 & 48.1 & 48.3 & 8.0 & 8.1 & 0.6 & 0.6 & 6.0 & 7.3 & 27.7 & 26.8 & 10.0 & 9.8 \\ 
  &9 & 1.8 & 2.0 & 4.2 & 4.1 & 49.4 & 49.5 & 8.2 & 8.2 & 0.5 & 0.6 & 5.8 & 7.0 & 29.3 & 27.9 & 10.0 & 9.8 \\ 
  &10 & 1.7 & 1.9 & 3.8 & 4.2 & 50.8 & 50.6 & 7.9 & 8.1 & 0.8 & 0.7 & 5.7 & 7.3 & 30.1 & 28.8 & 10.3 & 10.0 \\ 
  &11 & 1.7 & 1.7 & 4.0 & 4.0 & 52.0 & 51.7 & 8.2 & 8.3 & 0.8 & 0.6 & 5.9 & 7.3 & 30.6 & 29.0 & 10.2 & 10.0 \\ 
  &12 & 1.8 & 1.6 & 4.0 & 4.0 & 52.9 & 52.8 & 8.2 & 8.3 & 0.8 & 0.8 & 6.0 & 7.4 & 31.3 & 29.6 & 10.4 & 10.1 \\ 
  &13 & 1.8 & 1.5 & 3.9 & 4.1 & 53.4 & 53.5 & 8.6 & 8.6 & 0.8 & 0.7 & 6.2 & 7.6 & 31.8 & 30.8 & 10.8 & 10.4 \\ \hline
 & MV & 1.7 & 1.8 & 3.8 & 3.6 & 48.9 & 49.1 & 7.9 & 8.0 & 0.8 & 0.6 & 5.9 & 7.3 & 27.6 & 27.0 & 9.8 & 9.4 \\ \hline
  &$l_{MV}$ & 8.0 & 8.0 & 7.3 & 6.7 & 8.7 & 8.5 & 6.9 & 6.7 & 8.6 & 8.4 & 7.0 & 6.9 & 8.3 & 8.0 & 7.2 & 7.2 \\ \hline\hline
     \multirow{19}{*}{400} &1& 0.0 & 0.0 & 6.5 & 5.7 & 65.9 & 65.2 & 5.3 & 5.3 & 0.0 & 0.0 & 3.8 & 4.6 & 31.1 & 29.2 & 10.9 & 10.1 \\ 
  &2 & 0.5 & 0.4 & 5.8 & 5.3 & 24.9 & 24.9 & 5.5 & 5.7 & 0.0 & 0.0 & 3.8 & 4.7 & 14.1 & 13.8 & 8.9 & 8.3 \\ 
  &3 & 0.7 & 0.8 & 5.8 & 5.3 & 37.8 & 37.5 & 5.4 & 5.3 & 0.0 & 0.0 & 4.0 & 4.8 & 16.2 & 15.7 & 9.0 & 8.4 \\ 
  &4 & 0.8 & 0.9 & 5.3 & 5.0 & 28.3 & 28.3 & 5.5 & 5.5 & 0.0 & 0.2 & 4.4 & 4.9 & 14.8 & 14.8 & 8.6 & 8.2 \\ 
  &5 & 1.5 & 1.4 & 5.2 & 4.9 & 33.5 & 33.2 & 5.5 & 5.5 & 0.2 & 0.3 & 4.2 & 4.9 & 16.1 & 15.8 & 9.0 & 8.8 \\ 
  &6 & 1.5 & 1.6 & 5.0 & 5.1 & 30.9 & 31.0 & 5.5 & 5.5 & 0.3 & 0.6 & 4.3 & 5.0 & 16.3 & 16.2 & 8.8 & 8.4 \\ 
  &7 & 1.8 & 1.8 & 4.8 & 4.8 & 33.4 & 33.1 & 5.6 & 5.5 & 0.4 & 0.7 & 4.2 & 4.9 & 16.5 & 16.4 & 9.1 & 8.5 \\ 
  &8 & 2.1 & 2.0 & 5.1 & 4.8 & 32.9 & 33.0 & 5.7 & 5.5 & 0.7 & 1.1 & 4.3 & 5.1 & 17.3 & 16.8 & 8.8 & 8.6 \\ 
  &9 & 2.1 & 2.2 & 4.9 & 4.8 & 34.4 & 34.3 & 5.5 & 5.5 & 1.0 & 1.2 & 4.3 & 4.9 & 17.9 & 17.5 & 9.3 & 8.8 \\ 
  &10 & 2.4 & 2.3 & 5.1 & 4.9 & 34.8 & 34.8 & 5.3 & 5.4 & 1.1 & 1.4 & 4.3 & 5.2 & 18.4 & 18.1 & 9.3 & 8.9 \\ 
  &11& 2.3 & 2.4 & 5.1 & 5.1 & 35.9 & 35.9 & 5.8 & 5.6 & 1.2 & 1.6 & 4.5 & 5.4 & 18.9 & 18.6 & 9.3 & 8.9 \\ 
  &12 & 2.5 & 2.5 & 5.1 & 4.9 & 36.7 & 36.6 & 5.9 & 5.9 & 1.5 & 1.7 & 4.8 & 5.1 & 19.5 & 19.1 & 9.9 & 9.3 \\ 
  &13 & 2.6 & 2.4 & 5.1 & 5.0 & 37.1 & 37.3 & 5.8 & 5.8 & 1.8 & 1.8 & 4.7 & 5.4 & 20.4 & 19.9 & 9.8 & 9.2 \\ 
  &14 & 2.5 & 2.5 & 5.0 & 4.9 & 38.0 & 37.9 & 5.5 & 5.6 & 1.7 & 2.1 & 4.6 & 5.3 & 20.9 & 20.1 & 10.1 & 9.3 \\ 
  &15 & 2.5 & 2.5 & 5.1 & 4.9 & 38.6 & 38.6 & 5.7 & 5.7 & 1.8 & 2.0 & 4.8 & 5.5 & 21.3 & 20.7 & 10.0 & 9.2 \\ 
  &16 & 2.5 & 2.4 & 5.3 & 4.9 & 39.1 & 39.5 & 5.9 & 5.9 & 1.9 & 2.1 & 4.8 & 5.6 & 21.8 & 21.4 & 10.4 & 9.8 \\ 
  &17& 2.5 & 2.6 & 5.0 & 5.2 & 40.3 & 40.0 & 5.8 & 5.8 & 2.1 & 2.1 & 4.8 & 5.7 & 22.4 & 21.8 & 10.4 & 9.8 \\ \hline
  &MV & 2.1 & 2.1 & 5.2 & 4.8 & 36.0 & 35.4 & 5.5 & 5.5 & 1.2 & 1.4 & 4.4 & 4.9 & 18.6 & 18.3 & 9.8 & 8.9 \\ \hline
  &$l_{MV}$ & 10.2 & 10.3 & 9.6 & 8.8 & 11.5 & 11.5 & 8.4 & 8.3 & 10.9 & 10.9 & 8.6 & 8.6 & 10.1 & 9.8 & 9.2 & 9.0 \\ 
   \hline
\end{tabular}
\end{table}

\begin{table}[ht]
\centering\footnotesize\caption{Empirical Sizes for RDWB with $l$ chosen by the MV method and the deterministic choice (DC)  $l=\lf 6(n/100)^{1/4}\rf$, based on
2000 Monte-Carlo replications and 1000 Bootstrap replications under $\rho=1$ for all $({\rm MA}_{i,j})$ and $({\rm AR}_{i,j})$ models.
The nominal level is 5\%.}\label{table:RR1}
 \begin{tabular}{r@{\hskip5pt}r|r@{\hskip5pt}r|r@{\hskip5pt}r|r@{\hskip5pt}r|r@{\hskip5pt}r||r@{\hskip5pt}r|r@{\hskip5pt}r|r@{\hskip5pt}r|r@{\hskip5pt}r}
  \hline
 & &\multicolumn{8}{c||}{MA models}&\multicolumn{8}{c}{AR models}\\\cline{3-18}
&&  \multicolumn{4}{c|}{$n=100$}&\multicolumn{4}{c||}{$n=400$}&  \multicolumn{4}{c|}{$n=100$}&\multicolumn{4}{c}{$n=400$}\\\cline{3-18}
&&\multicolumn{2}{c|}{$\mathbf{T}_n$}&\multicolumn{2}{c|}{$\mathbf{t}_n$}&\multicolumn{2}{c|}{$\mathbf{T}_n$}&\multicolumn{2}{c||}{$\mathbf{t}_n$}&\multicolumn{2}{c|}{$\mathbf{T}_n$}&\multicolumn{2}{c|}{$\mathbf{t}_n$}&\multicolumn{2}{c|}{$\mathbf{T}_n$}&\multicolumn{2}{c}{$\mathbf{t}_n$}\\\cline{3-18}
 $i$&$j$& DC & MV & DC & MV & DC & MV & DC & MV  & DC & MV & DC & MV & DC & MV & DC & MV\\
  \hline
\multirow{5}{*}{1} &1 & 4.7 & 4.9 & 4.7 & 4.7 & 4.5 & 5.1 & 4.4 & 4.9 & 4.0 & 3.6 & 4.2 & 3.8 & 3.7 & 3.6 & 3.9 & 3.6 \\ 
&  2 & 5.0 & 4.9 & 4.9 & 5.0 & 4.5 & 4.5 & 4.5 & 4.4 & 4.6 & 4.6 & 4.7 & 5.0 & 4.0 & 3.8 & 4.1 & 4.0 \\ 
 & 3 & 4.9 & 5.1 & 5.3 & 5.6 & 5.9 & 6.2 & 5.9 & 5.9 & 5.9 & 5.9 & 7.2 & 7.3 & 4.3 & 4.4 & 5.1 & 4.9 \\ 
  &4 & 5.3 & 5.2 & 5.3 & 5.4 & 5.9 & 6.7 & 6.0 & 6.2 & 2.8 & 2.9 & 2.9 & 3.2 & 5.6 & 5.9 & 5.6 & 5.8 \\ 
  &5 & 4.1 & 4.4 & 4.0 & 4.2 & 4.9 & 4.7 & 5.0 & 5.0 & 3.5 & 3.7 & 3.8 & 4.0 & 4.1 & 4.2 & 4.0 & 4.0 \\ \hline
  \multirow{5}{*}{2} &1 & 19.8 & 19.8 & 20.0 & 19.9 & 12.2 & 11.8 & 12.0 & 11.8 & 7.5 & 7.9 & 7.7 & 8.0 & 5.7 & 5.5 & 5.5 & 5.5 \\ 
  &2& 22.3 & 22.3 & 22.5 & 22.4 & 10.6 & 10.2 & 10.7 & 10.2 & 7.3 & 7.8 & 7.2 & 8.0 & 5.5 & 5.3 & 5.5 & 5.3 \\ 
  &3 & 21.4 & 22.0 & 20.6 & 21.3 & 12.4 & 13.0 & 12.4 & 12.8 & 11.0 & 11.6 & 10.6 & 11.2 & 9.8 & 9.7 & 9.8 & 9.4 \\ 
  &4 & 18.4 & 18.9 & 18.3 & 18.8 & 11.1 & 11.2 & 11.2 & 11.3 & 6.6 & 7.6 & 6.7 & 7.4 & 6.8 & 7.2 & 6.8 & 7.2 \\ 
  &5 & 23.4 & 23.4 & 23.5 & 23.6 & 12.8 & 13.3 & 12.8 & 13.4 & 9.0 & 9.2 & 9.0 & 9.2 & 6.0 & 6.2 & 6.0 & 6.4 \\ \hline
  \multirow{5}{*}{3} &1 & 5.0 & 5.2 & 5.0 & 5.1 & 4.7 & 4.5 & 4.5 & 4.5 & 4.0 & 4.3 & 4.3 & 4.8 & 4.8 & 4.9 & 4.9 & 5.1 \\ 
  &2 & 5.2 & 5.5 & 5.0 & 5.2 & 4.8 & 5.0 & 5.1 & 4.9 & 4.1 & 4.0 & 4.1 & 4.6 & 3.5 & 3.6 & 3.6 & 3.6 \\ 
  &3 & 6.1 & 6.2 & 6.3 & 6.6 & 7.4 & 7.4 & 7.1 & 7.6 & 6.8 & 6.9 & 8.9 & 8.9 & 4.9 & 5.0 & 5.7 & 5.7 \\ 
  &4 & 5.5 & 5.9 & 5.8 & 6.6 & 6.2 & 6.5 & 6.3 & 6.5 & 3.3 & 3.8 & 3.5 & 3.6 & 5.5 & 5.7 & 5.3 & 5.5 \\ 
  &5 & 6.0 & 5.7 & 5.8 & 5.9 & 5.2 & 5.9 & 5.2 & 5.2 & 3.5 & 3.4 & 4.0 & 3.9 & 4.9 & 4.8 & 4.7 & 4.6 \\ \hline
  \multirow{5}{*}{4} &1 & 5.0 & 5.0 & 4.9 & 4.8 & 4.6 & 4.5 & 4.7 & 4.6 & 3.6 & 3.8 & 3.7 & 3.6 & 5.1 & 5.2 & 4.8 & 4.8 \\ 
  &2 & 5.2 & 5.2 & 5.0 & 5.2 & 4.0 & 4.2 & 4.2 & 4.2 & 3.9 & 4.0 & 3.9 & 4.2 & 5.4 & 5.1 & 5.2 & 5.6 \\ 
  &3 & 5.3 & 5.6 & 6.6 & 6.5 & 7.7 & 8.4 & 7.5 & 7.4 & 6.0 & 6.2 & 11.3 & 11.7 & 7.4 & 7.5 & 8.6 & 8.8 \\ 
  &4 & 4.8 & 4.5 & 4.9 & 4.8 & 6.2 & 6.2 & 6.2 & 6.2 & 3.6 & 3.6 & 3.6 & 3.5 & 5.4 & 5.8 & 5.1 & 5.2 \\ 
  &5 & 4.7 & 4.9 & 4.2 & 4.3 & 4.9 & 4.7 & 4.7 & 4.8 & 3.6 & 4.0 & 3.8 & 4.0 & 5.4 & 5.8 & 5.1 & 5.5 \\ \hline
  \multirow{5}{*}{5} &1 & 6.2 & 6.2 & 6.4 & 6.6 & 5.7 & 5.7 & 5.2 & 5.8 & 5.2 & 5.5 & 5.3 & 5.5 & 5.5 & 5.5 & 5.0 & 5.5 \\ 
  &2 & 5.1 & 4.9 & 5.2 & 5.0 & 6.2 & 6.5 & 6.5 & 6.7 & 3.9 & 3.8 & 3.7 & 3.8 & 6.0 & 6.5 & 6.3 & 6.7 \\ 
  &3 & 7.3 & 7.4 & 9.2 & 9.2 & 10.1 & 10.3 & 10.0 & 10.3 & 6.9 & 7.0 & 14.1 & 13.9 & 9.2 & 9.6 & 10.5 & 10.9 \\ 
  &4 & 5.5 & 5.4 & 5.5 & 5.3 & 5.7 & 6.0 & 5.9 & 6.0 & 4.5 & 4.9 & 4.2 & 4.7 & 5.8 & 6.0 & 5.5 & 5.5 \\ 
  &5 & 5.3 & 5.9 & 5.5 & 5.5 & 5.1 & 5.1 & 5.1 & 5.3 & 4.2 & 4.5 & 4.3 & 4.5 & 6.2 & 6.4 & 6.4 & 6.6 \\ \hline
  \multirow{5}{*}{6} &1 & 10.0 & 10.4 & 9.5 & 9.8 & 8.0 & 8.5 & 8.1 & 8.2 & 6.3 & 6.2 & 6.2 & 6.3 & 5.5 & 5.8 & 5.5 & 5.7 \\ 
  &2 & 9.8 & 9.6 & 9.9 & 9.5 & 6.8 & 7.2 & 7.0 & 7.0 & 6.5 & 6.6 & 6.5 & 6.5 & 4.8 & 4.6 & 4.8 & 4.8 \\ 
  &3 & 12.1 & 12.3 & 12.2 & 12.2 & 11.6 & 12.0 & 11.1 & 11.8 & 9.3 & 9.8 & 9.3 & 9.4 & 8.8 & 9.8 & 8.6 & 8.9 \\ 
  &4 & 8.6 & 8.4 & 8.6 & 8.9 & 6.7 & 6.6 & 6.7 & 6.6 & 6.4 & 6.8 & 6.4 & 6.7 & 5.9 & 5.8 & 5.8 & 5.9 \\ 
  &5 & 9.8 & 10.0 & 9.7 & 10.1 & 6.7 & 6.9 & 6.7 & 7.0 & 6.2 & 6.3 & 6.5 & 6.3 & 5.1 & 5.0 & 5.0 & 5.0 \\ 
   \hline
\end{tabular}
\end{table}

\section{Power Curves for All DGPs}\label{append:power}
\begin{figure}[h!]\centering
\resizebox{1\textwidth}{!}{\includegraphics{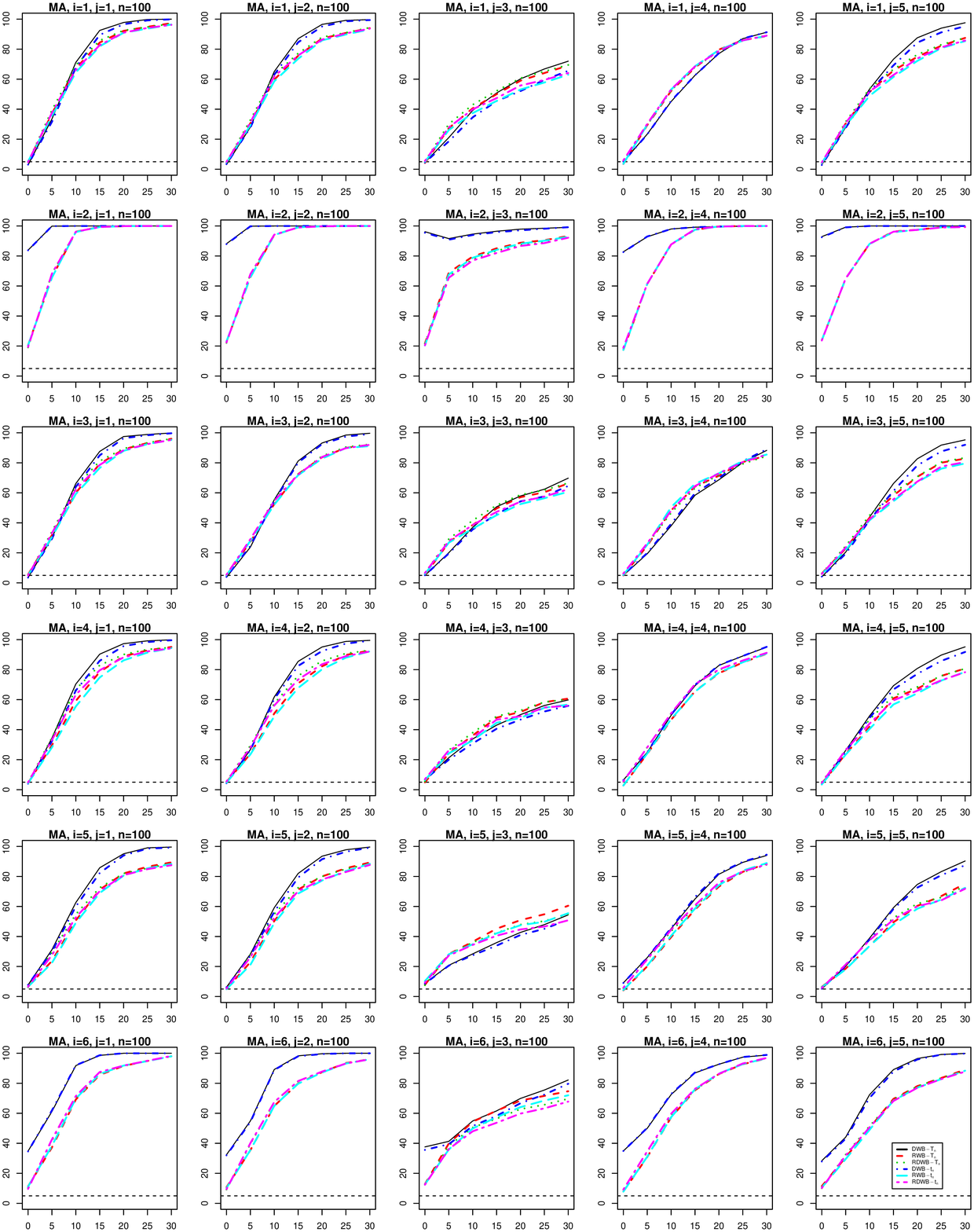}}
\caption{ Rejection frequencies (\%) versus $-c$, where $\rho=1+c/n$ for DWB, RWB, and RDWB unit root tests in MA models.
The sample size is $n=100$ and the nominal level is 5\%.}
\label{MA50-n100}
\end{figure}

\begin{figure}[h!]\centering\scriptsize
\resizebox{1\textwidth}{!}{\includegraphics{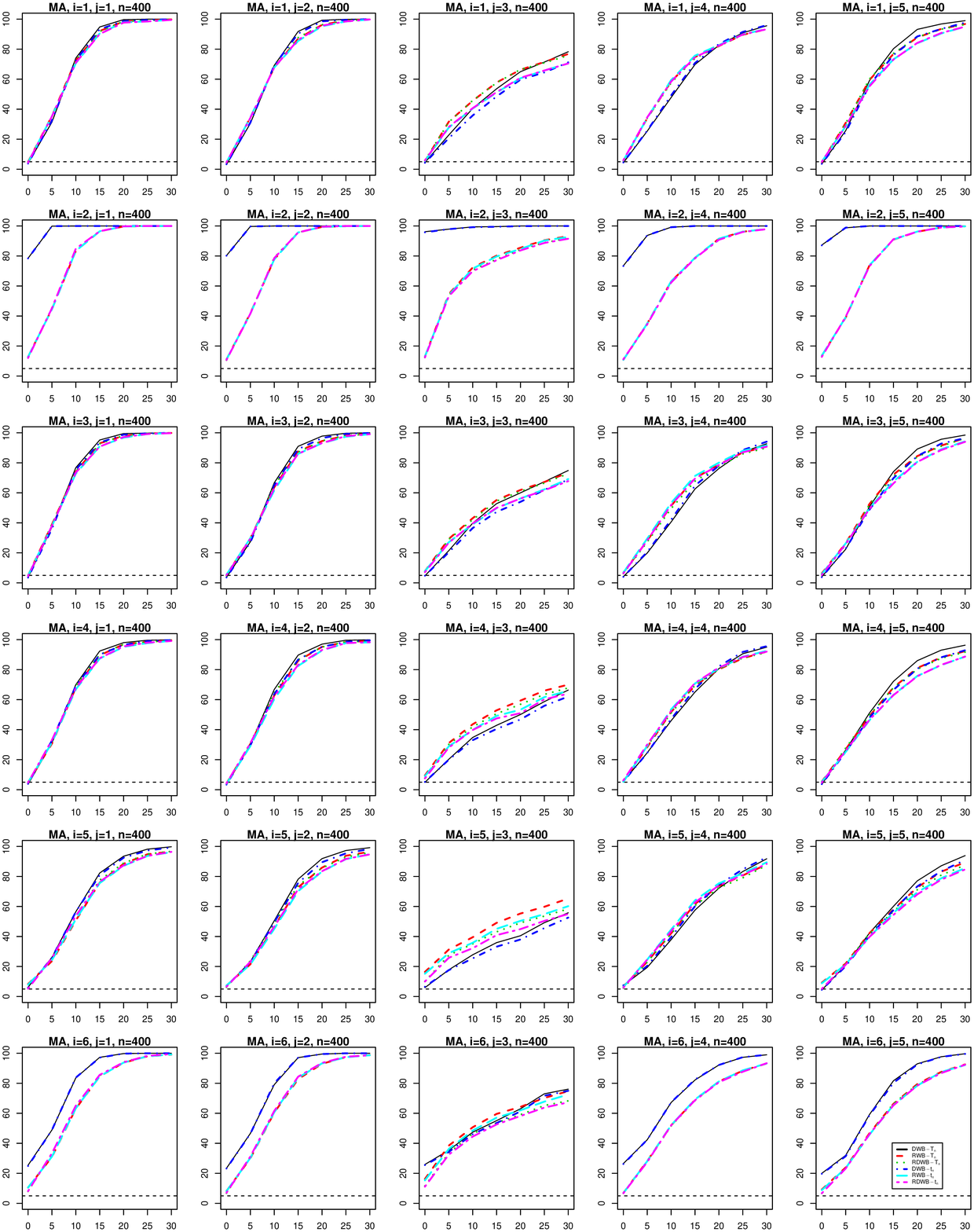}}
\caption{  Rejection frequencies (\%) versus $-c$, where $\rho=1+c/n$ for DWB, RWB, and RDWB unit root tests in MA models.
The sample size is $n=400$ and the nominal level is 5\%.}
\label{MA50-n400}
\end{figure}

\begin{figure}[h!]\centering
\resizebox{1\textwidth}{!}{\includegraphics{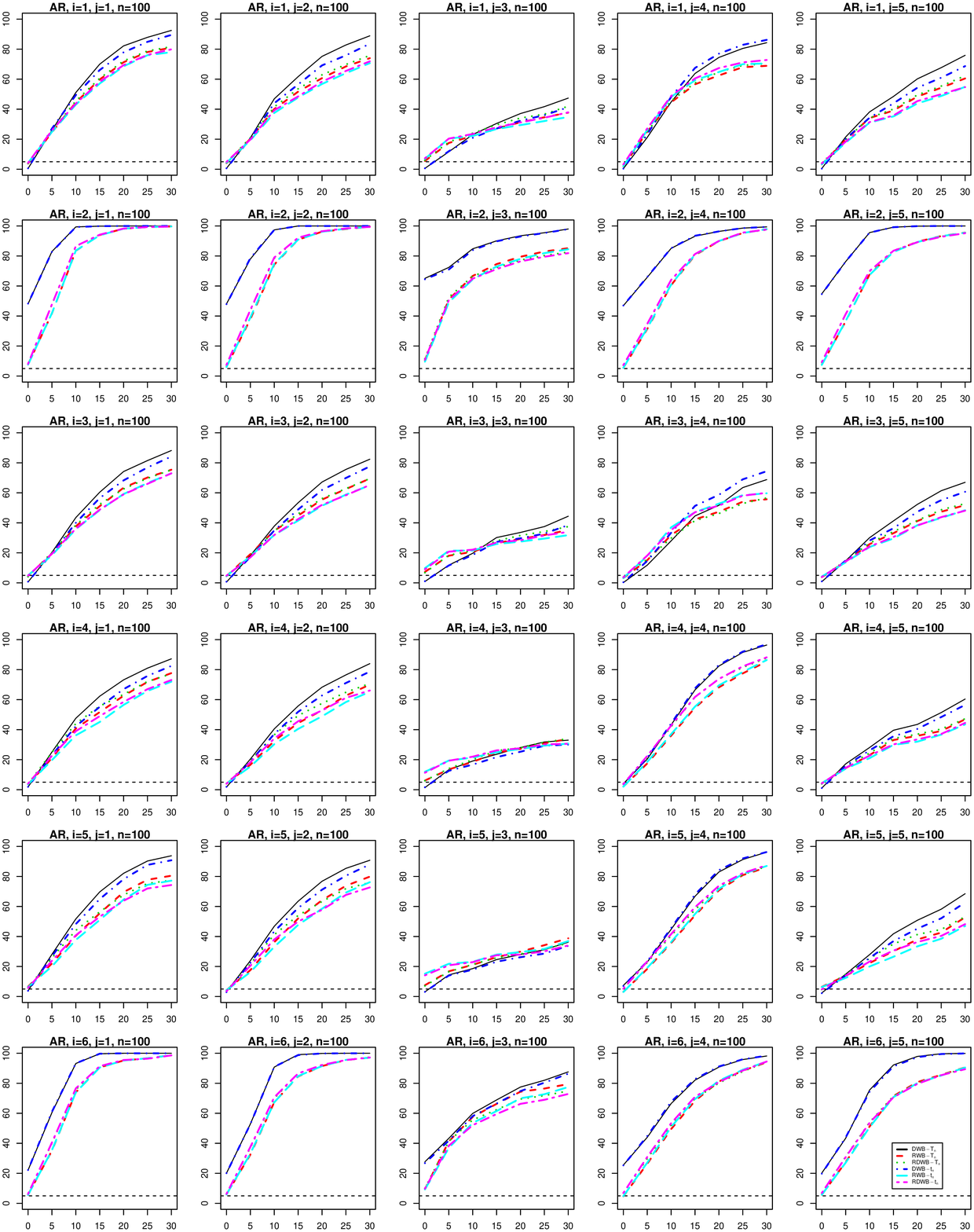}}
\caption{ Rejection frequencies (\%) versus $-c$, where $\rho=1+c/n$ for DWB, RWB, and RDWB unit root tests in AR models.
The sample size is $n=100$ and the nominal level is 5\%.}
\label{AR50-n100}
\end{figure}

\begin{figure}[h!]\centering\scriptsize
\resizebox{1\textwidth}{!}{\includegraphics{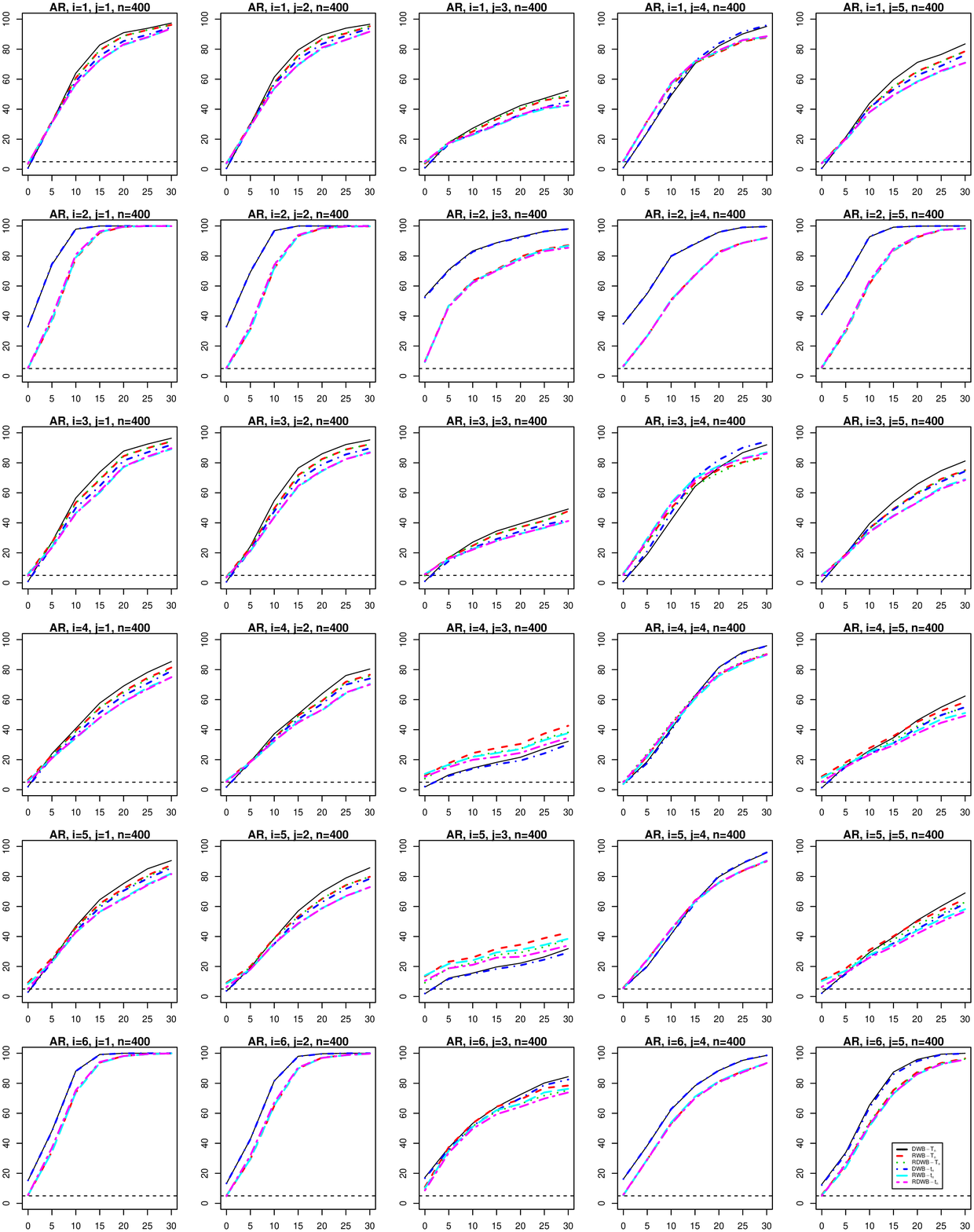}}
\caption{  Rejection frequencies (\%) versus $-c$, where $\rho=1+c/n$ for DWB, RWB, and RDWB unit root tests in AR models.
The sample size is $n=400$ and the nominal level is 5\%.}
\label{AR50-n400}
\end{figure}

Figures \ref{MA50-n100}-\ref{AR50-n400} present all power curves for the DWB, RWB, and RDWB methods and for $\mathbf{T}_n$ and $\mathbf{t}_n$ statistics.

\bibliography{References}{}
\bibliographystyle{chicago}